\documentclass[a4paper,11pt]{article}
\pdfoutput=1 % if your are submitting a pdflatex (i.e. if you have
\usepackage{jheppub} % for details on the use of the package, please
\makeatletter
\DeclareRobustCommand*{\bfseries}{%
  \not@math@alphabet\bfseries\mathbf
  \fontseries\bfdefault\selectfont
  \boldmath
}
\makeatother

\usepackage{calc}
\usepackage{rotating}
\usepackage[english]{babel}
\usepackage[utf8]{inputenc}
\usepackage{graphicx}
\usepackage{subfig}
\usepackage{float}
\usepackage{amsmath}
\usepackage{amssymb}
\usepackage{amsthm}
\usepackage{latexsym}
\usepackage{dcolumn}
\usepackage{hyperref}
\newcommand{\newc}{\newcommand*}

\long\def\begincomment#1\endcomment{%
        \begingroup\sf\baselineskip12pt#1\endgroup}

\newc{\etal}{\textrm{et al.}} 
\newc{\eg}{\textrm{e.g.}} 
\newc{\ie}{\textrm{i.e.}}
\newc{\etc}{\textrm{etc}}
\newc\vs{\textrm{vs.}}
\newc{\cl}{\rm {C.L.}}
\newc{\ev}{\ensuremath{\,\mathrm{eV}}}
\newc{\kev}{\ensuremath{\,\mathrm{keV}}}
\newc{\mev}{\ensuremath{\,\mathrm{MeV}}}
\newc{\gev}{\ensuremath{\,\mathrm{GeV}}}
\newc{\tev}{\ensuremath{\,\mathrm{TeV}}}
\newc{\MeV}{\mev} 
\newc{\TeV}{\tev}
\newc{\invpb}{\ensuremath{/\text{pb}}}
\newc{\invfb}{\ensuremath{/\text{fb}}}
\newc\nb{\ensuremath{\,\mathrm{nb}}} \newc\pb{\ensuremath{\,\mathrm{pb}}} \newc\fb{\ensuremath{\,\mathrm{fb}}}
\newc\pc{\ensuremath{\,\mathrm{pc}}}
\newc\kpc{\ensuremath{\,\mathrm{kpc}}}
\newc\mpc{\ensuremath{\,\mathrm{Mpc}}}
\newc\ps{\ensuremath{\,\mathrm{ps}}} 
\newc\cmeter{\ensuremath{\,\mathrm{cm}}} 
\newc\meter{\ensuremath{\,\mathrm{m}}} 
\newc\kmeter{\ensuremath{\,\mathrm{km}}}
\newc\second{\ensuremath{\,\mathrm{s}}}
\newc\msecond{\ensuremath{\,\mathrm{ms}}}
\newc\nsecond{\ensuremath{\,\mathrm{ns}}}
\newc\psecond{\ensuremath{\,\mathrm{ps}}}
\newc{\chisqmin}{\ensuremath{\chi^2_{\mathrm{min}}}}
\newc{\Delchisq}{\ensuremath{\Delta\chi^2}}
\newc{\chisq}{\ensuremath{\chi^2}}
\newc{\like}{\ensuremath{\mathcal{L}}}
\newc\lsim{\ensuremath{\mathrel{\rlap{\lower4pt\hbox{\hskip1pt$\sim$}}\raise1pt\hbox{$<$}}}}
\newc\gsim{\ensuremath{\mathrel{\rlap{\lower4pt\hbox{\hskip1pt$\sim$}}\raise1pt\hbox{$>$}}}}
\newc{\VEV}[1]{\ensuremath{\langle #1 \rangle}}
\newc{\dl}{\ensuremath{\stackrel{\leftarrow}{D}}}
\newc{\dr}{\ensuremath{\stackrel{\rightarrow}{D}}}
\newc{\bcenter}{\begin{center}}   
\newc{\ecenter}{\end{center}}
\newc{\bfl}{\begin{flushleft}}    
\newc{\efl}{\end{flushleft}}
\newc{\bfr}{\begin{flushright}}   
\newc{\efr}{\end{flushright}}

\newc{\bi}{\begin{itemize}}
\newc{\ei}{\end{itemize}}
\newc{\bed}{\begin{description}}
\newc{\eed}{\end{description}}
\newc{\ben}{\begin{enumerate}}
\newc{\een}{\end{enumerate}}

\newc{\be}{\begin{equation}}
\newc{\ee}{\end{equation}}
\newc{\bea}{\begin{eqnarray}}
\newc{\eea}{\end{eqnarray}}
\newc{\bfle}{\begin{flalign}}
\newc{\efle}{\end{flalign}}
\newc{\ra}{\rightarrow}
\newc{\alphas}{\ensuremath{\alpha_s}}
\newc{\alphatwo}{\ensuremath{\alpha_2}}
\newc{\alphaone}{\ensuremath{\alpha_1}}
\newc{\alphai}[1]{\ensuremath{\alpha_{#1}}}
\newc{\alphaem}{\ensuremath{\alpha_{\mathrm{em}}}}
\newc{\alphaeff}{\ensuremath{\alpha_{\mathrm{eff}}}}
\newc{\sineff}{\ensuremath{\sin \theta_{\mathrm{eff}}}}
\newc{\sinsqeff}{\ensuremath{\sin^2 \theta_{\mathrm{eff}}}}
\newc{\dalphahad}{\ensuremath{\Delta \alpha_{\mathrm{had}}}}
\newc{\yt}{\ensuremath{h_t}} \newc{\yb}{\ensuremath{h_b}} \newc{\ytau}{\ensuremath{h_{\tau}}}
\newc\mz{\ensuremath{M_Z}} 
\newc\mw{\ensuremath{m_W}}
\newc\mZ{\mz}        \newc\mW{\mw}
\newc\mhsm{\ensuremath{ m_{H_{\mathrm{SM}}}}}
\newc{\mtop}{\ensuremath{ m_t}}               \newc{\mtpole}{\ensuremath{ M_t}}
\newc{\mbottom}{\ensuremath{ m_b}} 
\newc{\mtau}{\ensuremath{ m_{\tau}}}
\newc{\mt}{\mtpole}
\newc{\mb}{\mbottom} 
\newc{\rtwogg}{\ensuremath{R_{h_2}(\gamma\gamma)}}
\newc{\rtwozz}{\ensuremath{R_{h_2}(ZZ)}}
\newc{\ronegg}{\ensuremath{R_{h_1}(\gamma\gamma)}}
\newc{\ronezz}{\ensuremath{R_{h_1}(ZZ)}}
\newc{\rsiggg}{\ensuremath{R_{h_\textrm{sig}}(\gamma\gamma)}}
\newc{\rsigzz}{\ensuremath{R_{h_\textrm{sig}}(ZZ)}}
\newc{\llbar}{\ensuremath{\ell\bar{\ell}}}
\newc{\tauptaum}{\ensuremath{ \tau^+\tau^-}}
\newc{\qqbar}{\ensuremath{ q\bar{q}}} \newc{\ppbar}{\ensuremath{ p\bar{p}}}
\newc{\bbbar}{\ensuremath{ b\bar{b}}} \newc{\ttbar}{\ensuremath{ t\bar{t}}}
\newc{\ffbar}{\ensuremath{ f\bar{f}}} \newc{\tautaubar}{\ensuremath{ \tau\bar{\tau}}}

\newc{\mchi}{\ensuremath{m_\neutone}}
\newc{\squark}{\ensuremath{\tilde{q}}}
\newc{\slepton}{\ensuremath{\tilde{l}}}
\newc{\gluino}{\ensuremath{\tilde{g}}} 
\newc{\mgluino}{\ensuremath{{m_{\gluino}}}}
\newc{\wino}{\ensuremath{\tilde{W}}} 
\newc{\mwino}{\ensuremath{{m_{\wino}}}}
\newc{\tone}{\ensuremath{{\tilde{t}_1}}}
\newc{\Hone}{\ensuremath{{\tilde{H}_{1}}}}
\newc{\Htwo}{\ensuremath{{\tilde{H}_{2}}}}
\newc{\Hhtwo}{\ensuremath{{H_{2}}}}
\newc{\qli}{\ensuremath{{\tilde{Q}_{i}}}}
\newc{\uri}{\ensuremath{{\tilde{u}_{i}}}}
\newc{\dri}{\ensuremath{{\tilde{d}_{i}}}}
\newc{\lli}{\ensuremath{{\tilde{L}_{i}}}}
\newc{\eri}{\ensuremath{{\tilde{e}_{i}}}}
\newc{\sthw}{\ensuremath{ \sin\theta_W}}              \newc{\cthw}{\ensuremath{\cos\theta_W}}
\newc{\tanthw}{\ensuremath{ \tan\theta_W}}              \newc{\cotthw}{\ensuremath{\cot\theta_W}}
\newc{\ssqthw}{\ensuremath{\sin^2 \theta_W}}
\newc{\msbar}{\ensuremath{\overline{MS}}} \newc{\drbar}{\ensuremath{\overline{DR}}}
\newc{\mtmtsmmsbar}{\ensuremath{ m_t(m_t)^{\msbar}_{{\mathrm{SM}}}}}
\newc{\mtmtsmdrbar}{\ensuremath{ m_t(m_t)^{\drbar}_{{\mathrm{SM}}}}}
\newc{\mtmtmssmdrbar}{\ensuremath{ m_t(m_t)^{\drbar}_{{\mathrm{SUSY}}}}}
\newc{\mbmbmsbar}{\ensuremath{ m_b(m_b)^{\msbar} }}
\newc{\mbmbsmmsbar}{\ensuremath{ m_b(m_b)^{\msbar}_{{\mathrm{SM}}}}}
\newc{\mbmzsmmsbar}{\ensuremath{ m_b(\mz)^{\msbar}_{{\mathrm{SM}}}}}
\newc{\mbmzsmdrbar}{\ensuremath{ m_b(\mz)^{\drbar}_{{\mathrm{SM}}}}}
\newc{\mbmzmssmdrbar}{\ensuremath{ m_b(\mz)^{\drbar}_{{\mathrm{SUSY}}}}}
\newc{\mtaumzsmmsbar}{\ensuremath{ m_{\tau}(\mz)^{\msbar}_{{\mathrm{SM}}}}}
\newc{\mtaumzsmdrbar}{\ensuremath{ m_{\tau}(\mz)^{\drbar}_{{\mathrm{SM}}}}}
\newc{\mtaumzmssmdrbar}{\ensuremath{ m_{\tau}(\mz)^{\drbar}_{{\mathrm{SUSY}}}}}
\newc{\alphasmzms}{\ensuremath{\alpha_s(M_Z)^{\overline{MS}}}}
\newc{\alphaimzms}[1]{\ensuremath{\alpha_{#1}(M_Z)^{\overline{MS}}}}
\newc{\alphaemmz}{\ensuremath{\alpha_{\mathrm{em}}(M_Z)^{\overline{MS}}}}
\newc{\mzero}{\ensuremath{{m_0}}}
\newc{\mhalf}{\ensuremath{ m_{1/2}}}
\newc{\tanb}{\ensuremath{\tan\beta}}
\newc{\azero}{\ensuremath{ A_0}}
\newc{\signmu}{\ensuremath{\rm{sgn}\,\mu}}
\newc{\atau}{\ensuremath{{A_{\tau}}}}
\newc{\mueff}{\ensuremath{\mu_{\rm{eff}}}}
\newc{\lam}{\ensuremath{{\lambda}}}
\newc{\kap}{\ensuremath{{\kappa}}}
\newc{\alam}{\ensuremath{{A_{\lambda}}}}
\newc{\akap}{\ensuremath{{A_{\kappa}}}}
\newc{\hs}{\ensuremath{ H_s}}      
\newc{\mhs}{\ensuremath{ m_{H_s}}} 
\newc{\mgut}{\ensuremath{ M_{\rm GUT}}}
\newc{\mvl}{\ensuremath{ M_{\rm VL}}}
\newc{\gut}{\ensuremath{{\rm GUT}}}
\newc{\mplanck}{\ensuremath{ M_{\rm P}}}      \newc{\mpl}{\ensuremath{ M_{\rm Pl}}}
\newc{\msusy}{\ensuremath{ M_{\rm SUSY}}}      \newc{\ms}{\ensuremath{ M_{\rm S}}}
 \newc{\hu}{\ensuremath{ H_u}}       \newc{\hd}{\ensuremath{ H_d}}
 \newc{\mhu}{\ensuremath{ m_{H_u}}}       \newc{\mhd}{\ensuremath{ m_{H_d}}}
 \newc{\mhuew}{\ensuremath{ m^{\ast}_{H_u}}}       \newc{\mhdew}{\ensuremath{ m^{\ast}_{H_d}}}
 \newc{\mhuewsq}{\ensuremath{ m^{\ast\, 2}_{H_u}}}       \newc{\mhdewsq}{\ensuremath{ m^{\ast\, 2}_{H_d}}}
 \newc{\mhl}{\ensuremath{m_\hl}} 
 \newc{\mhone}{\ensuremath{m_{h_1}}} 
 \newc{\mhtwo}{\ensuremath{m_{h_2}}} 
 \newc{\mhi}{\ensuremath{m_{\tilde{h}}}} 
 \newc{\mul}{\ensuremath{m_{\tilde{u}_L}}} 
 \newc{\mtone}{\ensuremath{m_{\tilde{t}_1}}} 
 \newc{\ma}{\ensuremath{m_A}} 
 \newc{\mH}{\ensuremath{m_H}} 
 \newc{\maone}{\ensuremath{m_{a_1}}} 
 \newc{\matwo}{\ensuremath{m_{a_2}}}
 \newc{\hone}{\ensuremath{h_1}}
 \newc{\htwo}{\ensuremath{h_2}}
 \newc{\aone}{\ensuremath{a_1}}
 \newc{\atwo}{\ensuremath{a_2}}
 \newc{\mqthree}{\ensuremath{m_{\tilde{Q}_3}^2}}
 \newc{\muthree}{\ensuremath{m_{\tilde{u}_3}^2}}
 \newc{\mqli}{\ensuremath{m_{\tilde{Q}_{i}}}}
 \newc{\muri}{\ensuremath{m_{\tilde{u}_{i}}}}
 \newc{\mdri}{\ensuremath{m_{\tilde{d}_{i}}}}
 \newc{\mlli}{\ensuremath{m_{\tilde{L}_{i}}}}
 \newc{\meri}{\ensuremath{m_{\tilde{e}_{i}}}}
 \newc{\ts}{\ensuremath{T_{SUSY}}}

\newc{\sigsip}{\ensuremath{\sigma^{\rm SI}_{p}}}	\newc{\sigsin}{\ensuremath{\sigma^{\rm SI}_{n}}}
\newc{\sigsdp}{\ensuremath{\sigma^{\rm SD}_{p}}}	\newc{\sigsdn}{\ensuremath{\sigma^{\rm SD}_{n}}}
\newc{\sigsi}{\ensuremath{\sigma^{\rm SI}}}	\newc{\sigsd}{\ensuremath{\sigma^{\rm SD}}}
\newc{\abund}{\ensuremath{ \Omega h^2}}
\newc{\omegadm}{\ensuremath{ \Omega_{{\rm DM}}}}     \newc{\abunddm}{\ensuremath{ \Omega_{{\rm DM}} h^2}} 
\newc{\omegam}{\ensuremath{ \Omega_{{\rm m}}}}       \newc{\abundm}{\ensuremath{ \Omega_{{\rm m}} h^2}}
\newc{\omegab}{\ensuremath{ \Omega_{{\rm b}}}}	\newc{\abundb}{\ensuremath{ \Omega_{{\rm b}} h^2}}
\newc{\omegatot}{\ensuremath{ \Omega_{{\rm TOT}}}}
\newc{\omegacdm}{\ensuremath{ \Omega_{{\rm CDM}}}}   \newc{\abundcdm}{\ensuremath{ \Omega_{{\rm CDM}} h^2}}
\newc{\omegalambda}{\ensuremath{ \Omega_{\Lambda}}} \newc{\abundlambda}{\ensuremath{ \Omega_{\Lambda} h^2}}
\newc{\omegarad}{\ensuremath{ \Omega_{{\rm rad}}}}  \newc{\abundrad}{\ensuremath{ \Omega_{{\rm rad}} h^2}}
\newc{\rhocrit}{\ensuremath{ \rho_{\rm crit}}}
\newc{\rhochi}{\ensuremath{ \rho_{\chi}}}
\newc{\abunchi}{\ensuremath{\Omega_\chi h^2}}
\newc{\abundlsp}{\ensuremath{\Omega_{\rm LSP}h^2}}
\newc{\amu}{\ensuremath{ a_{\mu}}}        \newc{\amususy}{\ensuremath{ a_{\mu}^{\mathrm{SUSY}}}}
\newc{\amuexpt}{\ensuremath{ a_{\mu}^{\mathrm{expt}}}}        \newc{\amusm}{\ensuremath{ a_{\mu}^{\mathrm{SM}}}}
\newc\deltaamu{\ensuremath{\Delta a_{\mu}}} \newc{\deltaamususy}{\ensuremath{\delta a_{\mu}^{\mathrm{SUSY}}}}
\newc\gmtwo{\ensuremath{ (g-2)_{\mu}}} 
\newc{\deltagmtwomususy}{\ensuremath{\delta\left(g-2\right)_{\mu}^{\mathrm{SUSY}}}}
\newc{\deltagmtwomu}{\ensuremath{\delta\left(g-2\right)_{\mu}}}
\newc{\deltagmtwoe}{\ensuremath{\delta\left(g-2\right)_{e}}}
\newc\BR{\ensuremath{\rm BR}}
\newc\bsgamma{\ensuremath{ b\rightarrow s \gamma }}
\newc\bxsgamma{\ensuremath{\overline{B}\rightarrow X_{s}\gamma}}
\newc\brbsgamma{\ensuremath{\BR\left(\bsgamma\right)}}
\newc\brbxsgamma{\ensuremath{\BR\left(\bxsgamma\right)}}
\newc\bsmumu{\ensuremath{B_s\to\mu^+\mu^-}}
\newc\brbsmumu{\ensuremath{\BR\left(B_s\to\mu^+\mu^-\right)}}
\newc\bdmmumu{\ensuremath{\overline{B}_d\to\mu^+\mu^-}}
\newc\bbbarmix{\ensuremath{\overline{B}_s\mbox{-}B_s}}      % B_s mixing
\newc\delmbs{\ensuremath{\Delta M_{B_s}}}
\newc{\butaunu}{\ensuremath{B_u \rightarrow \tau \nu}}
\newc{\brbutaunu}{\ensuremath{\BR\left(B_u \rightarrow \tau \nu\right)}}
\newcommand*{\reftable}[1]{Table~\ref{#1}}         
       
\newcommand*{\reffig}[1]{Fig.~\ref{#1}}
 
        \newcommand*{\refeq}[1]{Eq.~(\ref{#1})}
     \newcommand*{\refsec}[1]{Sec.~\ref{#1}}

\newcommand*{\refeqs}[2]{Eqs.~(\ref{#1})-(\ref{#2})}
\newcommand*{\neutone}{\ensuremath{\tilde{\chi}^0_1}}

\let\oldcite\cite
\renewcommand*{\cite}{~\oldcite}

\newcommand*{\hl}{\ensuremath{h}}

\newcommand*{\tr}{\textrm{Tr}}

\newcommand*{\eps}{\ensuremath{\epsilon}}
\restylefloat{figure}

\title{Naturally small neutrino mass with asymptotic safety and gravitational-wave signatures}

\author{Abhishek Chikkaballi,}
\author{Kamila Kowalska}
\author{and Enrico Maria Sessolo}

\affiliation{National Centre for Nuclear Research,\\
Pasteura 7, 02-093 Warsaw, Poland}

\emailAdd{abhishek.chikkaballiramalingegowda@ncbj.gov.pl}
\emailAdd{kamila.kowalska@ncbj.gov.pl}
\emailAdd{enrico.sessolo@ncbj.gov.pl}

\abstract{We revisit the dynamical generation of an arbitrarily small neutrino Yukawa coupling in the Standard Model with trans-Planckian asymptotic safety and apply the same mechanism to the gauged $B-L$ model. We show that thanks to the presence of additional irrelevant couplings,
the described neutrino-mass generation in the $B-L$ model is potentially more in line with existing theoretical calculations in quantum gravity. 
Interestingly, the model can accommodate, in full naturalness and without extensions, the possibility of purely Dirac, pseudo-Dirac, and Majorana neutrinos with any see-saw scale. We investigate eventual distinctive signatures of these cases in the detection of gravitational waves from first-order phase transitions. We find that, while it is easy to produce a signal observable in new-generation interferometers, its discriminating features are washed out by the strong dependence of the gravitational-wave spectrum on the relevant parameters of the scalar potential.}

\begin{document}
\maketitle
%\flushbottom

\setcounter{footnote}{0}

%%%%%%%%%%%%%%%%%%%%%%%%%%%%%
\section{Introduction\label{sec:intro}}
%%%%%%%%%%%%%%%%%%%%%%%%%%%%%

A large amount 
of atmospheric, reactor, and accelerator data have robustly shown 
that neutrinos have a mass, 
and that their mass is much smaller than the masses of the other fermions 
of the Standard Model~(SM). 
If (Dirac) neutrino masses were generated via the Higgs mechanism, they would require a minuscule Yukawa 
coupling, of the order of $10^{-13}$, lower by several orders of magnitude than the other 
SM Yukawa couplings, which
range between $\sim 10^{-5}$ and $1$. To deal with such uncomfortably and potentially unnaturally small values, numerous 
new physics~(NP) constructions have been developed in recent decades
with the goal of dynamically generating the neutrino mass. 
Perhaps the most famous of those constructions is the see-saw mechanism\cite{Minkowski:1977sc,Gell-Mann:1979vob,Yanagida:1979as,Glashow:1979nm,Mohapatra:1980yp,Schechter:1981cv,Schechter:1980gr}, although radiative models also lend a popular alternative, see, \textit{e.g.}, Refs.\cite{Cai:2017jrq,Klein:2019iws} for reviews. 

Recently\cite{Kowalska:2022ypk} (see also\cite{Eichhorn:2022vgp}), 
some of us proposed yet another way of obtaining dynamically a naturally 
small neutrino mass, which does not have to 
be suppressed by a large Majorana scale, like in the see-saw mechanism, nor is it parameterized by the small spontaneous breaking of lepton-number symmetry, like in the inverse see-saw model\cite{Deppisch:2004fa,Abada:2014vea,Lindner:2014oea}.
In Ref.\cite{Kowalska:2022ypk}, the trans-Planckian renormalization group~(RG) flow of the neutrino Yukawa coupling develops a Gaussian infrared~(IR)-attractive fixed point. The neutrino can naturally be a Dirac particle, 
because its Yukawa coupling will be exponentially suppressed. 
In the SM with the addition of three right-handed neutrinos~(SMRHN) it was shown that such a mechanism is consistent with all low-energy data on neutrino masses and mixing and that it favors the normal, rather than inverted, hierarchical ordering.

The construction of Ref.\cite{Kowalska:2022ypk} finds its motivation in the vast body of work 
pointing to the existence of asymptotically safe quantum gravity. 
Asymptotic safety~(AS) is the property of a 
quantum field theory to develop
fixed points of the RG flow of the action\cite{inbookWS}. Following the development of functional renormalization group (FRG) techniques a few decades ago\cite{WETTERICH199390,Morris:1993qb}, it was shown in several papers that AS can 
arise quite naturally in quantum gravity and provide the key ingredient for the non-perturbative renormalizability of the theory. Fixed points were identified initially for the rescaled Newton coupling and the cosmological constant in the Einstein-Hilbert truncation of the effective action\cite{Reuter:1996cp,Lauscher:2001ya,Reuter:2001ag}, 
and later confirmed in the presence of gravitational operators of increasing mass dimension\cite{Lauscher:2002sq,Litim:2003vp,Codello:2006in,Machado:2007ea,Codello:2008vh,Benedetti:2009rx,Dietz:2012ic,Falls:2013bv,Falls:2014tra}, and of matter-field operators\cite{Oda:2015sma,Hamada:2017rvn,Christiansen:2017cxa}. 

The properties of asymptotically safe quantum gravity may also influence 
particle physics in four space-time dimensions, as not only the gravitational action but the full system of gravity and matter may feature ultraviolet~(UV) fixed points in the energy regime where gravitational interactions become strong\cite{Robinson:2005fj,Pietrykowski:2006xy,Toms:2007sk,Tang:2008ah,Toms:2008dq,Rodigast:2009zj,Zanusso:2009bs,Daum:2009dn,Daum:2010bc,Folkerts:2011jz,Eichhorn:2016esv,Eichhorn:2017eht}. 
A trans-Planckian fixed point may thus induce some
specific boundary conditions for some of the \emph{a priori} 
free couplings of the matter Lagrangian, 
as long as they correspond to \textit{irrelevant} directions in theory space. 
Early ``successes'' of AS applied to particle physics are a gravity-driven solution to the triviality problem in U(1) gauge theories\cite{Harst:2011zx,Christiansen:2017gtg,Eichhorn:2017lry}; a ballpark prediction for the value of the Higgs mass (more precisely, of the quartic coupling of the Higgs potential) obtained a few 
years ahead of its discovery\cite{Shaposhnikov:2009pv} (see also Refs.\cite{Eichhorn:2017als,Kwapisz:2019wrl,Eichhorn:2021tsx}); and the retroactive ``postdiction'' of the top-mass
value\cite{Eichhorn:2017ylw}.

The simple ingredient beneath the construction of Ref.\cite{Kowalska:2022ypk} 
is that the trans-Planckian renormalization group equations~(RGEs) should accommodate a negative critical exponent for the Gaussian fixed point of the neutrino Yukawa coupling. Such a feature should ideally emerge from a first-principle calculation based on the FRG. It turns out, however, that at least in the SMRHN it may not be easy to obtain full consistency between the quantum-gravity calculation and a phenomenologically viable neutrino-mass generation. The reason ultimately lies in an inherent lack of free couplings in the matter theory. 

Let us clarify this point by recalling that the effects of gravity on the SM Yukawa RGEs 
are universal, so that 
they cannot be responsible for any qualitative behavior differentiating one type of Yukawa coupling from the others. In other words, any feature specific to the neutrino Yukawa coupling (but not shared by the others) must be driven by the SM-like terms of the trans-Planckian RGEs, rather than by effects originating in the purely gravitational action. As we shall see in \refsec{sec:mech}, the non-interactive IR-attractive fixed point of the neutrino Yukawa coupling can only be reached in the SMRHN in the presence of an interactive IR-attractive fixed point of the hypercharge gauge coupling $g_Y$. Since the value of the hypercharge coupling is very well measured at low energy, a first-principle calculation based on the FRG would have to produce an extremely precise desired outcome or otherwise spoil 
the entire low-energy phenomenology.

The problem just described 
can be avoided by replacing the dynamical ``pull'' exerted on the RG flow by the IR-attractive hypercharge fixed point with an equivalent effect due to other couplings that are not well-measured yet, trading thus a constraint for a prediction. We consider in this paper perhaps the simplest and most natural extension of the SMRHN, the well-known gauged $B-L$ model\cite{Jenkins:1987ue,Buchmuller:1991ce},  which extends the SM gauge group 
with an abelian U(1)$_{B-L}$ symmetry. The gauge coupling $g_{B-L}$ and kinetic mixing $g_{\epsilon}$ can generate a negative critical exponent for the neutrino Yukawa coupling in the same way as $g_Y$ 
does in the~SMRHN.

Being anomaly-free, the $B-L$ model naturally accommodates the three right-handed neutrino spinor fields. By only featuring gauge symmetries, moreover, it allows one to bypass problems potentially associated with continuous global symmetries in quantum gravity.\footnote{There are indications that asymptotically safe gravity preserves global symmetries, 
at least under all the truncations investigated in the context of the FRG\cite{Eichhorn:2017eht}. An apparent discrepancy with general arguments pointing to the violation of global symmetries in quantum gravity might be resolved in AS by the existence of black hole remnants\cite{Falls:2010he}, which may potentially provide protection against the disappearance of conserved global charges\cite{Aharonov:1987tp,Banks:1992ba}.}  The $B-L$ model also allows for the generation of a Majorana mass term from spontaneous symmetry breaking. Such a feature may seem to blunt the need for a dynamical mechanism alternative to the see-saw. However, we will show that pseudo-Dirac and even fully Dirac neutrino masses can naturally emerge when the $B-L$ model is embedded in~AS. 

Finally, by being endowed with a NP scalar field, the model provides a natural framework for the spontaneous generation of intermediate scales, either directly or via dimensional transmutation with the Coleman-Weinberg mechanism\cite{PhysRevD.7.1888}. 
Assuming the latter applies, it is then interesting
to compute potential 
gravitational-wave~(GW) signatures from first-order phase transitions~(FOPTs)\cite{PhysRevD.45.4514,PhysRevLett.69.2026,Kosowsky:1992vn,Kamionkowski:1993fg} (see also Ref.\cite{Athron:2023xlk} for a recent comprehensive review). 
We investigate them in this work, with the ultimate hope of associating
some of their features to the dynamical generation of
a Majorana or a Dirac neutrino mass in the context of AS.

The paper is organized as follows. In \refsec{sec:FRG} we review some general notions of trans-Planckian AS, which is the framework we adopt for our neutrino-mass generation mechanism. The details of this mechanism are recalled in \refsec{sec:mech}, where we show that it applies equally well to the SMRHN and the gauged $B-L$ model. We also discuss the obstacles one encounters when confronting the former with an asymptotically safe UV completion based on quantum gravity. The trans-Planckian boundary conditions of the $B-L$ model are given in \refsec{sec:FP_BL}, where we also discuss the scalar potential. GW signatures from FOPTs are investigated in \refsec{sec:GW}. Finally, we derive our conclusions in \refsec{sec:summary}. Appendices are dedicated to the explicit form of the RGEs, the explicit form of the thermally corrected effective potential, and a brief review of the GW generation from FOPTs.    

%%%%%%%%%%%%%%%%%%%%%%%%%%%%%%%%%%%%%%%%%%%%%%%%
\section{General notions of asymptotic safety\label{sec:FRG}}
%%%%%%%%%%%%%%%%%%%%%%%%%%%%%%%%%%%%%%%%%%%%%%%%

The scale-dependence of all Lagrangian couplings is encoded in the RG flow. In AS, quantum gravity effects kick in at about the Planck scale, where 
the flow of the gravitational action develops dynamically a fixed point. 
Let us consider a (renormalizable) matter theory with gauge and Yukawa interactions. 
The RGEs receive modifications above the Planck scale that look like
\bea
\frac{d g_i}{d t}&=&\beta_i^{(\textrm{matter})}-f_g\, g_i \label{eq:betag} \\
\frac{d y_j}{d t}&=&\beta_j^{(\textrm{matter})}-f_y\, y_j \label{eq:betay}\,,
\eea
where we indicate the renormalization scale with $t=\ln\mu$\,,
$g_i$ and $y_j$ (with $i,j=1,2,3\dots$) 
are the set of gauge and Yukawa couplings, respectively,
and the original beta functions (without gravity) 
are indicated schematically with $\beta_{i,j}^{(\textrm{matter})}$. 

The trans-Planckian gravitational corrections $f_g$ and $f_y$ are universal, in the sense
that they multiply linearly all matter couplings of the same kind, in agreement
with the expectation that gravity should not distinguish the internal degrees of freedom of the matter theory.
The $f_g$ and $f_y$ coefficients depend on the fixed points of the operators of the gravitational action, and can be computed using the techniques of the FRG. Their computation is subject to extremely large uncertainties, which relate to the choice of truncation in the gravity/matter action, to the selected renormalization scheme, to the gauge-fixing parameters, and other effects\cite{Reuter:2001ag,Lauscher:2002sq,Percacci:2002ie,Percacci:2003jz,Codello:2007bd,Benedetti:2009rx,Narain:2009qa,Dona:2013qba,Falls:2017lst,Falls:2018ylp}. A generic functional form for $f_g$ and $f_y$ is (and is possibly bound to remain) unknown. 
What is established with some confidence 
is that $f_g$ is likely not negative, 
irrespective of the chosen RG scheme\cite{Folkerts:2011jz,Christiansen:2017cxa}. 
This is encouraging, as
$f_g>0$ will preserve asymptotic freedom 
in the non-abelian gauge sector of the SM. 
No fundamental constraints currently 
apply instead to the leading-order gravitational term $f_y$.
The gravitational contribution to the Yukawa coupling was investigated in a set of simplified models\cite{Rodigast:2009zj,Zanusso:2009bs,Oda:2015sma,Eichhorn:2016esv}, 
but no general results and definite conclusions regarding the size and sign of $f_y$ are available (but see Ref.\cite{Pastor-Gutierrez:2022nki} for the most recent determination of $f_y$ in the SM). 

Explicit forms of $f_g$ and $f_y$ exist in the literature. 
To give but one example,
it was found in Refs.\cite{Eichhorn:2017ylw,Eichhorn:2017lry} that,
for a theory with gauge and Yukawa couplings in the matter Lagrangian, an FRG calculation in the Einstein-Hilbert truncation of the gravity action, 
with Litim-type regulator and $\alpha=0,\beta=1$ gauge-fixing choice, yields
\be\label{eq:fgfyFRG}
f_g=\tilde{G}^{\ast}\frac{1-4\tilde{\Lambda}^{\ast}}{4\pi\left(1-2\tilde{\Lambda}^{\ast} \right)^2}\,, \quad \quad 
f_y=-\tilde{G}^{\ast}\frac{96+\tilde{\Lambda}^{\ast}\left(-235+103\tilde{\Lambda}^{\ast}+56\tilde{\Lambda}^{\ast 2}\right)}{12\pi \left(3-10\tilde{\Lambda}^{\ast} +8\tilde{\Lambda}^{\ast 2} \right)^2}\,,
\ee
where $\tilde{\Lambda}=\Lambda/\mu^2$, $\tilde{G}=G\,\mu^2$ are the dimensionless cosmological and Newton constant, which parameterize the Einstein-Hilbert action, 
and we have indicated the trans-Planckian (interactive) fixed-point values with an asterisk. 

A trans-Planckian fixed point of the system of Eqs.~(\ref{eq:betag}) and (\ref{eq:betay}) 
is a set $\{g_i^\ast,y_j^\ast\}$, 
corresponding to a zero of the beta functions: $\beta_{i(j)}^{(\textrm{matter})}(g_i^\ast,y_j^\ast)-f_{g(y)}\,g_i^{\ast}(y_j^{\ast})=0$.
The RGEs of couplings $\{\alpha_k\}\equiv\{g_i,y_j\}$ are then linearized
around the fixed point to derive the stability matrix $M_{ij}$\,, which is defined as
\be\label{stab}
M_{ij}=\partial\beta_i/\partial\alpha_j|_{\{\alpha^{\ast}_k\}}\,.
\ee
Eigenvalues of the stability matrix define the opposite of 
critical exponents $\theta_i$, 
which characterize the power-law evolution of the matter 
couplings in the vicinity of the fixed point. 
If $\theta_i$ is positive, the corresponding eigendirection is dubbed as \textit{relevant} and UV-attractive. All RG trajectories along this direction will asymptotically reach the fixed point and, 
as a consequence, a deviation of a relevant coupling from the fixed point introduces a free parameter in the theory (this freedom can be used to adjust the coupling at the high scale so that it matches 
an eventual measurement at the low scale). If $\theta_i$ is negative, 
the corresponding eigendirection is dubbed as \textit{irrelevant} and IR-attractive. 
There exists in this case only one trajectory that the coupling's flow can follow in its run to the low scale, thus potentially providing a clear prediction for its value at the experimentally accessible scale. Finally, $\theta_i=0$ corresponds to a \textit{marginal} eigendirection. The RG flow along this direction is logarithmically slow and one ought to go beyond the linear approximation provided by the stability matrix 
to decide whether a fixed point is UV-attractive or IR-attractive.

For the purposes of this paper, it is enough to work with $\beta_{i,j}^{(\textrm{matter})}$ at one loop. A thorough quantitative analysis of the uncertainties introduced by neglecting higher-order contributions was performed in Ref.\cite{Kotlarski:2023mmr}. It was shown there that such uncertainties remain very small, especially when considering couplings that are perturbative along the entire RG~flow. 

Note also that at the first order in perturbation theory the parameters of the scalar potential somewhat ``decouple'' from the gauge-Yukawa system, as they can only affect
\refeq{eq:betag} from the third-loop level up, and \refeq{eq:betay} from the second loop.
An expression similar to Eqs.~(\ref{eq:betag}) and (\ref{eq:betay}) potentially applies to the quartic couplings of the scalar potential as well, with the gravitational corrections parameterized as $f_{\lam}$. However, whether gravitational corrections to the running couplings of the scalar potential are multiplicative or not remains a model-dependent issue, as some truncations of the gravitational action can generate additive contributions to the matter beta functions of the scalar potential\cite{Eichhorn:2020kca}. We will come back to the detailed analysis of 
the scalar potential of the $B-L$ model in the context of AS in \refsec{sec:FP_BL}.

We close this section recalling that predictions from trans-Planckian AS in particle physics systems have been investigated in depth in many recent papers. Reference\cite{Eichhorn:2018whv} attempted to extract a gravity-driven prediction of the 
top/bottom mass ratio of the SM. Possible imprints of asymptotically safe quantum gravity in the flavor structure of the SM and, in particular, the Cabibbo-Kobayashi-Maskawa matrix, were sought in Ref.\cite{Alkofer:2020vtb}. An equivalent analysis for the Pontecorvo-Maki-Nakagawa-Sakata (PMNS) matrix elements can be found in Ref.\cite{Kowalska:2022ypk}. The impact of asymptotically safe quantum-gravity calculations on the RGEs of the Majorana mass term was 
investigated in detail in Refs.\cite{DeBrito:2019rrh,Hamada:2020vnf,Domenech:2020yjf}.
Predictions were also extracted for 
several NP models in relation to neutrino masses\cite{Grabowski:2018fjj}, flavor anomalies\cite{Kowalska:2020gie,Chikkaballi:2022urc},
the muon anomalous magnetic moment\cite{Kowalska:2020zve},
the relic abundance of dark matter\cite{Reichert:2019car,Eichhorn:2020kca,Kowalska:2020zve}, baryon number\cite{Boos:2022jvc,Boos:2022pyq}, as well as axion models\cite{deBrito:2021akp} and 
GWs\cite{Eichhorn:2023gat}.

%%%%%%%%%%%%%%%%%%%%%%%%%%%%%%%%%%%%%%%%%%%%%%%%%
\section{Small Yukawa couplings from UV fixed points\label{sec:mech}}
%%%%%%%%%%%%%%%%%%%%%%%%%%%%%%%%%%%%%%%%%%%%%%%%

\subsection{General setup\label{sec:GS}}

In a gauge-Yukawa theory embedded in trans-Planckian AS, it is possible to concoct a dynamical mechanism that makes some Yukawa couplings naturally small\cite{Kowalska:2022ypk,Eichhorn:2022vgp}. If there exists an IR-attractive, Gaussian fixed point for those Yukawa couplings, their flow from a different,  UV-attractive fixed point will asymptotically tend to zero as they approach the Planck scale from above.

We exemplify this pattern by considering 
a simple generic system of matter 
RGEs, comprising 
one (abelian) gauge coupling $g_Y$, and two Yukawa couplings, $y_X$ and $y_Z$. In the deep trans-Planckian regime the system takes the form of Eqs.~(\ref{eq:betag}) and (\ref{eq:betay}),
\bea
\frac{d g_Y}{d t}&=&\frac{b_Y}{16 \pi^2} g_Y^3-f_g\, g_Y\label{eq:gy}\\
\frac{d y_X}{d t}&=&\frac{y_X}{16\pi^2}\left[ \alpha_X\, y_X^2+\alpha_Z y_Z^2-\alpha_Y\, g_Y^2\right]-f_y\, y_X\label{eq:yx} \\
\frac{d y_Z}{d t}&=&\frac{y_Z}{16\pi^2}\left[ \alpha'_X\, y_X^2+\alpha'_Z y_Z^2-\alpha'_Y\, g_Y^2\right]-f_y\, y_Z\,,\label{eq:yz}
\eea
where $b_Y, \alpha^{(\prime)}_X, \alpha^{(\prime)}_Y, \alpha^{(\prime)}_Z \geq 0$ are one-loop 
coefficients. The reader may think of Eqs.~(\ref{eq:gy})-(\ref{eq:yz}) as the system of 
hypercharge/top Yukawa/neutrino Yukawa coupling in the SMRHN, 
after all other couplings have been set to their (UV-attractive) Gaussian fixed point. The 
discussion however is generic and can be applied to any gauge-Yukawa
system\cite{Kowalska:2022ypk}.

Let us set the beta functions to zero and select a solution where $y_Z$ develops a Gaussian fixed point,
$y_Z^{\ast}=0$. We expect such solution to be IR-attractive, which means that  
the corresponding critical exponent $\theta_Z$ should be negative: 
\be\label{eq:irr}
16\, \pi^2 \theta_Z\approx -\left(\alpha_X'\, y_X^{\ast 2}  -\alpha_Y'\,g_Y^{\ast 2}- 16 \pi^2 f_y\right) < 0\,.
\ee
Equation~(\ref{eq:irr}) imposes a condition on the quantum gravity parameter $f_y$\cite{Kowalska:2022ypk}. 

On the other hand, the system (\ref{eq:gy})-(\ref{eq:yz}) 
may also feature a predictive (irrelevant) fixed point $y_X^{\ast}\neq 0$. This means  
that one could follow the unique trajectory of $y_X(t)$ to the low energy and extract its value at the electroweak symmetry-breaking (EWSB) scale. If $y_X$ is then a SM Yukawa coupling, the overall consistency of the AS framework implies that $f_y$ can only assume those specific values 
allowing $y_X^{\ast}$ to be in agreement with a low energy determination of $y_X(t)$. 
In other words, in \refeq{eq:irr} 
one can trade $f_y$, which should emerge from a calculation with the FRG, with 
$y_X^{\ast}$, which should be consistent, within uncertainties, with low energy observations. We get
\be\label{eq:irr1}
16\, \pi^2 \theta_Z\approx -\left(\alpha_Y-\alpha^{\prime}_Y\right)g_Y^{\ast 2}
+\left(\alpha_X-\alpha^{\prime}_X\right)y_X^{\ast 2}.
\ee
If $y_X$ and $y_Z$ represent, respectively, the top quark and neutrino Yukawa couplings in the SMRHN, one finds $\alpha_Y>\alpha'_Y$, $\alpha_X>\alpha^{\prime}_X$ (see, \textit{e.g.}, Appendix A of Ref.\cite{Kowalska:2022ypk}), 
so that \refeq{eq:irr1} can only be negative if $g_Y^{\ast}\neq 0$. 

The dynamical flow of $y_Z(t)$ towards the trans-Planckian IR can finally be extracted by integrating \refeq{eq:yz}.
After replacing $f_y\to (\alpha_X y_X^{\ast 2}-\alpha_Y g_Y^{\ast 2})/16\pi^2$,
$y_Z(t)$ is expressed in terms of $y_X^{\ast}$ and $g_Y^{\ast}$, 
plus an arbitrary constant $\kappa$ setting 
the boundary condition at the Planck scale. One gets
\be\label{eq:analytic}
y_Z(t,\kappa)=\left[\frac{c_Y\, g_Y^{\ast 2}-c_X\, y_X^{\ast 2}}{e^{-\left(c_Y\, g_Y^{\ast 2}-c_X\, y_X^{\ast 2}\right)\left(t/8\pi^2-2 \kappa\right)}+\alpha_Z'}\right]^{1/2},
\ee
where we have defined $c_Y= \alpha_Y-\alpha'_Y$ and $c_X= \alpha_X-\alpha'_X$\,. 
Imposing $\theta_Z<0$ in \refeq{eq:irr1}
implies that $y_Z(t,\kappa)$ is monotonically increasing with $t$ in 
the trans-Planckian regime, and that its value 
at the Planck scale is set by its ``distance'' from $16\pi^2 \kappa$.
As expected, $y_Z$ 
can reach arbitrarily small values without fine tuning, being 
parameterized exclusively by the integration constant $\kappa$.

The mechanism just described applies to any gauge-Yukawa particle physics model embedded in asymptotically safe quantum gravity, as long as the corresponding RGEs take the form of Eqs.~(\ref{eq:gy})-(\ref{eq:yz}). In the SMRHN
this mechanism can give 
rise to a Dirac neutrino
mass without fine tuning after EWSB. Moreover, the asymptotically safe SMRHN 
turns out to 
be consistent with 
all the existing data on mass-squared differences and mixing angles, if the normal ordering of neutrino masses is 
assumed\cite{Kowalska:2022ypk}. 

Some concerns may arise when pondering 
the consistency of this mechanism with quantum-gravity calculations of 
$f_g$ and $f_y$ based on the FRG and this is particularly true in the SMRHN. 
Let us recall that by imposing $g_Y^{\ast}\neq 0$ 
along an irrelevant direction -- as we did below \refeq{eq:irr1} -- 
we imply that the RG flow of $g_Y(t)$, followed from the fixed point down to low energies, 
yields a specific prediction for the hypercharge gauge coupling.  
This requires in turn that only one value of $f_g$ is allowed to 
emerge from the FRG calculation:
\be\label{eq:fg_1l}
f_g \approx \frac{b_Y\, g_Y^{\ast 2}(M_{\textrm{Pl}})}{16\, \pi^2}\,.
\ee

The numerical value of \refeq{eq:fg_1l} ought to be computed very precisely, more precisely than $f_y$
since the uncertainties on the experimental determination 
of $g_Y$ are smaller than those on, \textit{e.g.}, the $\overline{MS}$ value of $y_t(M_t)$ and other Yukawa couplings. 
Even considering that 
FRG calculations are marred by large theoretical uncertainties, it may seem exceedingly constraining
that such a specific outcome ought to emerge from the deep UV construction.  

A simple way out comes from generalizing the dynamical generation of small Yukawa couplings to models less dependent on precisely measured quantities. This allows us to modify the system of Eqs.~(\ref{eq:gy})-(\ref{eq:yz}) in two possible ways. One can either add some extra irrelevant gauge couplings to the 
system, one can add extra Yukawa couplings, or both. Adding additional gauge couplings, $g_{i=1,2,3...}$\,, 
will induce a straightforward modification of \refeq{eq:irr1},
\be\label{eq:blmod}
16\, \pi^2 \theta_Z\approx -\sum_{i,j=1,2,3...}\left(\alpha_{ij}-\alpha^{\prime}_{ij}\right)g_i^{\ast} g_j^{\ast}
+\left(\alpha_X-\alpha^{\prime}_X\right)y_X^{\ast 2},
\ee
with new coefficients $\alpha^{(\prime)}_{ij}$\,. One may thus select a fixed point with a UV attractive $g_Y^{\ast}=0$ to let
the hypercharge gauge coupling effectively become a free parameter of the system. The
results of an FRG computation of $f_g$ will not need to be correlated 
with the $g_Y(M_t)$ value at low energies, and instead will provide some welcome 
predictions for NP couplings that have not been measured yet. 

Alternatively, one may add to Eqs.~(\ref{eq:gy})-(\ref{eq:yz}) the 
beta function of an extra 
Yukawa coupling, $y_W$, 
which will develop an irrelevant interactive fixed point,  $y_W^{\ast}\neq 0$.
The RGE system thus becomes   
\bea
\frac{d g_Y}{d t}&=&\frac{b_Y}{16 \pi^2} g_Y^3-f_g\, g_Y\label{eq:gy1}\\
\frac{d y_X}{d t}&=&\frac{y_X}{16\pi^2}\left[ \alpha_X\, y_X^2+\alpha_Z y_Z^2+\alpha_W y_W^2 -\alpha_Y\, g_Y^2\right]-f_y\, y_X\\
\frac{d y_Z}{d t}&=&\frac{y_Z}{16\pi^2}\left[ \alpha'_X\, y_X^2+\alpha'_Z y_Z^2 +\alpha'_W y_W^2
-\alpha'_Y\, g_Y^2\right]-f_y\, y_Z\\
\frac{d y_W}{d t}&=&\frac{y_W}{16\pi^2}\left[ \alpha''_X\, y_X^2+\alpha''_Z y_Z^2 +\alpha''_W y_W^2
-\alpha''_Y\, g_Y^2\right]-f_y\, y_W\,,\label{eq:yw}
\eea
with new available parameters $\alpha''_X, \alpha''_Y, \alpha''_Z, \alpha''_W > 0$.
Selecting a fixed point with $g_Y^{\ast}=0$, will imply a modification of \refeq{eq:irr1}: 
\be\label{eq:irr2}
16\, \pi^2 \theta_Z\approx -\frac{\alpha''_{W}\left(\alpha_X-\alpha'_X \right)+\alpha_W \left(\alpha'_X-\alpha''_X\right)+\alpha'_W\left(\alpha''_X-\alpha_X \right)}{\alpha_W-\alpha''_W} y_X^{\ast 2}\,.
\ee
Equation~(\ref{eq:irr2}) can be made negative, given appropriate coefficients in \refeq{eq:yw}. 
As before, the results of the FRG calculation of $f_g$ will cease to be correlated 
with known values, and one obtains a prediction for $y_W^{\ast}$ that can be tested in future experiments.

As was anticipated in \refsec{sec:intro}, both solutions are implemented straightforwardly if instead of the SMRHN one embeds the gauged $B-L$ model in trans-Planckian AS. 

%%%%%%%%%%%%%%%%%%%%%%%%%%%%%%%%%%
\subsection{Trans-Planckian features of neutrino mass models}\label{sec:models}

Let us briefly summarize in this subsection the features of the SMRHN and the gauged $B-L$ model, showing that both constructions give rise to equivalent predictions for the neutrino Yukawa coupling.

\paragraph{The SMRHN} The SM 
is enriched with three Weyl spinors, $\nu_{R,i=1,2,3}$, 
that are singlets under the gauge symmetry group. The Yukawa part of the Lagrangian features a set of new Yukawa couplings, 
\be\label{mass:dir}
\mathcal{L}_D=-y_\nu^{ij} \nu_{R,i}\, (H^c)^{\dag} L_j + \textrm{H.c.}\,,
\ee  
where $L_j$, and $H$ are the SM lepton and Higgs boson SU(2)$_L$ doublets, $H^c\equiv i \sigma_2 H^{\ast}$ is the charged conjugate doublet, and a sum over SM generations is implied. The new Yukawa term does not violate lepton-number symmetry and the left-handed neutrino component of $L_i$ 
can be combined with a right-handed counterpart, $\nu^{\dag}_{R,i}$, 
to form three Dirac fermions once the Higgs field develops its vev, $v_H$, upon EWSB. The Dirac mass is generated through the Higgs mechanism as $m_D\sim y_\nu v_H /\sqrt{2}$.

Since the right-handed neutrinos are singlets of the SM gauge group, they may additionally 
acquire a Majorana mass,
\be\label{mass:maj}
\mathcal{L}_M=-\frac{1}{2}M_N^{ij}\nu_{R,i}\,\nu_{R,j}+\textrm{H.c.}\,,
\ee
which violates the conservation of lepton number and can be forbidden if the latter is promoted to being a symmetry of the theory. A possibly more natural alternative to the Dirac-type generation of neutrino masses is thus the see-saw mechanism\cite{Minkowski:1977sc,Gell-Mann:1979vob,Yanagida:1979as,Glashow:1979nm,Mohapatra:1980yp,Schechter:1981cv,Schechter:1980gr}. 
In its simplest formulation, Type-I,  one obtains the well-known neutrino mass matrix by combining \refeq{mass:dir} and \refeq{mass:maj}: 
\be\label{eq:m_mat}
m_\nu=\left(\begin{array}{cc}
0 & m_D^T \\
m_D & M_N
\end{array}\right)\,.
\ee
If the eigenvalues of the Majorana mass matrix $M_N$ are much larger than typical $m_D$ values, the diagonalization of $m_\nu$ leads to three light Majorana neutrinos with mass $\sim y^2_\nu v_H^2/(\sqrt{2}M_N)$, and heavy Majorana neutrinos with mass $\sim M_N$. The full set of gauge and Yukawa 
RGEs of the SMRHN -- including the PMNS matrix elements -- can be found, \textit{e.g.}, in Appendix A of Ref.\cite{Kowalska:2022ypk}.

\begin{figure}[t]
	\centering%
 	    \subfloat[]{%
		\includegraphics[width=0.45\textwidth]{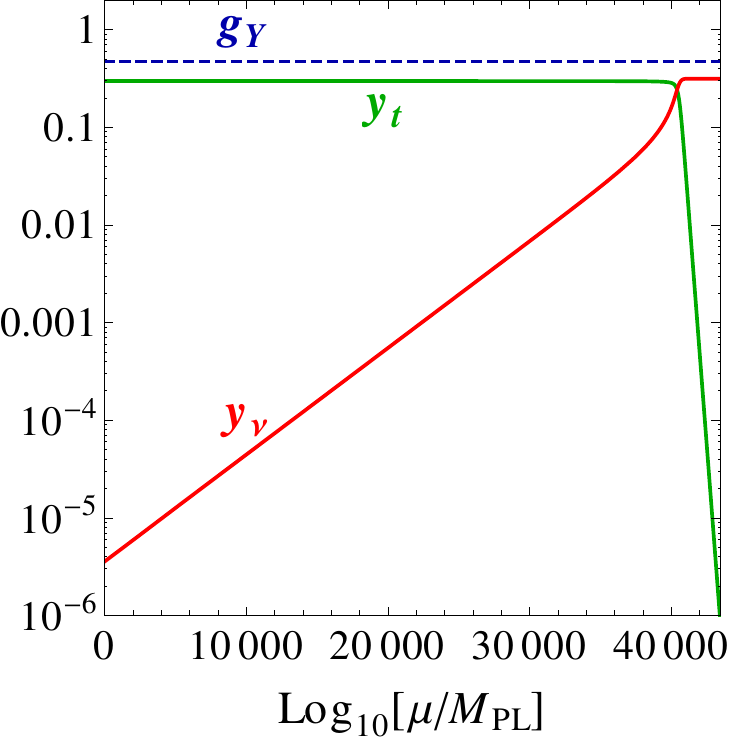}}
  		\hspace{0.8cm}
    	\subfloat[]{%
  		\includegraphics[width=0.45\textwidth]{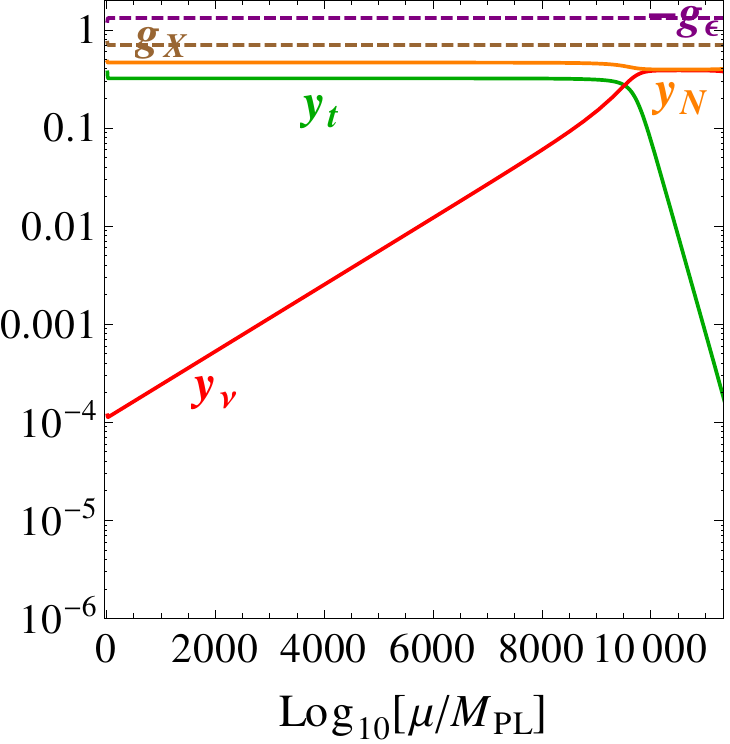}}
\caption{(a) An example of trans-Planckian trajectories for the RGE system composed of $g_Y$ (blue, dashed), $y_t$ (green, solid), and $y_{\nu}$ (red, solid) in the SMRHN with $f_g=0.0097$, $f_y=0.0005$. Neutrino Yukawa coupling $y_{\nu}$ can be made arbitrarily small at the Planck scale by adjusting the integration constant $\kappa$ in \refeq{eq:analytic}. (b)~The same in the gauged $B-L$ model. We set $f_g=0.05$ and $f_y=-0.005$. Besides $y_{\nu}$ and $y_{t}$, we plot the trans-Planckian flow of $g_X$ (brown, dashed), $-g_{\epsilon}$ (purple, dashed) and $y_N$ (orange, solid). Note that $g_Y$ (not shown) is here a relevant parameter.}
\label{fig:mechanism}
\end{figure}

In the context of trans-Planckian AS, the eventual size of $f_y$ effectively selects one or the other type of neutrino mass generation (Dirac or Majorana), thus creating an interesting link with the quantum gravity computation. Referring to \refeq{eq:irr}, let us select $\alpha'_{X=t}=3$, $\alpha'_Y=3/4$ and have $y_{X=t}^{\ast}\neq 0$ 
belong to the range
consistent with the low-energy determination of the top pole mass.\footnote{\label{fn:top} In Ref.\cite{Kowalska:2022ypk} this was loosely determined to be $0.94\lesssim y_t(M_t)\lesssim 1$, leading to $-10^{-4}\lesssim f_y\lesssim 10^{-3}$.} Then, $f_y\lesssim 8.2 \times 10^{-4}$
is required to get the critical exponent $\theta_Z<0$ and one generates an arbitrarily small Yukawa coupling $y_{Z=\nu}$ following its trans-Planckian flow towards an IR-attractive Gaussian fixed point, cf.~\refeq{eq:analytic}. The corresponding 
RG flow is shown in \reffig{fig:mechanism}(a). One might say that a first-principle calculation yielding $f_y$ in the range above will naturally favor the presence of Dirac and/or pseudo-Dirac neutrinos, or even Majorana neutrinos with a low see-saw scale. Note that, in order for this behavior to take place, it is also necessary that  $f_g=0.0097$ to a very good approximation, cf.~\refeq{eq:fg_1l} with 
$g_Y^{\ast}(M_{\textrm{Pl}})=0.47$ (at one loop) and the discussion of 
\refsec{sec:GS}. 

%%%%%%%%%%%%%%%%%%%%%%%%%%%%%%%%%%%%%%%%%%%%%%%%%%%%%%%%%%%%%%%%%%%%%%%
%%%%%%%%%%%%%%%%%%%%%%%%%%%%%%%%%%%%%%%%%%%%%%%%%%%%%%%%%%%%%%%%%%%%%

Conversely,  $f_y> 8.2 \times 10^{-4}$ will lead to $\theta_Z>0$ in \refeq{eq:irr}, \textit{i.e.}, a relevant $y_{\nu}^{\ast}=0$. One also finds an irrelevant (predictive) $y_{\nu}^{\ast}\neq 0$ which is $\mathcal{O}(1)$ in size\cite{Kowalska:2022ypk}, 
thus favoring a Majorana neutrino with a high see-saw scale. 

%%%%%%%%%%%%%%%%%%%%%%%%%%%%%%%%%%%%%%%%%%
\paragraph{Gauged $\boldsymbol{B-L}$ model} 
The SM symmetry is extended by an abelian gauge group U(1)$_{B-L}$, with gauge coupling $g_{B-L}$. 
The particle content is extended with a SM-singlet complex scalar field 
$S$, whose vev $v_S$ spontaneously breaks U(1)$_{B-L}$. The abelian charges of the SM and NP fields can be found, \textit{e.g.}, in Refs.\cite{Coriano:2015sea,Lyonnet:2016xiz}. 

The Yukawa part of the Lagrangian includes \refeq{mass:dir}, whereas \refeq{mass:maj}
is replaced by 
\be\label{mass:BL}
\mathcal{L}_M= - y_{N}^{ij} S\, \nu_{R,i}\,\nu_{R,j}+\textrm{H.c.}\,,
\ee
in terms of a new Yukawa coupling matrix in flavor space, $y_N^{ij}$. The vev $v_S$ generates the Majorana mass upon spontaneous breaking of U(1)$_{B-L}$\,. As we shall see in \refsec{sec:scal_pot}, $v_S\gg v_H$, so that one can work in the basis where the Majorana mass is diagonal. If the boundary conditions from AS require that they are irrelevant, all three diagonal couplings will be equal, $y_N^{ii}\equiv y_N$.

The abelian gauge part of the Lagrangian takes the form
\bea\label{eq:lagE}
\mathcal{L}&\supset& -\frac{1}{4}B_{\mu\nu}B^{\mu\nu}-\frac{1}{4}X_{\mu\nu}X^{\mu\nu}-\frac{\epsilon}{2} B_{\mu\nu}X^{\mu\nu}\nonumber\\
& &\qquad  +i\bar{f}\left(\partial^\mu-i g_Y Q_Y \tilde{B}^\mu-i g_{B-L} Q_{B-L} \tilde{X}^\mu\right)\gamma_\mu f\,,
\eea
where 
$\tilde{B}^\mu$ and $\tilde{X}^\mu$ are the gauge bosons of U(1)$_Y$ and U(1)$_{B-L}$, respectively, 
and $B_{\mu\nu}$ and $X_{\mu\nu}$ are the corresponding field strength tensors. 
Kinetic mixing $\epsilon$ is typically generated between the two abelian groups, as
SM fermions $f$ transform under both symmetry factors with charges $Q_Y$, $Q_{B-L}$.

It is convenient to work in a basis in which the gauge fields are canonically normalized. This is typically achieved by a rotation\cite{Holdom:1985ag,Babu:1996vt}
\be
\begin{pmatrix}
\tilde{B}^\mu\\
\tilde{X}^\mu
\end{pmatrix}
=\begin{pmatrix}
1 && -\epsilon/\sqrt{(1-\epsilon^2)} \\
0 && 1/\sqrt{(1-\epsilon^2)}  
\end{pmatrix}
\begin{pmatrix}
V^\mu\\
D^\mu
\end{pmatrix}\,,
\ee
which parameterizes the gauge interaction vertices of Lagrangian~(\ref{eq:lagE}) in terms of a SM gauge boson $V^{\mu}$
and a NP gauge boson $D^{\mu}$:
\be\label{mix:vertex}
(Q_Y\,,Q_{B-L})
\left(\begin{array}{cc}
g_{Y} & 0 \\ 0  & g_{B-L} 
\end{array} \right)
\left(\begin{array}{c}
\tilde{B}^\mu\\ \tilde{X}^\mu
\end{array} \right)\quad \to \quad
(Q_Y\,,Q_{B-L})
\left(\begin{array}{cc}
g_{Y} & g_{\epsilon} \\ 0  & g_X 
\end{array} \right)
\left(\begin{array}{c}
V^\mu\\ D^\mu
\end{array} \right)\,.
\ee
The elements $g_Y$, $g_X$, and $g_{\epsilon}$ 
are related to the original couplings as
\be\label{eq:gcoups}
g_Y\to g_Y,\quad g_X=\frac{g_{B-L}}{\sqrt{1-{\epsilon}^2}}\,,\quad g_{\epsilon}=-\frac{\epsilon\, g_Y}{\sqrt{1-\epsilon^2}}\,.
\ee

The RGEs of the gauged $B-L$ model are given at one loop in Appendix~\ref{app:rges}. One can refer to \refeq{eq:blmod} and \refeq{eq:irr2}, with appropriate numerical coefficients, to show that the RG flow of the neutrino Yukawa coupling admits a Gaussian irrelevant fixed point driven by the irrelevant fixed points of new gauge and Yukawa couplings, $g_X^{\ast}\neq 0$,  $g_{\epsilon}^{\ast}\neq 0$, and
$y_{N}^{\ast}\neq 0$. 
We show the trans-Planckian flow of the couplings in \reffig{fig:mechanism}(b). Note that the behavior of $y_{\nu}$ mimics exactly the SMRHN case, while $g_Y$ can originate from a relevant fixed point and remain free independently of the value of $f_g$\,.

%%%%%%%%%%%%%%%%%%%%%%%%%%%%%%%%%%%%%%%%%%%%%%%%%%%%%%%%%%%%%%%%%%%%%%%%
\subsection{Possible connections to the FRG \label{sec:par_sp}}

The freedom of adjusting the value of $f_g$ arbitrarily without spoiling the dynamical generation of a small neutrino Yukawa coupling becomes valuable when confronting the phenomenologically viable parameter space with existing computations from the FRG. As was discussed in \refsec{sec:FRG}, calculations of $f_g$ and $f_y$ from first principles are marred by large theory uncertainties\cite{Reuter:2001ag,Lauscher:2002sq,Percacci:2002ie,Percacci:2003jz,Codello:2007bd,Benedetti:2009rx,Narain:2009qa,Dona:2013qba,Falls:2017lst,Falls:2018ylp}. We can nonetheless refer to some of the explicit existing cases in the literature and
make the point that, even considering those uncertainties, the $B-L$ model likely provides a more flexible framework than the SMRHN to match a UV calculation with the low-scale phenomenology.  

\begin{figure}[t]
	\centering%
	    \subfloat[]{%
		\includegraphics[width=0.45\textwidth]{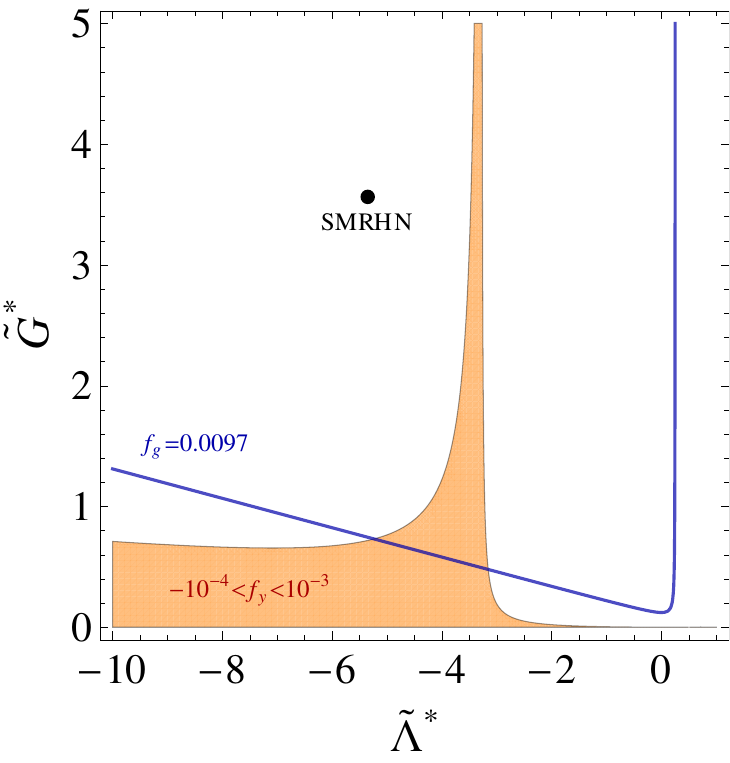}}
		\hspace{0.8cm}
		\subfloat[]{%
		\includegraphics[width=0.45\textwidth]{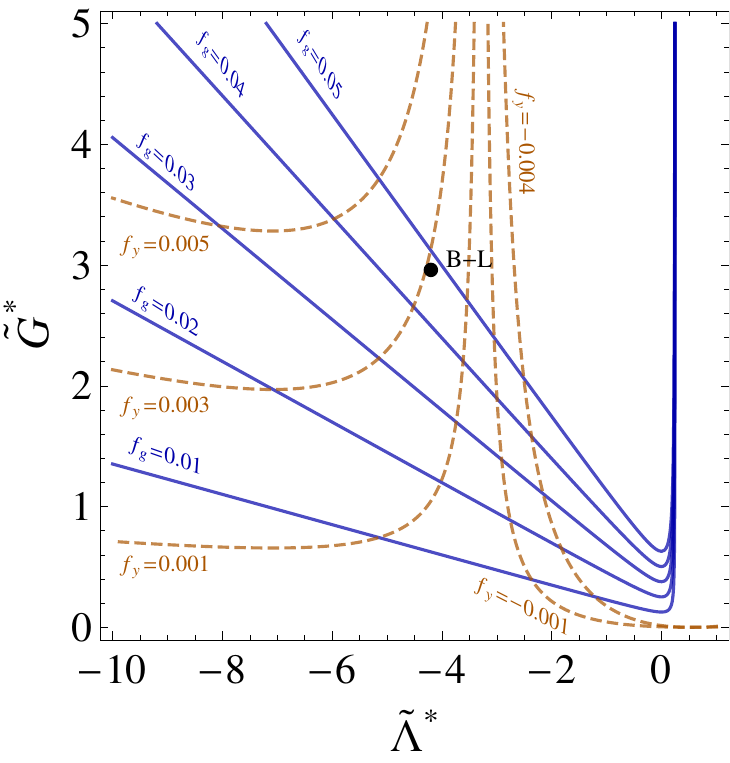}}
\caption{(a) The region of $(\tilde{\Lambda}^{\ast},\tilde{G}^{\ast})$ parameter space consistent with a small neutrino Yukawa coupling in the SMRHN. Blue solid line shows $f_g=0.0097$, which is required for generating an irrelevant $g_Y^{\ast}\neq 0$. Orange region is consistent with the $\overline{MS}$ value of the top mass. Black dot shows the outcome of a calculation with FRG techniques\cite{Eichhorn:2017lry}. (b) The same in the $B-L$ model. All contours of $f_g$ (solid blue) are consistent with a small neutrino Yukawa coupling. $f_y$ contours (dashed brown) will be subject to constraints described in detail in \refsec{sec:FP_BL}. Black dot shows the outcome of a 
calculation with FRG techniques\cite{Eichhorn:2017lry}.}
\label{fig:par_space}
\end{figure}

Let us return to  
the explicit $f_g$, $f_y$ computations of Refs.\cite{Eichhorn:2017ylw,Eichhorn:2017lry}, which were recalled in \refeq{eq:fgfyFRG}. In \reffig{fig:par_space}(a) we show in the $(\tilde{\Lambda}^{\ast},\tilde{G}^{\ast})$ plane the parameter space consistent with the generation of a small neutrino Yukawa coupling and phenomenological constraints in the SMRHN. The solid blue line corresponds to 
$f_g=0.0097$, cf.~\refeq{eq:fg_1l}, and the shaded (orange) region  corresponds to the requirement of 
having the correct top mass, cf.~footnote~\ref{fn:top}.

%%%%%%%%%%%%%%%%%%%%%%%%%%%%%%%%%%%%%%%%%%%%%%%%%%%%%%%%%%%%%%%%%%%%%%%
%%%%%%%%%%%%%%%%%%%%%%%%%%%%%%%%%%%%%%%%%%%%%%%%%%%%%%%%%%%%%%%%%%%%%

Fixed point values $(\tilde{\Lambda}^{\ast},\tilde{G}^{\ast})$ are zeros of the gravity beta functions, which are computed 
with FRG techniques. In the Einstein-Hilbert truncation and $\beta=1\,,\alpha=0$ gauge, they read\cite{Eichhorn:2017lry}
\bea
\frac{d\tilde{G}}{dt}&=&2\tilde{G}+\frac{\tilde{G}^2}{6\pi}\left(2 N_D+N_S-4 N_V\right)-\frac{\tilde{G}^2}{6\pi}\left(14+\frac{6}{1-2\tilde{\Lambda}}+\frac{9}{(1-2\tilde{\Lambda})^2}\right)\label{eq:betaG}\\
\frac{d\tilde{\Lambda}}{dt}&=&-2\tilde{\Lambda}+\frac{\tilde{G}}{4\pi}\left(N_S-4 N_D+2 N_V\right)
+\frac{\tilde{G}\tilde{\Lambda}}{6\pi}\left(N_S+2 N_D-4 N_V\right)-\frac{3\tilde{G}}{2\pi}-\frac{7\tilde{G}\tilde{\Lambda}}{3\pi}\nonumber\\
 & &+\frac{7\tilde{G}}{4\pi(1-2\tilde{\Lambda})}-\frac{3\tilde{G}}{4\pi (1-2\tilde{\Lambda})^2}\,.\label{eq:betaL}
\eea
Equations (\ref{eq:betaG}) and (\ref{eq:betaL}) depend on the number of matter fields in the theory: real scalars~($N_S$), Dirac fermions~($N_D$),
and vector gauge bosons~($N_V$). In the SMRHN, $N_S=4$, $N_D=24$, $N_V=12$ imply the fixed point 
\be
\textrm{SMRHN:}\quad \tilde{G}^{\ast}=3.56\,, \quad \tilde{\Lambda}^{\ast}=-5.35\,,
\ee
which is indicated as a black dot in \reffig{fig:par_space}(a). If \refeq{eq:fgfyFRG} is taken at face value one gets $f_g=0.046$, $f_y=0.0050$. Even taking into account that the given 
explicit forms of Eqs.~(\ref{eq:fgfyFRG}), (\ref{eq:betaG}), and (\ref{eq:betaL}) are all sensitive to 
the chosen action truncation and gauge\cite{Dona:2013qba}, and that one cannot \textit{a priori} exclude that, 
after all sources of theoretical uncertainty were accounted for, the final expression may well allow for the black 
dot to sit right inside the phenomenologically preferred region, 
it is nonetheless disappointing 
that in the SMRHN the dynamical generation of a small Yukawa coupling depends strongly on effects that are not currently under control.      

Conversely, one can recast the above discussion in the framework of the gauged $B-L$ model. Now,
$N_S=6$, $N_D=24$, $N_V=13$ yield the fixed point
\be
B-L:\quad \tilde{G}^{\ast}=2.96\,, \quad \tilde{\Lambda}^{\ast}=-4.20\,,
\ee
which corresponds to $f_g=0.047$, $f_y=0.0028$, and
is indicated as a black dot in \reffig{fig:par_space}(b). We draw in \reffig{fig:par_space}(b) as solid blue contours some sample values of $f_g$, which are \textit{all} currently allowed by phenomenological constraints and give rise to different predictions for the $B-L$ gauge coupling and kinetic mixing. Brown dashed lines show the contours of selected values of $f_y$. As we shall see in \refsec{sec:FP_BL}, the question of which of those values can lead to a top-quark mass determination in agreement with observations depends on the size of $f_g$ and has to be addressed on a case-by-case basis. Nevertheless, it appears that the spectrum of possibilities for the eventual outcome of an FRG calculation opens up significantly in the $B-L$ model with respect to the SMRHN.

%%%%%%%%%%%%%%%%%%%%%%%%%%%%%%%%%%
\section{Boundary conditions of the $\boldsymbol{B-L}$ model\label{sec:FP_BL}}
%%%%%%%%%%%%%%%%%%%%%%%%%%%%%%%%%%

\subsection{Fixed points of the gauge-Yukawa system}

We discuss in this section the trans-Planckian fixed points of the $B-L$ gauge-Yukawa RGE system, whose explicit form can be found in Appendix~\ref{app:rges}. We limit our discussion to real fixed points consistent with the dynamical generation of an arbitrarily small Yukawa coupling for the neutrino.
We reiterate that in the $B-L$ model this is a viable possibility for all values of $f_g\geq 0.0097$. 

%%%%%%%%%%%%%%%%%%%%%%%%%%%%%%%%%%%%%%%%%%%%%%%%%%%%%%%%%%%%%%%%
\begin{table}[t]
\footnotesize
\begin{center}
\begin{tabular}{|c|c|c|c|}
\hline
$f_g$ & $g_Y^{\ast}$ & Other Abelian & Dynamical Mechanism \\
\hline
0.0097 & 0.47 (irr.) & $g_X^{\ast}=g_{\eps}^{\ast}=0$ (rel.)  & yes  \\
 &  & $g_X^{\ast}=0.44$, $g_{\eps}^{\ast}=-0.34$ (irr.)  & no \\
\hline
$f_g > 0.0097$ & 0 (rel.) & $g_X^{\ast}=g_{\eps}^{\ast}=0$ (rel.) & no \\
 &  & $12 g_X^{\ast 2}+\frac{32}{3} g_X^{\ast} g_{\epsilon}^{\ast}+\frac{41}{6} g_{\eps}^{\ast 2}-16 \pi^2 f_g =0$ (irr.) & yes \\
\hline
\end{tabular}
\caption{Trans-Planckian fixed points of the abelian gauge sector of the $B-L$ model, as a function of $f_g$. Fixed points can be either relevant (rel.), \textit{i.e.}~UV-attractive, or irrelevant (irr.), \textit{i.e.}~IR-attractive.}
\label{tab:mod_kin}
\end{center}
\end{table}
%%%%%%%%%%%%%%%%%%%%%%%%%%%%%%%%%%%%%%%%%%%%%%%%%%%%%%%%%%%%%%%%%%%%

We summarize in \reftable{tab:mod_kin} the fixed points of the abelian gauge sector corresponding to different $f_g$ values. The first line features the case equivalent to the SMRHN: $f_g$ adopts the specific value making $g_Y^{\ast}$ irrelevant, while the other two gauge couplings are relevant and become free parameters of the theory. They can thus be adjusted to fit all existing constraints. 

The second line shows the fixed point with three irrelevant interactive gauge couplings. In this case, however, the gauge-Yukawa system does not admit an irrelevant fixed point with $y_{\nu}^{\ast}=0$ if $y_t^{\ast}$ is required to be real.

Increasing the value of $f_g$ allows one to find a relevant Gaussian solution $g_Y^{\ast}=0$. However, no irrelevant fixed point with $y_{\nu}^{\ast}=0$ can be identified when all three abelian gauge couplings correspond to relevant directions (line~3 of \reftable{tab:mod_kin}). Thus, the fixed points of the two abelian NP couplings $g_X$ and $g_{\epsilon}$ must be irrelevant (line~4). It turns out that they are not independent, but instead belong to an ellipse, parameterized by $f_g$:
\be\label{eq:ellipse}
12 g_X^{\ast 2}+\frac{32}{3} g_X^{\ast} g_{\epsilon}^{\ast}+\frac{41}{6} g_{\eps}^{\ast 2}-16 \pi^2 f_g =0\,.
\ee

Let us focus on the set of fixed points~(\ref{eq:ellipse}). As the size of the ellipse is determined by $f_g$, a larger $f_g$ (perhaps from the FRG) will result in larger values for the fixed points of the $B-L$ gauge coupling and kinetic mixing. Additionally, the full RGE system is subject to bounds from the Yukawa sector, which constrain the size of $f_y$ on a case-by-case basis. Let us recall that the Yukawa sector of the $B-L$ model consists of several couplings, for which we seek the following trans-Planckian irrelevant fixed points:
\be\label{eq:yuk_fp}
y_t^\ast\neq 0\,,\qquad y_\nu^\ast=0\,, \qquad y_N^{\ast} =0 \;\; {\rm or}\;\; y_N^{\ast} \neq0\,.
\ee
Note that we assume $y_t^\ast\neq 0$ in \refeq{eq:yuk_fp}. This is required to obtain a negative critical exponent for the neutrino Yukawa couplings, as can be inferred by plugging into Eqs.~(\ref{eq:gy1})-(\ref{eq:yw})
the numerical coefficients corresponding to the RGEs of the $B-L$ model. The fixed points of the three sterile neutrino Yukawa couplings are assumed to be degenerate, $y_N^{ii\ast}= y_N^{\ast}$, in both the Gaussian and interactive case. Similarly $y_\nu^{ii\ast}= y_\nu^{\ast}$.
The choice $y_N^{\ast}=0$ would correspond to the absence of the Majorana mass term in \refeq{mass:BL}, thus leading to purely Dirac light neutrinos.

The trans-Planckian fixed-point value of the top quark Yukawa coupling is directly determined by $f_y$ as  
\be\label{eq:topfp}
y_t^\ast=\sqrt{\frac{8g_X^{\ast 2}+20 g_X^\ast g^\ast_\epsilon+17g_\epsilon^{\ast 2}%+17g_Y^{\ast 2}
+192 \pi^2 f_y }{54}}\,.
\ee
We impose $0.255\leq y_t^\ast\leq 0.325$. If we decouple gravity sharply at $M_{\textrm{Pl}}=10^{19}\gev$ the chosen range leads to the low-scale Yukawa coupling being consistent, within a few \gev\ uncertainty,
with the $\overline{MS}$ top quark mass. Such constraint loosely links the range of $f_y$ to the value of $f_g$, which is what determines the size of $g_X^{\ast}$ and $g_{\epsilon}^{\ast}$ in \refeq{eq:topfp}. 
One should thus take notice that in the $B-L$ model $f_g$ and $f_y$ are not entirely independent from one another, as they are correlated by the low-scale phenomenology.

We impose several other conditions on the Yukawa sector:
\begin{itemize}
\item $y_t^\ast$ has to be real, \textit{i.e.}, $17 g_\epsilon^{\ast 2} + 20 g_\epsilon^\ast g_X^\ast + 8 g_X^{\ast 2} + 192 \pi^2  f_y>0$. This also guarantees that the top Yukawa coupling is irrelevant, $\theta_t<0$.
\item $y^\ast_{\nu}=0$ has to be irrelevant, \textit{i.e.}, $\theta_\nu<0$.
\item if $y_N^{\ast}\neq 0$, it has to be real, \textit{i.e.}, $3g_X^{\ast 2}+ 8 \pi^2 f_y >0$. This also guarantees that $y_N$ is irrelevant, $\theta_N<0$. 
\item if $y_N^{\ast}= 0$, it has to be irrelevant, \textit{i.e.},
$\theta_N=3g_X^{\ast 2}+ 8 \pi^2  f_y<0$.
\end{itemize}

\begin{figure}[t]
	\centering%
    \subfloat[]{%
		\includegraphics[width=0.4\textwidth]{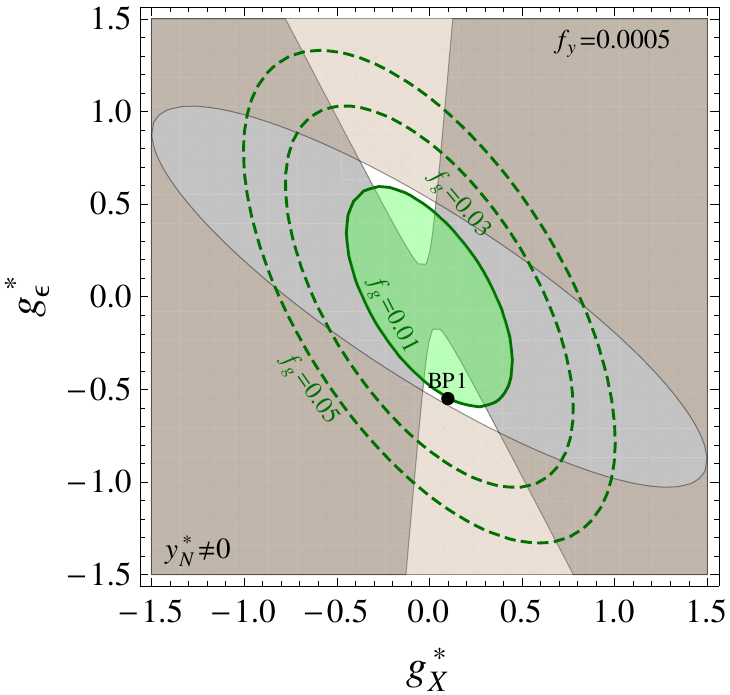}}
      \hspace{0.2cm}
    \subfloat[]{%
		\includegraphics[width=0.4\textwidth]{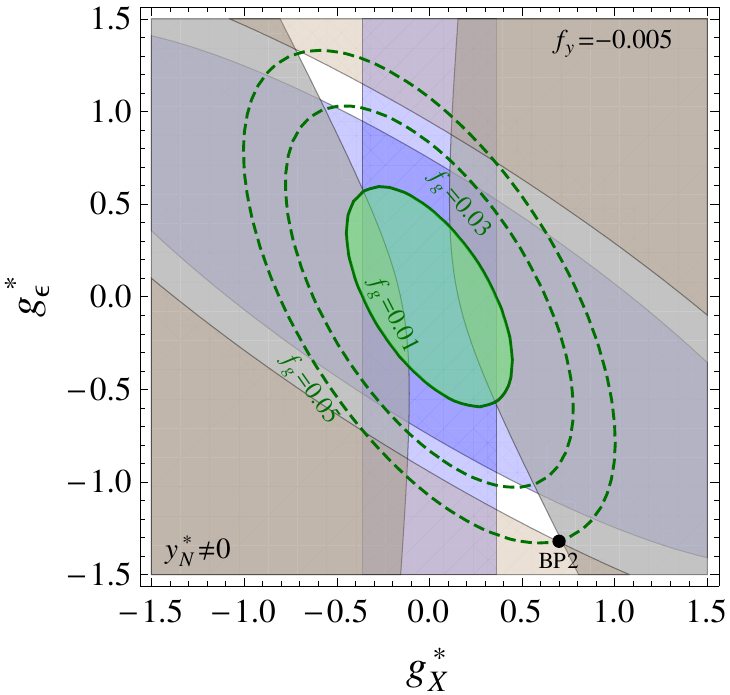}}\\
    \subfloat[]{%
   	\includegraphics[width=0.4\textwidth]{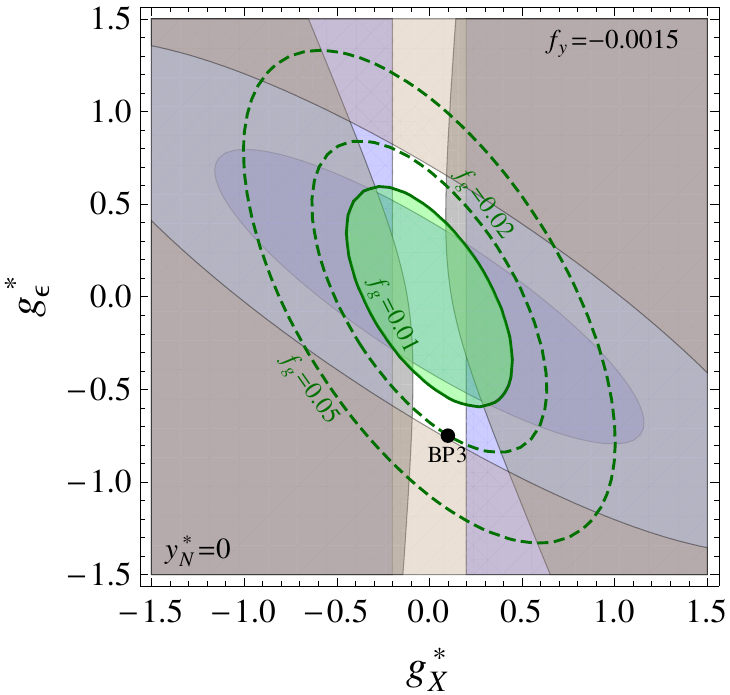}}
      \hspace{0.2cm}
    \subfloat[]{%
	\includegraphics[width=0.4\textwidth]{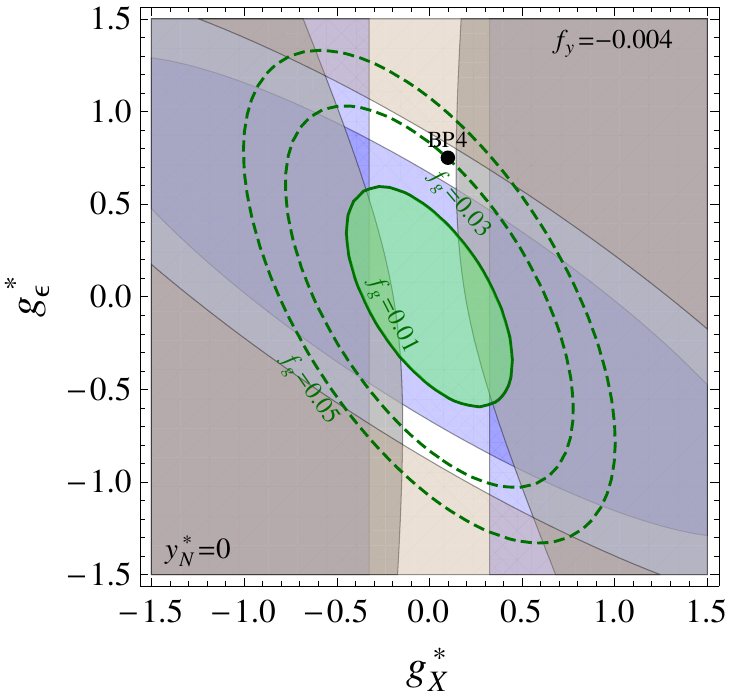}}  
\caption{(a) Dashed green ellipses trace the fixed-point solutions of \refeq{eq:ellipse} for different values of $f_g$. The inner green region, $f_g< 0.0097$, is excluded by the low-scale determination of the hypercharge gauge coupling. For $f_y=0.0005$, 
brown shading excludes the parameter space which does not allow the matching of the correct value of the top mass at the low scale, and the
gray region indicates the parameter space in which critical exponent $\theta_\nu>0$ at
$y_\nu^\ast=0$, \textit{i.e.}, there is no dynamical generation of a small Yukawa coupling. 
The white region is phenomenologically viable and we indicate with a black dot benchmark point BP1 (see also \reftable{tab:bench}). (b) Same as (a), but $f_y=-0.005$. 
The elliptical and vertical blue shaded regions indicate, respectively,  the values of the gauge couplings for which $y_t^\ast$ and $y_N^\ast$ are imaginary. Black dot indicates benchmark point BP2. (c) Same as (a), but $f_y=-0.0015$. The elliptical and vertical blue shaded regions indicate, respectively, the values of the gauge couplings for which $y_t^\ast$ is imaginary and $y_N$ is relevant. Black dot indicates benchmark point BP3. (d) Same as in (c), but $f_y=-0.004$. Black dot indicates benchmark point BP4.}
\label{fig:ellipseYNN}
\end{figure}

We show in \reffig{fig:ellipseYNN} 
the regions of the $(g_X^{\ast},g_{\epsilon}^{\ast})$ plane subject to the constraints listed above. The color scheme is described in detail in the caption. We select (a)~$f_y=0.0005$, (b)~$f_y=-0.005$, (c)~$f_y=-0.0015$, and (d)~$f_y=-0.004$. In all panels, black dots mark the position of selected benchmark points. We report their  characteristic features in \reftable{tab:bench}.

%%%%%%%%%%%%%%%%%%%%%%%%%%%%%%%%%%%%%%%%%%%%%%%%%%%%%%%%%%%%%%%%%%%%%%%
%%%%%%%%%%%%%%%%%%%%%%%%%%%%%%%%%%%%%%%%%%%%%%%%%%%%%%%%%%%%%%%%%%%%%

The last three columns of \reftable{tab:bench} show the predicted values of irrelevant  couplings $g_X$, $g_{\epsilon}$, and $y_N$ at three sub-Planckian scales of interest, $10^5\gev$, $10^7\gev$, and $10^9\gev$. 
Those are our chosen reference scales for the analysis of gravitational wave signatures in 
\refsec{sec:GW}.\footnote{Note that low-scale predictions originating from the trans-Planckian fixed-point analysis are derived under the assumption that the RGEs are not altered by large couplings below the Planck scale. Therefore, any additional low-scale interaction ought to be feeble. This is a commonly adopted approach in phenomenological studies with AS, as was discussed in Refs.\cite{Chikkaballi:2022urc,Kotlarski:2023mmr}.}

%%%%%%%%%%%%%%%%%%%%%%%%%%%%%%%%%%%%%%%%%%%%%%%%%%%%%%%%%%%%%%%%
\begin{table}[t]
\footnotesize
\begin{center}
\begin{tabular}{|c|c|c|c|c|c|c|c|c|}
\hline
 & $f_g$ &  $f_y$ & $g_X^{\ast}$ & $g_{\epsilon}^{\ast}$ & $y_N^{\ast}$ & $g_X\,(10^{5,7,9}\gev)$ &  $g_{\epsilon}\,(10^{5,7,9}\gev)$ & $y_N\,(10^{5,7,9}\gev)$ \\
\hline
BP1 & 0.01 & 0.0005 & 0.10 & $-0.55$ & 0.12 & 0.29, 0.29, 0.30 & $-0.26$, $-0.27$, $-0.28$ & 0.16, 0.16, 0.16 \\
\hline
BP2 & 0.05 & $-0.005$ & 0.70 & $-1.32$ & 0.47 & 0.40, 0.41, 0.44 & $-0.52$, $-0.56$, $-0.61$ & 0.42, 0.44, 0.45 \\
\hline
BP3 & 0.02 & $-0.0015$ & 0.10 & $-0.75$ & 0.0 & 0.12, 0.12, 0.12 & $-0.33$, $-0.35$, $-0.37$ & 0.0 \\
\hline
BP4 & 0.03 & $-0.004$ & 0.10 & 0.75 & 0.0 & 0.09, 0.09, 0.09 & 0.23, 0.25, 0.28 & 0.0 \\
\hline
\end{tabular}
\caption{The values of $f_g$ and $f_y$, trans-Planckian fixed points of the irrelevant couplings (indicated with an asterisk), and predicted values of those couplings at three low scales of reference for the four benchmark points selected in this study. All four points admit irrelevant $y_{\nu}^{\ast}=0$.}
\label{tab:bench}
\end{center}
\end{table}
%%%%%%%%%%%%%%%%%%%%%%%%%%%%%%%%%%%%%%%%%%%%%%%%%%%%%%%%%%%%%%%%%%%%

BP1 and BP2 feature $y_N^{\ast}\neq 0$, which is of order~1 in size. Equation~(\ref{mass:BL}) implies in this case that the Majorana mass scale is $M_N=\sqrt{2} y_N v_S$. It is a canonically relevant parameter of the theory and can thus be chosen anywhere, as long as it is in agreement with phenomenological constraints on the scalar potential. Note that the see-saw mechanism can be here invoked to give mass to the active neutrinos, $m_{\nu} \sim y^2_\nu v_H^2/(\sqrt{2}M_N)$. The theory is consistent with
AS whatever the Majorana mass scale is, 
since the correct size of the neutrino Yukawa coupling can be generated dynamically in the trans-Planckian flow.

On the other hand, BP3 and BP4 in \reftable{tab:bench} feature $y_N^{\ast}= 0$ along irrelevant directions, similarly to $y_{\nu}^{\ast}$. 
These cases allow for the interesting possibility that the sterile-neutrino Yukawa coupling sits tight at the irrelevant Gaussian fixed point $y_N^{\ast}= 0$. The Majorana mass is never generated, and its absence is protected along the entire RG 
flow by quantum scale invariance. The theory thus supports Dirac neutrinos with mass  $m_{\nu}\sim y_\nu v_H /\sqrt{2}$\,, where the required minuscule Yukawa coupling is generated dynamically.

In alternative, the theory might originate from a UV-attractive $y_N$ fixed point, and dynamically flow towards the IR-attractive one, thus generating -- besides an arbitrarily small $y_{\nu}$ -- also an arbitrarily small $y_{N}$. 
This case supports the existence 
of a Majorana mass, but the latter may be naturally decoupled from the size of $v_S$ and the constraints on the scalar potential. Thus, BP3 and BP4 may additionally provide a natural framework for accommodating 
the phenomenologically 
interesting possibility 
of pseudo-Dirac neutrinos\cite{Lindner:2014oea}.

%%%%%%%%%%%%%%%%%%%%%%%%%%%%%%%%%%%%%%%%%%%%%%%%%%%%%%%%%%%%%%%%%%%%%
\subsection{Scalar potential\label{sec:scal_pot}}

The tree-level scalar potential of the gauged $B-L$ model is given by
\be\label{eq:scapot}
V(H,S)=m_H^2\,H^{\dag} H +m_S^2\,  S^{\dag} S + \lam_1 \left(H^{\dag} H\right)^2+\lam_2 \left(S^{\dag} S\right)^2+\lam_3 \left(H^{\dag} H\right)  \left(S^{\dag} S\right)\,,
\ee
where $H$ is the SM-like Higgs SU(2)$_L$ doublet, which is neutral under U(1)$_{B-L}$, and $S$ is a complex scalar SM singlet, charged under U(1)$_{B-L}$ with $Q_S=2$\,. 
The spontaneous breaking of U(1)$_{B-L}$ generates the 
mass of the abelian $Z'$ gauge boson, which is approximately proportional to the vev along the $S$ direction: $m_{Z'}\approx 2\,g_X v_S$.

The benchmark points in \reftable{tab:bench} are all characterized by large kinetic mixing,
\be
\epsilon=\frac{g_{\epsilon}}{\sqrt{g_Y^2+g_{\epsilon}^2}}\approx 0.5-0.8\,.
\ee
As a direct consequence, $m_{Z'}$ is bounded from below by direct LHC constraints on high-mass dilepton resonance searches. The most recent measurements by ATLAS\cite{ATLAS:2019erb} and CMS\cite{CMS:2021ctt}, based on the 140\invfb\ data set in proton–proton collisions at the centre-of-mass energy of $\sqrt{s}=13\tev$ was numerically recast to the $(m_{Z'},\epsilon)$ plane in Fig.~3 of Ref.\cite{Chikkaballi:2022urc}. Since in our case the $Z'$ gauge boson couples directly to the quarks of the first two generations, the actual lower bound on the $Z'$ mass is stronger than for a dark gauge 
boson coupling to the quarks only through the kinetic mixing. For the four benchmark points in \reftable{tab:bench} one finds $m_{Z'}\gsim 6\tev$, which implies the following bounds on the vev of $S$:
\bea
\textrm{BP1:} & & v_S\gsim 10\tev\label{eq:bou1}\\
\textrm{BP2:} & & v_S\gsim 7.5\tev\label{eq:bou2}\\
\textrm{BP3:} & & v_S\gsim 25\tev\label{eq:bou3}\\
\textrm{BP4:} & & v_S\gsim 33\tev\,.\label{eq:bou4}
\eea
We have checked that the LHC direct measurements are currently more constraining than the bounds from the $\rho_0$ precision parameter\cite{Workman:2022ynf}, 
which require, for the same kinetic mixing,
$m_{Z'}\gsim 2\tev$.

Since $v_S\gg v_H=246\gev$, the two directions of the scalar potential effectively decouple. The vev $v_S$ may
arise from the presence of a large mass $m_S^2$ in \refeq{eq:scapot}. 
However, in this work we rather decide to investigate the well-known 
possibility that the scalar potential of the $B-L$ model 
develop its vevs from dimensional transmutation\cite{Hempfling:1996ht,Sher:1996ib,Nishino:2004kb,
Meissner:2006zh,Iso:2009ss}, 
through the usual Coleman-Weinberg mechanism. In particular, we next discuss whether the possibility of developing a radiatively generated minimum 
is consistent with the benchmark points in \reftable{tab:bench} and with AS in general.

Let us define $\phi\equiv \sqrt{2}\,\textrm{Re} (S)$, 
and project the potential to the $\phi$ direction. The corresponding, RGE-improved 
Coleman-Weinberg potential in 
the $B-L$ model reads (cf.~Appendix~\ref{sec:thermal})
\begin{multline}\label{eq:CWrgimpr}
V(\phi)=\frac{1}{2}m^2_S(t)\phi^2+\frac{1}{4}\lam_2(t)\, \phi^4\\
+\frac{1}{128\, \pi^2}\left[20 \lam_2^2(t)+96\, g_X^4(t)-48\, y_{N}^4(t) \right]\phi^4\left(-\frac{25}{6}+\ln \frac{\phi^2}{\mu^2}\right)\,, 
\end{multline}
where $t=\ln \mu$ is the renormalization scale. The potential can develop a minimum due to a large finite 1-loop contribution. If at the scale $\mu$
one finds 
\be
\lam_2\approx \frac{1}{55}\left(12\pi^2-2\sqrt{-3630\, g_X^4+1815\, y_N^4 +36\,\pi^4+330\, \pi^2\, m_S^2/\mu^2}\right),
\ee the minimum resides at $v_S \approx \mu$. 

Constraints apply from EWSB,
\bea
M_h^2&=&-2 m_H^2-\lam_3 v_S^2\label{eq:Higgs}\\
v_H^2&=&\frac{-2 m_H^2-\lam_3 v_S^2}{2 \lam_1}\,,\label{eq:vevH}
\eea
where $M_h$ is the Higgs mass and $v_H$ is the SM Higgs doublet vev. For $m_H^2\,, v_H^2\ll v_S^2$, these typically require $|\lam_3(v_S)|\ll \lam_1(v_S)$\,. 

It is interesting to investigate whether the Coleman-Weinberg construction is consistent with boundary conditions originating from trans-Planckian AS. A complementary analysis of the $B-L$ scalar potential in AS can be found, \textit{e.g.}, in Ref.\cite{Reichert:2019car}. 

Let us define the dimensionless running parameters $\tilde{m}_H^2=m_H^2/\mu^2$ and $\tilde{m}_S^2=m_S^2/\mu^2$\,.
The RGEs of the scalar potential are modified in the trans-Planckian regime with a ``correction'' due to the gravity fixed points, 
in analogy to Eqs.~(\ref{eq:betag}) and (\ref{eq:betay}). Following several studies in the literature\cite{Wetterich:2016uxm,Pawlowski:2018ixd,Wetterich:2019zdo} one writes
\bea
\frac{d\lam_1}{d t}&= &4\,\eta_1 \lam_1+\beta_{\lam_1,\textrm{add}}(g_i,y_j,\lam_k)-f_{\lam}\lam_1+\beta_{\lam_1,\textrm{grav}}\label{eq:lam1}\\
\frac{d\lam_2}{d t}&= &4\,\eta_2 \lam_2+\beta_{\lam_2,\textrm{add}}(g_i,y_j,\lam_k)-f_{\lam}\lam_2+\beta_{\lam_2,\textrm{grav}}\label{eq:lam2}\\
\frac{d\lam_3}{d t}&= &2\left(\eta_1+\eta_2 \right) \lam_3+\beta_{\lam_3,\textrm{add}}(g_i,y_j,\lam_k)-f_{\lam}\lam_3+\beta_{\lam_3,\textrm{grav}}\label{eq:lam3}\\
\frac{d \tilde{m}_H^2}{d t}&= &\left(-2+2\, \eta_1\right)\tilde{m}_H^2+\beta_{\tilde{m}_H^2,\textrm{add}}(\tilde{m}_H^2,\tilde{m}_S^2)-f_{\lam}\tilde{m}_H^2\label{eq:RGEmH}\\
\frac{d \tilde{m}_S^2}{d t}&= &\left(-2+2\, \eta_2\right)\tilde{m}_S^2+\beta_{\tilde{m}_S^2,\textrm{add}}(\tilde{m}_H^2,\tilde{m}_S^2)-f_{\lam}\tilde{m}_S^2\label{eq:RGEmS}\,,
\eea
where the matter anomalous dimensions $\eta_1$ and $\eta_2$ are given in Eqs.~(\ref{eq:eta1}), (\ref{eq:eta2}) of Appendix~\ref{app:rges}. 
The terms with an ``add'' subscript parameterize additive contributions to the matter beta functions not included in the anomalous dimensions, which are given explicitly in Eqs.~(\ref{eq:badd1})-(\ref{eq:baddS}). $f_{\lam}$ is the universal multiplicative correction analogous to $f_g$ and $f_y$, which typically depends on the fixed points of the gravitational action. Finally, the terms with a ``grav'' subscript parameterize additive contributions potentially arising from non-minimal direct couplings of the scalar potential to gravitational operators, see, \textit{e.g.}, the truncation introduced in Ref.\cite{Eichhorn:2020kca}.

One may categorize the outcome of the eventual FRG calculation of $f_{\lam}$ in three broad classes inducing different qualitative behavior:\smallskip

\textbf{Case A.} $f_{\lam}\ll -2$\,. Under this condition Eqs.~(\ref{eq:RGEmH}) and (\ref{eq:RGEmS}) admit the Gaussian fixed point 
$\tilde{m}_H^{2\ast}=0$, $\tilde{m}_S^{2\ast}=0$, fully irrelevant. The 
potential of \refeq{eq:scapot} becomes thus scale-invariant and it remains so at all scales, protected by quantum scale symmetry.
It was shown in 
Refs.\cite{Wetterich:2016uxm,Pawlowski:2018ixd,Wetterich:2019zdo} that
$f_{\lam}\ll -2$ may emerge in FRG calculation of the Higgs potential and Einstein-Hilbert truncation of the gravitational action, for certain values of the running Planck mass. 

In the context of the $B-L$ model with Coleman-Weinberg potential, 
conformal symmetry makes the model very predictive. 
After solving Eqs.~(\ref{eq:Higgs}) and (\ref{eq:vevH}) for $\lam_1$, $\lam_3$ 
with $M_h=125\gev$, $v_H=246\gev$,
one derives the values of the couplings at the scale of reference $v_S$.
For example, at $v_S=10^5\gev$,
\be\label{eq:bc_down}
\lam_1\left(10^5\gev\right)\approx 0.05\,, \quad \lam_3\left(10^5\gev\right)\approx -1.5 \times 10^{-6}\,.
\ee
On the other hand, one can solve Eqs.~(\ref{eq:lam1})-(\ref{eq:lam3}) to obtain the boundary conditions at the Planck scale in terms of the fixed points of \reftable{tab:bench}. In the absence of non-minimal gravitational (``grav'') contributions one finds
\bea
\lam_1^{\ast}&\approx &-\frac{\beta_{\lam_1,\textrm{add}}(g_i^{\ast},y_j^{\ast},\lam_k^{\ast})}{|f_{\lam}|}\, \approx\,-\frac{\frac{3}{8}g_{\epsilon}^{\ast 4}-6\, y_{t}^{\ast 4}}{16\pi^2 |f_{\lam}|}\nonumber \\
\lam_2^{\ast}&\approx &-\frac{\beta_{\lam_2,\textrm{add}}(g_i^{\ast},y_j^{\ast},\lam_k^{\ast})}{|f_{\lam}|}\, \approx\, -\frac{96\, g_{X}^{\ast 4}-48\, y_{N}^{\ast 4}}{16\pi^2 |f_{\lam}|}\nonumber \\
\lam_3^{\ast}&\approx &-\frac{\beta_{\lam_3,\textrm{add}}(g_i^{\ast},y_j^{\ast},\lam_k^{\ast})}{|f_{\lam}|}\, \approx\, -\frac{ 12\, g_{X}^{\ast 2} g_{\epsilon}^{\ast 2}}{16\pi^2 |f_{\lam}|}\,,\label{eq:bc_Pl}
\eea
all of them along irrelevant directions. Since the fixed points~(\ref{eq:bc_Pl}) are suppressed by large $|f_{\lam}|$, $\lam_{1}^{\ast}$, $\lam_{2}^{\ast}$, and $\lam_{3}^{\ast}$ are predicted to be 
extremely close to zero at the Planck scale and, as a consequence,
the boundary conditions (\ref{eq:bc_Pl}) are not consistent with Eqs.~(\ref{eq:bc_down}) after following the RG flow of the quartic couplings to IR energies. 

Even more damning for this case is perhaps the fact 
that following the flow of $\lam_2$ from the Planck scale to the low energy 
one obtains large negative values (\textit{e.g.}, $\lam_2=-0.56$ at $v_S=10^5\gev$),
which end up destabilizing the 
scalar potential from below in the $S$ direction (see also Ref.\cite{Reichert:2019car}, where the same conclusion was reached). These considerations make us conclude that Case~A may be 
phenomenologically viable only in the presence of some specific truncations of the gravitational action\cite{Eichhorn:2020kca}.
In particular, a substantial $\beta_{\lam_2,\textrm{grav}}<0$ at the fixed point will be required to match the low-energy phenomenology, implying, possibly, some fine tuning.\smallskip
  
\textbf{Case B.} $-2 \lesssim f_{\lam} < 0$. An eventual outcome of the FRG calculation within this range 
would imply that the fixed point at $\tilde{m}_H^{2\ast}=0$, $\tilde{m}_S^{2\ast}=0$ is fully relevant, while the
fixed points~(\ref{eq:bc_Pl}) of the quartic couplings are irrelevant. As such, the latter remain predictions of the theory at all scales, whereas the former are effectively free parameters. At the low scale, $m_H^2$ can be determined by solving \refeq{eq:Higgs} once $\lam_3$ is known. However, no realistic value of $f_{\lam}<0$ can 
generate a fixed-point value $\lam_2^{\ast}>0$ large enough to avoid the destabilization of the potential at low energies. As we concluded in Case~A, in Case~B as well one has to rely on the non-minimal contributions to the beta function parameterized by $\beta_{\lam_2,\textrm{grav}}$.\smallskip

\textbf{Case C.} $f_{\lam}\gg 0$. In this case the fixed points of the scalar potential are relevant. Masses and quartic couplings cannot be predicted from UV considerations, 
as any adopted value is eventually consistent with AS. It was shown in 
Ref.\cite{Pawlowski:2018ixd} that $f_{\lam}>0$ cannot be an
outcome of the FRG calculation with the action comprising the SM Higgs potential and gravity 
in the Einstein-Hilbert truncation, independently of the fixed-point value of the gravitational parameters and of the choice of regulator. 

On the other hand, it was recently shown in Ref.\cite{Pastor-Gutierrez:2022nki} 
that the situation may be different if one keeps track 
in the FRG calculation of several 
higher-order operators arising from a Taylor expansion of the SM Higgs potential. One can then 
observe the emergence
of a second relevant direction at the Gaussian fixed point, in addition to the one typically associated with $m_H^{2\ast}=0$. This additional relevant direction, which remains hidden in calculation performed in a $\mathcal{O}(H^4)$ truncation, is indeed welcome as it allows a phenomenologically viable connection between the UV fixed point and the physical SM at low energies. For the purposes of this paper, 
which is phenomenological in spirit, we assume that a similar behavior can be extended 
to the $S$ direction of the $B-L$ scalar potential, effectively allowing us to consider 
$f_{\lam}\gg 0$ in  Eqs.~(\ref{eq:lam1})-(\ref{eq:lam3}) and connecting the asymptotically 
safe fixed point in the deep UV to any desired Planck-scale boundary condition for the quartic couplings.

%%%%%%%%%%%%%%%%%%%%%%%%%%%%%%%%%%
\section{Gravitational waves\label{sec:GW}}
%%%%%%%%%%%%%%%%%%%%%%%%%%%%%%%%%%

It has long been known that a GW signal from the $B-L$ FOPT 
can be strong enough to allow detection in new-generation interferometers\cite{Jinno:2016knw,Chao:2017ilw,Okada:2018xdh,Marzo:2018nov,Brdar:2018num,Hasegawa:2019amx}. It is therefore enticing to investigate predictions for GWs associated with the boundary conditions from trans-Planckian AS discussed in \refsec{sec:FP_BL}.

In the presence of a hot plasma in the early Universe, the effective scalar 
potential receives thermal corrections\cite{PhysRevD.9.3357,PhysRevD.9.3320}, which generate a thermal barrier between the false ($\phi=0$) and true ($\phi=v_S$) 
$B-L$ vacua (cf.~Appendix~\ref{sec:thermal}). Tunneling from the former to the latter can then proceed through bubble nucleation\cite{LINDE198137,LINDE1983421}, leading to the generation of GWs. 

The GW physics is governed by several parameters that mainly depend on the shape of the effective thermal potential and on the bubbles' profile. These are the latent heat $\alpha$, the nucleation speed $\beta$, and the reheating temperature $T_{\rm{rh}}$. We refer the reader to Appendix~\ref{sec:gw} for a brief review of the physics of FOPTs. The present-day GW signal is then characterized by the peak amplitude $\Omega^{\rm{peak}}(\alpha,\beta,T_{\rm{rh}})$ and the peak frequency $f^{\rm{peak}}(\alpha,\beta,T_{\rm{rh}})$,
\be
h^2\Omega_{\rm{GW}}(f)=h^2\Omega^{\rm{peak}}\times \mathcal{F}(f/f^{\rm{peak}})\,,
\ee
where $h=H_0/(100\,\rm{km/s/Mpc})$ is the present-day dimensionless Hubble parameter and $\mathcal{F}$ is a function of the frequency $f$. The explicit forms of the GW signal  amplitudes are collected in \refeqs{eq:gwampli}{eq:gwpeak} of Appendix~\ref{sec:gw}.

We calculate the GW spectra for the benchmark points listed in \reftable{tab:bench}. The tree-level scalar mass $m_S^2$ plays an important role in the phenomenological prediction.  In Case~A of \refsec{sec:scal_pot} we considered $f_{\lam}\ll -2$, so that 
$m_S^2=0$ is protected by quantum scale symmetry. Scale invariance is typically associated with strong supercooling, and may give rise to large GW signals\cite{Hambye:2013dgv,Hashino:2016rvx,Marzola:2017jzl,Kang:2020jeg,Dasgupta:2022isg}. 
The presence of Yukawa couplings $y_N\neq 0$ 
is another key factor determining the properties of the FOPT. The height of the thermal barrier is directly proportional to $y_N$ -- cf.~\refeq{eq:phimassN} and \refeq{eq:vthermal} in Appendix~\ref{sec:thermal} -- forcing the nucleation and percolation temperature to be lower than with $y_N=0$ and therefore enhancing the impact of supercooling. This effect is particularly prominent for BP2, were we find that $T_p<0.1\gev$, independently of the chosen value of $v_S$. 
At $T\approx 0.1\gev$ the QCD phase transition takes place, a case that we do not analyze in this study.  

We observe the same behavior in BP1 at $v_S=10^9\gev$. Conversely, for BP1 at $v_S=10^5,10^7\gev$, the non-zero value of $y_N$ sets the nucleation/percolation temperature at several~\gev.  At the same time, $y_N\neq 0$ makes the Coleman-Weinberg minimum shallower, so that the probability of tunneling at a given temperature is reduced. As a result, the nucleation termination condition, \refeq{eq:tnstop} in Appendix~\ref{sec:gw}, is not satisfied. 
Those effects combined 
lead us to the conclusion that no GW signal can be observed for BP1 and BP2 with $m_S^2=0$.

BP3 and BP4 are characterized by relatively small values of the $B-L$ coupling. The Coleman-Weinberg minimum is thus rather shallow, indicating low decay rate of the false vacuum, $S_3(T)/T\gg 100$. As a result, there is no FOPT in BP3 and BP4 when $m_S^2=0$ for all chosen values of $v_S$.

Meanwhile, we discussed in \refsec{sec:scal_pot} that the purely conformal case, $m_S^2=0$, may be in strong tension with the requirement of a stable (bounded from below) scalar potential at the chosen scales $v_S$, unless one considers non-minimal terms in the matter-gravity action, which are likely to involve some fine tuning. A more natural possibility in the quantum gravity setup is thus that the mass terms in \refeq{eq:scapot} remain relevant parameters, in agreement with their canonical scaling, a situation described in  
Case~B and Case~C of \refsec{sec:scal_pot}. No quantum scale symmetry can in those cases prevent the presence of tree-level masses in the Lagrangian. On the other hand, since $|\lam_3|\ll 1$, the mass parameters appear, to a good approximation, multiplicatively in their respective RGEs -- Eqs.~(\ref{eq:RGEmH}) and (\ref{eq:RGEmS}) -- and are thus technically natural. As a consequence, they can take any desired value. We consider here the situation where $m_S^2<0$ is small enough not to interfere with the generation of a thermal barrier -- the transition is still first order -- but large enough to enhance the decay rate of the false vacuum given in \refeq{eq:gammaft}, thus
triggering the phase transition at nucleation (and percolation) temperature larger than it would be required in the conformal case.  

%%%%%%%%%%%%%%%%%%%%%%%%%%%%%%%%%%%%%%%%%%%%%%%%%%%%%%%%%%%%%%%%%%%%%%%
 \begin{figure}[t]
	\centering%
     \subfloat[]{%
		\includegraphics[width=0.47\textwidth]{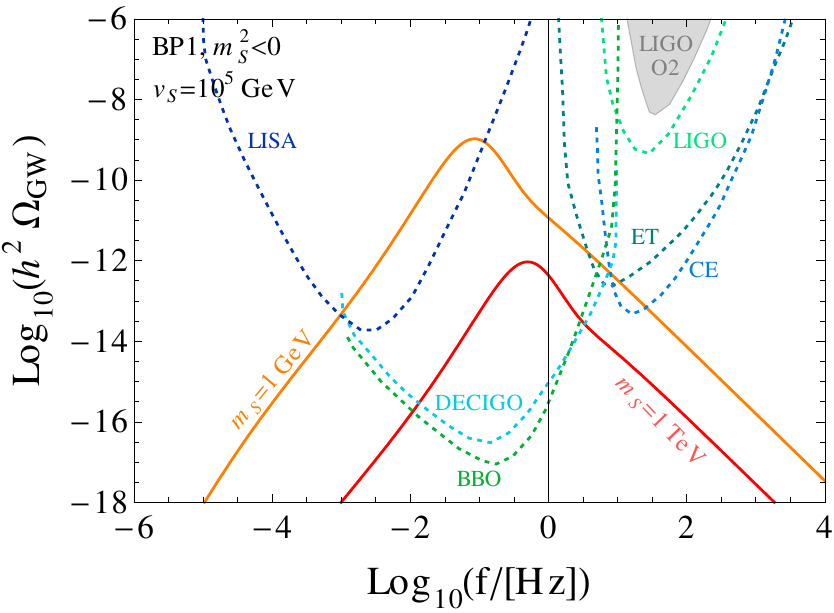}}
     \hspace{0.2cm}
     \subfloat[]{%
  		\includegraphics[width=0.47\textwidth]{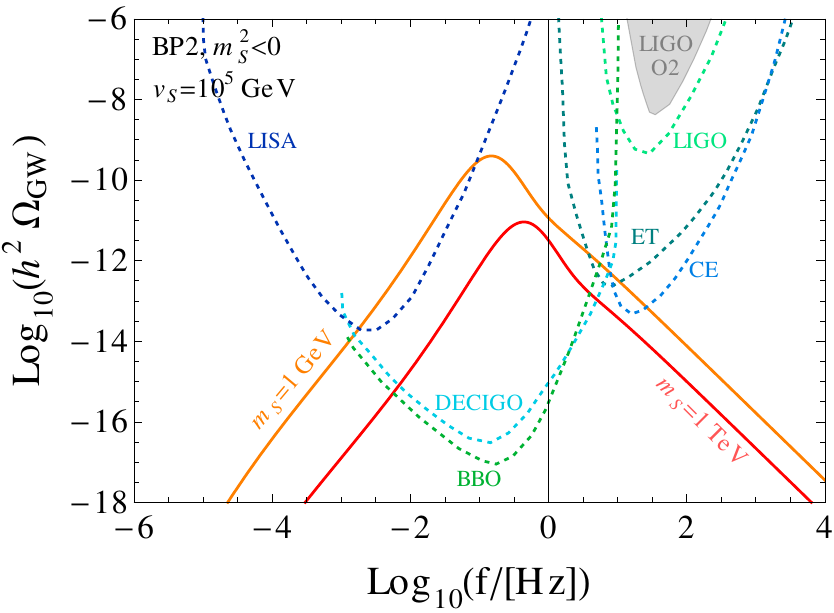}}\\
     \subfloat[]{%
    \includegraphics[width=0.47\textwidth]{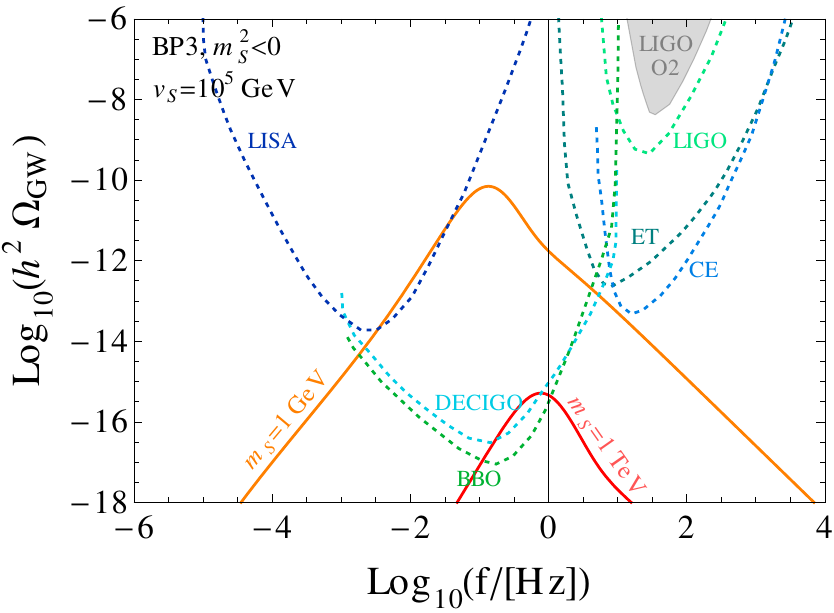}}
     \hspace{0.2cm}
      \subfloat[]{%
    \includegraphics[width=0.47\textwidth]{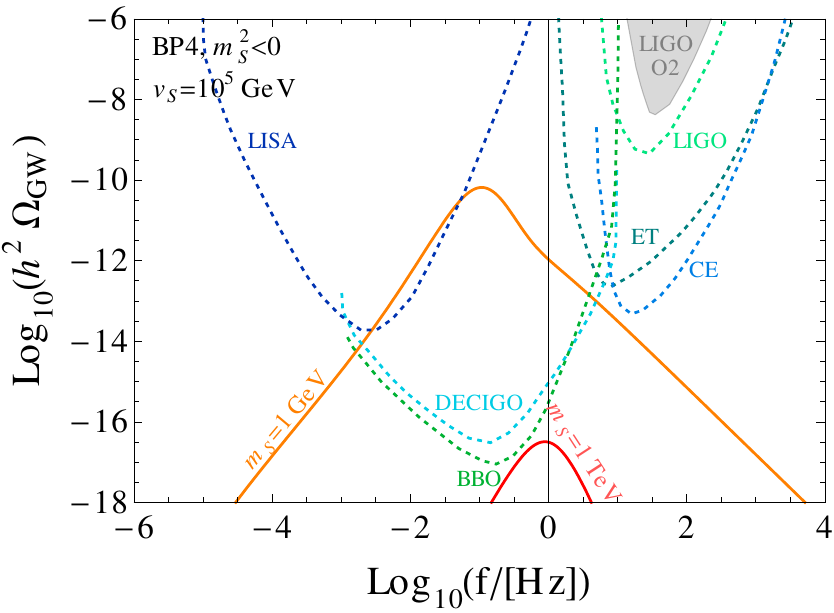}}
\caption{Gravitational-wave spectra of the benchmark points listed in \reftable{tab:bench}, for two selected values of the scalar mass parameter $m_S\equiv\sqrt{|m_S^2|}$ (solid lines). The scalar vev is set at $v_S=10^5\gev$. Also shown are the sensitivity curves for various GW interferometers (dotted lines).}
\label{fig:graviwaves2}
\end{figure}
%%%%%%%%%%%%%%%%%%%%%%%%%%%%%%%%%%%%%%%%%%%%%%%%%%%%%%%%%%%%%%%%%%%%%

We present in \reffig{fig:graviwaves2} the expected GW signal at $v_S= 10^5\gev$ 
for the four benchmark points in \reftable{tab:bench} given two selected values of the mass: $\sqrt{|m_S^2|}=1\gev$~(yellow) and  
$\sqrt{|m_S^2|}=1\tev$~(red). The signal is confronted with 
integrated sensitivity curves for the Big-Bang Observer (BBO)\cite{Crowder:2005nr,Corbin:2005ny}, Cosmic Explorer (CE)\cite{Reitze:2019iox}, Deci-Hertz Interferometer Gravitational-Wave Observatory (DECIGO)\cite{Seto:2001qf,Musha:2017usi}, Einstein Telescope (ET)\cite{Punturo:2010zz,Sathyaprakash:2012jk}, Laser Interferometer Gravitational-Wave Observatory (LIGO)\cite{Harry:2010zz,LIGOScientific:2014pky,LIGOScientific:2016wof} and Laser Interferometer Space Antenna (LISA)\cite{LISA:2017pwj,Baker:2019nia}, which are shown as dotted curves. Gray region in the upper part of the plot indicates the current exclusion bound by the LIGO-VIRGO O2 run\cite{Renzini:2019vmt,KAGRA:2021kbb}. 

Benchmark point BP1 is shown in panel~(a) and BP2 in panel~(b). Despite obtaining in the two cases quite different predictions from the fixed-point analysis, we observe similar GW amplitudes and frequencies. This is because the phase transitions are triggered by the mass term, so that they feature, given equivalent $m_S^2<0$, a similar, relatively fast nucleation speed $\beta$. Analogous behavior is observed for BP3 in panel~(c) and BP4 in panel~(d), with the biggest difference with respect to BP1 and BP2
being that the signal amplitude and frequency show greater sensitivity to the $m_S^2$ spread. This is due 
to the much lesser depth of the minimum in BP3 and BP4, which enhances the impact of a 
finite $m_S^2$ on the bounce action.
All in all, we observe strong similarities of signatures in the four cases. We are thus 
forced to conclude that different fixed points cannot be distinguished with a detection of GWs from FOPTs. 
Equally bleak are the prospects of distinguishing the Majorana vs.~Dirac nature of the neutrino with this method, since 
the fixed points with $y_N^{\ast}\neq 0$ and those with $y_N^{\ast}=0$ show very similar spectra, shaped in all cases by the $m_S^2$ relevant parameter.\footnote{The situation may be more optimistic if the dynamical mechanism for the generation of neutrino masses relies on a UV completion that gives rise to topological defects\cite{King:2023cgv}.}

%%%%%%%%%%%%%%%%%%%%%%%%%%%%%%%%%%%%%%%%%%%%%%%%%%%%%%%%%%%%%%%%%%%%%%%
 \begin{figure}[t]
	\centering%
   		\includegraphics[width=0.47\textwidth]{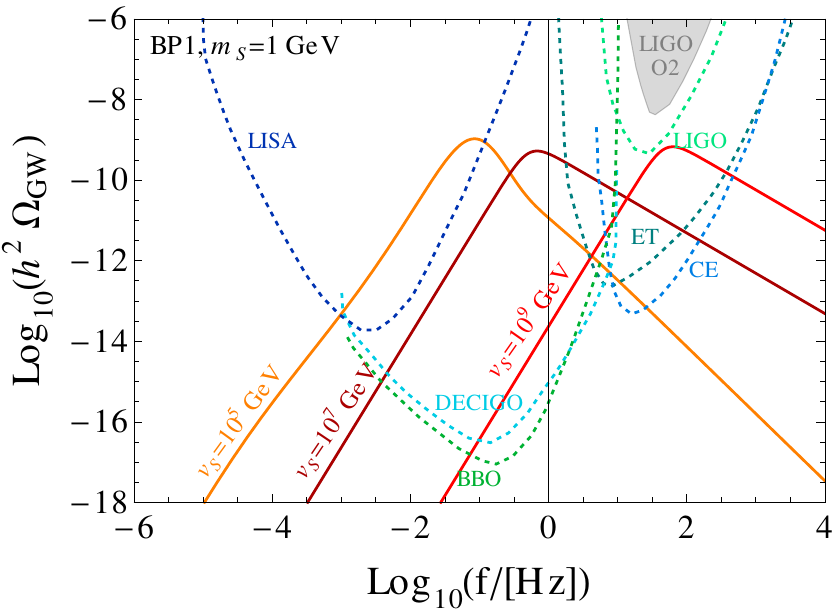}
\caption{Gravitational-wave spectra of benchmark point BP1, for increasing values of the scalar vev $v_S$ (solid lines). 
The scalar mass parameter is set at $m_S\equiv \sqrt{|m_S^2|}=1 \gev$.  
Also shown are the sensitivity curves for various GW interferometers (dotted lines).}
\label{fig:graviwaves}
\end{figure}
%%%%%%%%%%%%%%%%%%%%%%%%%%%%%%%%%%%%%%%%%%%%%%%%%%%%%%%%%%%%%%%%%%%%%

We finally show in \reffig{fig:graviwaves} the GW signal of BP1 with $\sqrt{|m_S^2|}=1\gev$ for all the $v_S$ values given in \reftable{tab:bench}. As is well known, larger vev induces a signal at larger frequency. More interestingly, one can see that while at $v_S=10^5\gev$ (orange) the main source of GW is sound waves and turbulence of the plasma, for larger values (brown and red) the main source is bubble collisions, which flatten the signal at higher frequencies. This is because, as the critical temperature increases with the vev, the percolation temperature remains set at approximately the same value, determined by the size of $m_S^2$. The overall effect is thus to increase the amount of supercooling with larger vevs.    

%%%%%%%%%%%%%%%%%%%%%%%%%%%%%%%%%%%%%%%%%%%%%%%%%%%%%%%%
\section{Conclusions\label{sec:summary}}
%%%%%%%%%%%%%%%%%%%%%%%%%%%%%%%%%%%%%%%%%%%%%%%%%%%%%%%%

In this paper, we have 
revisited the dynamical generation of an arbitrarily small neutrino Yukawa coupling based on the existence of Gaussian IR-attractive fixed points of the trans-Planckian RG flow, which was investigated first in Refs.\cite{Kowalska:2022ypk,Eichhorn:2022vgp}. While in the original studies 
the low-energy theory that is completed in the UV with boundary conditions consistent 
with asymptotically safe quantum gravity was
the SM with three right-handed neutrinos~(SMRHN), 
in this work we have focused on the well-known gauged $B-L$ model.

The $B-L$ model offers several advantages with respect to the SMRHN. Some are well established -- like requiring the existence of right-handed neutrino spinor fields based on the cancellation of gauge anomalies and the reliance on gauge rather than global or accidental symmetries -- others apply more directly to the realm of trans-Planckian AS and the quantum-gravity nature of the UV completion. 

On the one hand, we have shown that the $B-L$ model may justify  
a richer phenomenology in the context of neutrino mass-generation, since it seems to be able to accommodate quite naturally each and every feature that neutrinos may eventually show experimentally. 
Assuming in fact that there exists and IR-attractive fixed point at $y_{\nu}^{\ast}=0$, 
trajectories of the RG flow originating from an irrelevant $y_N^{\ast}=0$ will lead to purely Dirac neutrinos at the low scale; trajectories originating from a different, relevant $y_N^{\ast}$, whose trans-Planckian flow conjoins to the IR-attractive $y_N^{\ast}=0$, may lead to pseudo-Dirac neutrinos; trajectories originating in an irrelevant $y_N^{\ast}\neq 0$ will lead straightforwardly to Majorana neutrinos, and so on. The see-saw scale too, being a canonically relevant parameter of the Lagrangian, can freely assume any desired value.   

On the other hand, we have shown that this rich phenomenology may be found to be in better agreement with calculations of the quantum gravity UV completion, which are likely to be based on FRG techniques. 
In fact, in the SMRHN a very specific, potentially fine-tuned value of the gravity contribution $f_g$ is required, lest we risk 
failing to reproduce the measured low-scale hypercharge coupling. This is not the case in the $B-L$ model, as the dynamical role played by the irrelevant $g_Y$ fixed point in the SMRHN is here enacted 
through the fixed points of the $B-L$ gauge coupling and kinetic mixing. This gives us the ability to untie the results of the first-principle calculation in quantum gravity -- which are marred by significant theoretical uncertainties -- from the precise measurement of a well-known quantity at the low scale, opening up the spectrum of observable possibilities.    

Among the several interesting signatures of the model, we investigated in detail the generation of gravitational waves from FOPTs. 
For our four different benchmark points we found that, 
while it will be easy to observe a clear signal in future interferometers, it will prove extremely more 
challenging to be able to 
discern different fixed points --
their gauge and Yukawa coupling values and, more importantly, the Majorana vs.~Dirac nature of the neutrino -- from one another. 
This is because an explicit mass term in the effective scalar potential is necessary to trigger the $B-L$ phase transition. While this is a welcome feature for the theoretical consistency of the model -- scalar masses are relevant parameters within the trans-Planckian UV completion so that conformal symmetry is not a property enforced by RG running -- it also makes the GW spectrum extremely sensitive to parameters that cannot, by their own nature, be predicted from UV considerations. 
Even if a GW signal were to be detected, additional observations in complementary experimental venues 
--  for example, the detection of new spin-0 and spin-1 resonances in multilepton searches at future high-energy colliders -- will be required to extract unequivocally the shape of the potential and uncover the UV nature of these fixed points.   

%%%%%%%%%%%%%%%%%%%%%%%%%%%%%%%%%%%%%%%%%%%%%%%%%%%%%%%%%%%%%
\bigskip
\begin{center}
\textbf{ACKNOWLEDGMENTS}
\end{center}
\noindent 
We would like to thank Anish Ghoshal for many discussions and Wojciech Kotlarski for help with  \texttt{PyR@TE}. AC and EMS are supported in part by the National Science Centre (Poland) under the research Grant No.~2020/38/E/ST2/00126. KK is supported in part by the National Science Centre (Poland) under the research Grant No.~2017/26/E/ST2/00470. The use of the CIS computer cluster at the National Centre for Nuclear Research in Warsaw is gratefully
acknowledged.

\bigskip
%%%%%%%%%%%%%%%%%%%%%%%%%%%%%%%%%%%%%%%%%%%%%%%%%%%%%%%

%\bigskip
\appendix

%\section*{Appendices}
\addcontentsline{toc}{section}{Appendices}

\section{Renormalization group equations}
\label{app:rges}

In this appendix we present the RGEs of the gauged $B-L$ model, which we derive using \texttt{PyR@TE 3}\cite{Poole:2019kcm,Sartore:2020gou}. Capital letters indicate a coupling matrix in flavor space. We work at one loop, so that the trans-Planckian RGE of a generic gauge or Yukawa coupling $c_i$ takes the form
\be
\frac{d c_i}{dt}=\frac{1}{16 \pi^2}\beta^{(1)}(c_i)-f_{c}\,c_i\,,
\ee
where $f_c=f_g$ for all gauge couplings and $f_c=f_y$ for all Yukawa couplings.

%%%%%%%%%%%%%%%%%%%%%%%%%%%%%%%%%%%%%%%%%%%
\paragraph{Gauge couplings}
\be
\beta^{(1)}(g_{Y}) =\frac{41}{6} g_{Y}^{3}
\ee
\be
\beta^{(1)}(g_\epsilon) =
 \frac{32}{3} g_{Y}^{2} g_{X}
+ \frac{32}{3} g_{X} g_\epsilon^{2}
+ \frac{41}{3} g_{Y}^{2} g_\epsilon
+ 12 g_{X}^{2} g_\epsilon
+ \frac{41}{6} g_\epsilon^{3}
\ee
\be
\beta^{(1)}(g_{X}) =
 12 g_{X}^{3}
+ \frac{41}{6} g_{X} g_\epsilon^{2}
+ \frac{32}{3} g_{X}^{2} g_\epsilon
\ee
\be
\beta^{(1)}(g_2) =- \frac{19}{6} g_2^{3}
\ee
\be
\beta^{(1)}(g_3) =-7 g_3^{3}\,.
\ee
%%%%%%%%%%%%%%%%%%%%%%%%%%%%%%%%%%%%%%%%%%%
\paragraph{Yukawa couplings}
\begin{multline}
\beta^{(1)}(Y_u) =
 \frac{3}{2} Y_u Y_u^{\dagger} Y_u
+ 3 \tr\left(Y_u^{\dagger} Y_u \right) Y_u
+ \tr\left(Y_\nu^{\dagger} Y_\nu \right) Y_u\\
-  \frac{17}{12} g_{Y}^{2} Y_u
-  \frac{2}{3} g_{X}^{2} Y_u
-  \frac{5}{3} g_{X} g_\epsilon Y_u
-  \frac{17}{12} g_\epsilon^{2} Y_u
-  \frac{9}{4} g_2^{2} Y_u
- 8 g_3^{2} Y_u
\end{multline}
\begin{multline}
\beta^{(1)}(Y_\nu) =
 \frac{3}{2} Y_\nu Y_\nu^{\dagger} Y_\nu
+ 2 Y_\nu Y_N^{*} Y_N
+ 3 \tr\left(Y_u^{\dagger} Y_u \right) Y_\nu
+ \tr\left(Y_\nu^{\dagger} Y_\nu \right) Y_\nu\\
-  \frac{3}{4} g_{Y}^{2} Y_\nu
- 6 g_{X}^{2} Y_\nu
- 3 g_{X} g_\epsilon Y_\nu
-  \frac{3}{4} g_\epsilon^{2} Y_\nu
-  \frac{9}{4} g_2^{2} Y_\nu
\end{multline}
\be
\beta^{(1)}(Y_N) =
 Y_\nu^{T} Y_\nu^{\ast} Y_N
+ Y_N Y_\nu^{\dagger} Y_\nu
+ 4 Y_N Y_N^{\ast} Y_N
+ 2 \tr\left(Y_N^{\ast} Y_N \right) Y_N
- 6 g_{X}^{2} Y_N\,.
\ee
%%%%%%%%%%%%%%%%%%%%%%%%%%%%%%%%%%%%%%%%%%%
\paragraph{Anomalous dimensions and additive terms of the scalar potential}
\be\label{eq:eta1}
\eta_1=\frac{1}{16\pi^2}\left[-\frac{3}{4}g_Y^2-\frac{3}{4}g_{\epsilon}^2-\frac{9}{4}g_2^2+3\textrm{Tr}\left(Y_u^{\dag}Y_u\right)+\textrm{Tr}\left(Y_{\nu}^{\dag}Y_{\nu}\right) \right]
\ee
\be\label{eq:eta2}
\eta_2=\frac{1}{16\pi^2}\left[-12 g_X^2+2\textrm{Tr}\left(Y_N^{\ast}Y_N\right)\right]
\ee
\begin{multline}\label{eq:badd1}
\beta_{\lam_1,\textrm{add}}(g_i,y_j,\lam_k)=\frac{1}{16\pi^2}\left[24 \lam_1^2+\lam_3^2 +\frac{3}{8} g_Y^4+\frac{3}{4}g_Y^2 g_{\epsilon}^2+\frac{3}{8} g_{\epsilon}^4
+\frac{3}{4}g_Y^2 g_2^2+\frac{3}{4}g_2^2 g_{\epsilon}^2+\frac{9}{8} g_{2}^4\right.\\
\left. -6\textrm{Tr}\left(Y_u^{\dag}Y_u\right)^2-2\textrm{Tr}\left(Y_{\nu}^{\dag}Y_{\nu}\right)^2 \right]
\end{multline}
\be
\beta_{\lam_2,\textrm{add}}(g_i,y_j,\lam_k)=\frac{1}{16\pi^2}\left[20\lam_2^2+2 \lam_3^2+96\, g_X^4-16\textrm{Tr}\left(Y_N^{\ast}Y_N\right)^2\right]
\ee
\be
\beta_{\lam_3,\textrm{add}}(g_i,y_j,\lam_k)=\frac{1}{16\pi^2}\left[12\lam_1 \lam_3+8 \lam_2 \lam_3+4\lam_3^2 +12 g_X^2 g_{\epsilon}^2 -16 \textrm{Tr}\left( Y_{\nu}^{\dag}Y_{\nu} Y_N^{\ast} Y_N\right)\right]
\ee
\be
\beta_{\tilde{m}_H^2,\textrm{add}}(\tilde{m}_H^2,\tilde{m}_S^2)=\frac{1}{16\pi^2}\left(12 \lam_1 \tilde{m}_H^2+2 \lam_3 \tilde{m}_S^2 \right)
\ee
\be\label{eq:baddS}
\beta_{\tilde{m}_S^2,\textrm{add}}(\tilde{m}_H^2,\tilde{m}_S^2)=\frac{1}{16\pi^2}\left(4 \lam_3 \tilde{m}_H^2+8 \lam_2 \tilde{m}_S^2 \right)\,.
\ee

%%%%%%%%%%%%%%%%%%%%%%%%%%%%%%%%%
\section{Thermally corrected effective potential\label{sec:thermal}}

The tree-level scalar potential of the $B-L$ model  is defined in \refeq{eq:scapot}. 
Since the vev of the scalar $S$ is much larger than the Higgs vev, $vs\gg v_H$ (cf.~\refsec{sec:scal_pot}), the $B-L$ phase transition occurs along the $S$ direction. Thus, we can limit our analysis to the $S$-dependent part of $V(H,S)$. The symmetry of the potential only allows terms in powers of $S^\dagger S$ (the same goes for the SM Higgs scalar), it is then enough to consider the effective potential for the radial component $\phi=\sqrt{2}\,\textrm{Re}(S)$,
\be
S^\dagger S=\frac{1}{2}\phi^2.
\ee
The tree-level part of the effective potential reads
\be\label{eq:Vtree}
V_0(\phi)=\frac{1}{2}m_S^2\,\phi^2+\frac{1}{4}\lambda_2\,\phi^4\,.
\ee
The finite one-loop Coleman-Weinberg contributions are given by\cite{PhysRevD.7.1888}
\be
V_{1-\rm{loop}}(\phi)=\frac{1}{64\pi^2}\sum_J n_J\, m^4_J(\phi)\left[\log\frac{m_J^2(\phi)}{\mu^2}-C_J\right],
\ee
where $n_J=(-1)^{2s_J}\,Q_J\,N_J\,(2s_J+1)$, with $Q_J=1(2)$ for uncharged (charged) particles, $N_J=1(3)$ for uncolored (colored) particles, and $C_J=\frac{5}{6}\left(\frac{3}{2}\right)$ for vector bosons (fermions and scalars); $s_J$ is the particle spin and $m_J$ is the field-dependent mass of the particle. In the $B-L$ model the sum runs over the gauge boson $Z'$, three right-handed neutrinos $\nu_{R,i}$, the real scalar $\phi$ and the corresponding Goldstone boson $G$,
\bea\label{eq:potCW}
V_{1-\rm{loop}}(\phi)&=&\frac{3}{64\pi^2}m_{Z'}^4(\phi)\left[\log\frac{m_{Z'}^2(\phi)}{\mu^2}-\frac{5}{6}\right]-\frac{2}{64\pi^2}\sum_{i=1}^3 m_{\nu_{R,i}}^4(\phi)\left[\log\frac{m_{\nu_{R,i}}^2(\phi)}{\mu^2}-\frac{3}{2}\right]\nonumber\\
&+&\frac{1}{64\pi^2}m_{\phi}^4(\phi)\left[\log\frac{m_{\phi}^2(\phi)}{\mu^2}-\frac{3}{2}\right]+\frac{1}{64\pi^2}m_{G}^4(\phi)\left[\log\frac{m_{G}^2(\phi)}{\mu^2}-\frac{3}{2}\right]\,.
\eea
The $\phi$-dependent masses that enter \refeq{eq:potCW} are obtained from the second derivatives of \refeq{mass:BL}, \refeq{eq:lagE} and \refeq{eq:Vtree} and read 
\bea
m^2_{Z'}(\phi)&=&4\,g_X^2\,\phi^2\\
m_{\nu_{R,i}}^2(\phi)&=&2\,y_{N}^2\,\phi^2\label{eq:phimassN}\\
m_{\phi}^2(\phi)&=&3\,\lambda_2\,\phi^2+m_S^2\\
m_{G}^2(\phi)&=&\lambda_2\,\phi^2+m_S^2\,.
\eea
The finite-temperature one-loop corrections to the effective potential are given by\cite{Quiros:1999jp,Quiros:1994dr}
\be\label{eq:vthermal}
V_{\rm{thermal}}(\phi,T)=\frac{T^4}{2\pi^2}\sum_J n_J J_J\left(\frac{m_J^2(\phi)}{T^2}\right)\,,
\ee
and again the sum runs over $Z'$, $\nu_{R,i}$, $\phi$ and $G$. The thermal integrals $J_J$ are defined as
\be
J_{B,F}(y)=\int_0^\infty dx\, x^2\log\left(1\mp e^{-\sqrt{x^2+y}}\right)\,,
\ee
where $B$ and $F$ stands for bosons and fermions, respectively. 
Additionally, the resummed daisy diagrams with the bosonic degrees of freedom contribute as\cite{Arnold:1992rz}
\be
V_{\rm{daisy}}(\phi,T)=\frac{T}{12\pi}\sum_J k_J\left[m_J^3(\phi)-\mathcal{M}_J^3(\phi,T)\right],
\ee
where $k_{Z'}=k_{\phi}=k_{G}=1$ and thermally corrected (Debye) masses are defined as $\mathcal{M}_J^2(\phi,T)=m^2_J(\phi)+\Pi_J(\phi,T)$, with the temperature-dependent self-energies\cite{Marzo:2018nov}
\bea
\Pi_{Z'}(\phi,T)&=&4\,g_X^2\, T^2\\
\Pi_\phi(\phi,T)&=&\left(g_X^2+\frac{1}{3}\lambda_2+\frac{1}{2}y_{N}^2\right)T^2\\
\Pi_G(\phi,T)&=&\Pi_\phi(\phi,T)\,.
\eea
The total thermally-corrected effective potential for the scalar $\phi$ thus reads
\be\label{eq:fullVT}
V(\phi,T)=V_0(\phi)+V_{1-\rm{loop}}(\phi)+V_{\rm{thermal}}(\phi,T)+V_{\rm{daisy}}(\phi,T)\,.
\ee

Finally, to mitigate the dependence of our results on the renormalization scale, we implement the RG-improvement of the scalar potential $V(\phi,T)$. We replace all the couplings (collectively denoted with $\{\alpha_i\}$) with the running couplings in the $\overline{MS}$ scheme, $\{\alpha_i\}\to \{\alpha_i(t)\}$, and we introduce the field strength renormalization constant $\phi\to\sqrt{Z_\phi(t)}\,\phi$, with $t=\ln\,\mu$. After the counterterms are introduced to absorb the infinities arising at one loop, \refeq{eq:potCW} leads straightforwardly to \refeq{eq:CWrgimpr}.

%%%%%%%%%%%%%%%%%%%%%%%%%%%%%%%%%%%%%%%%%%%%%%%%%%%%
\section{Phase transition and gravitational waves\label{sec:gw}}

At high temperatures, the potential $V(\phi,T)$ is dominated by thermal corrections and it features a single minimum at $\langle \phi\rangle=0$. As the Universe cools down, the second minimum with $\langle \phi\rangle\neq 0$ is formed and at the critical temperature, $T_c$, the two minima reach the same depth. Below $T_c$, the symmetry-breaking minimum becomes the global one and as the temperature further decreases and a thermal barrier between $\langle \phi\rangle=0$ and $\langle \phi\rangle\neq 0$ keeps getting lower, tunneling from the false to the true vacuum may take place. A FOPT begins and bubbles of the symmetry-broken phase start to nucleate and grow in the sea of the false vacuum. Its decay rate is given by\cite{LINDE198137,LINDE1983421}
\be\label{eq:gammaft}
\Gamma(T)\approx T^4\left(\frac{S_3(T)}{2\pi T}\right)^{\frac{3}{2}}e^{-S_3(T)/T}\,.
\ee
In \refeq{eq:gammaft} $S_3(T)$ indicates the three-dimensional thermal bounce action along the tunneling path, which we calculate with \texttt{CosmoTransitions}\cite{Wainwright:2011kj}. One can now define the nucleation temperature, $T_n$, as the temperature at which at least one bubble per Hubble volume has nucleated\cite{PhysRevD.46.2384}, 
\be
N(T_n)=\int_{T_n}^{T_c}\frac{dT}{T}\frac{\Gamma(T)}{H(T)^4}=1\,.
\ee
The Hubble parameter $H(T)$ is defined, \textit{e.g.}, in Eq.~(4.10) of Ref.\cite{Kierkla:2022odc}.
The fraction of the true vacuum volume at a given temperature, $I(T)$, reads\cite{PhysRevD.23.876,PhysRevLett.44.631,Ellis:2018mja}
\be
I(T)=\frac{4\pi}{3}\int^{T_c}_{T}dT'\frac{\Gamma(T')}{T^{'4}H(T')}\left(\int_{T}^{T'} dT''\frac{1}{H(T'')}\right)^3\,.
\ee
As the number and the size of the true-vacuum bubbles increase, the bubble collisions start to take place and the GW signal is generated.  The percolation temperature is then defined as the temperature at which bubbles form an infinitely connected cluster, which corresponds to\cite{PhysRevD.46.2384,Rintoul_1997}
\be
I(T_p)=0.34\,.
\ee
It may happen that the
expansion rate of the Universe is faster that the growth of the bubbles. In such a case the percolation would never end and the phase transition would never be completed. To make sure that the entire false vacuum decays, its decay rate must exceed the expansion rate of the Universe. The corresponding condition for successful completion of the phase transition reads\cite{Ellis:2018mja,Kierkla:2022odc}
\be\label{eq:tnstop}
3+T\frac{dI(T)}{dT}\Big|_{T=T_p}<0\,.
\ee

The production of GWs from the phase transition occurs around the percolation temperature and all the parameters are evaluated at that temperature. The strength of the signal is determined by the 
latent heat released during the transition\cite{Ellis:2018mja}, 
\be
\alpha=\frac{\Delta V(T)-T\frac{\partial \Delta V(T)}{\partial T}}{\rho_R(T)}\Big|_{T_p}\,,
\ee
where $\Delta V$ is the effective potential difference between the false and the true vacuum, while $\rho_R$ denotes the density of radiation given by $\rho_R(T)=g_\ast T^4 \pi^2/30$, with $g_\ast$ being the number of degrees of freedom in the plasma.\footnote{See also Refs.\cite{Giese:2020rtr,Giese:2020znk} for a more rigorous definition of $\alpha$.} Another important quantity is the inverse time scale of the transition (in other words, the nucleation speed),
\be
\frac{\beta}{H_\ast}=T_p\frac{{\rm d}(S_3/T)}{{\rm d} T}\Big|_{T_p}\,.
\ee
After the phase transition ends, the energy stored in the vacuum is immediately turned into radiation. Reheating thus takes place, characterized by the temperature\cite{Ellis:2018mja,Eichhorn:2020upj}
\be
T_{\textrm{rh}}=T_p[1+\alpha(T_p)]^{1/4}\,.
\ee

The stochastic gravitational wave background produced at the time of the FOPT can originate from collisions of bubble walls ($\Omega_{\textrm{coll}}$), sound waves in the plasma ($\Omega_{\textrm{sw}}$), and magnetohydrodynamic turbulence in the plasma ($\Omega_{\textrm{turb}}$). The total amplitude of the  redshifted signal observed today is then given by
\be
h^2\Omega_{\rm{GW}} =h^2\Omega_{\textrm{coll}} +h^2\Omega_{\textrm{sw}} +h^2\Omega_{\textrm{turb}}\,,
\ee
where $h=H_0/(100\,\rm{km/s/Mpc})$ is the present-day value of the dimensionless Hubble parameter. The corresponding GW spectra are given in terms of the peak amplitudes $\Omega^{\rm{peak}}$ and the peak frequencies $f^{\rm{peak}}$ as\cite{PhysRevD.45.4514,PhysRevLett.69.2026,Kosowsky:1992vn,Kamionkowski:1993fg,Caprini:2007xq,Gogoberidze:2007an,Huber:2008hg,Kahniashvili:2008pe,Kahniashvili:2009mf,Caprini:2009fx,Caprini:2009yp,Espinosa:2010hh,Hindmarsh:2013xza,Giblin:2014qia,Kalaydzhyan:2014wca,Hindmarsh:2015qta,Caprini:2015zlo,Hindmarsh:2016lnk,Jaeckel:2016jlh,Hindmarsh:2017gnf}
\bea\label{eq:gwampli}
h^2\Omega_{\textrm{coll}}(f)&=&h^2\Omega_{\textrm{coll}}^{\rm{peak}}\left(\frac{f}{f^{\rm{peak}}_{\textrm{coll}}}\right)^{2.8}\left(\frac{3.8}{1+2.8\,(f/f^{\rm{peak}}_{\textrm{coll}})^{3.8}}\right)\nonumber\\
h^2\Omega_{\textrm{sw}}(f)&=&h^2\Omega_{\textrm{sw}}^{\rm{peak}}\left(\frac{f}{f^{\rm{peak}}_{\textrm{sw}}}\right)^{3}\left(\frac{7}{4+3\,(f/f^{\rm{peak}}_{\textrm{sw}})^{2}}\right)^{7/2}\nonumber\\
h^2\Omega_{\textrm{turb}}(f)&=&h^2\Omega_{\textrm{turb}}^{\rm{peak}}\left(\frac{f}{f^{\rm{peak}}_{\textrm{turb}}}\right)^{3}\left(\frac{1}{1+(f/f^{\rm{peak}}_{\textrm{turb}})}\right)^{11/3}\frac{1}{1+8\pi fa_0/a_\ast H_\ast}\,.
\eea
The red-shifted Hubble parameter reads 
\be\label{eq:hubp}
\frac{a_\ast}{a_0}H_\ast=1.65\times 10^{-5}\,\rm{Hz}\,\left(\frac{g_\ast}{100}\right)^{1/6}\left(\frac{T_\ast}{100\gev}\right).
\ee
In \refeq{eq:hubp}, $T_\ast$ is the temperature of the plasma at the time of the GW  production, after the transition has completed and reheating has taken place, which will be identified with the reheating temperature, $T_\ast=T_{\rm rh}$. 
The peak amplitude for each source of GW is given by
\bea\label{eq:gwpeak}
h^2\Omega_{\textrm{coll}}^{\rm{peak}}&=&1.67\times 10^{-5}\,\kappa^2_{\textrm{coll}}\left(\frac{\alpha}{1+\alpha}\right)^2\left(\frac{v_w}{\beta/H_\ast}\right)^2\left(\frac{100}{g_\ast}\right)^{1/3}\left(\frac{0.11v_w}{0.42+v^2_w}\right)\nonumber\\
h^2\Omega_{\textrm{sw}}^{\rm{peak}}&=&2.65\times 10^{-6}\,\kappa^2_{\textrm{sw}}\left(\frac{\alpha}{1+\alpha}\right)^2\left(\frac{v_w}{\beta/H_\ast}\right)\left(\frac{100}{g_\ast}\right)^{1/3}\nonumber\\
h^2\Omega_{\textrm{turb}}^{\rm{peak}}&=&3.35\times 10^{-4}\,\kappa^{3/2}_{\textrm{turb}}\left(\frac{\alpha}{1+\alpha}\right)^{3/2}\left(\frac{v_w}{\beta/H_\ast}\right)\left(\frac{100}{g_\ast}\right)^{1/3}\,,
\eea
where $v_w$ is the bubble wall velocity assumed to be equal to the speed of light, $v_w=1$ (the effects of a smaller wall velocity were discussed in Ref.\cite{Ai:2023see}). 
The parameters $\kappa_{\textrm{coll}}$,  $\kappa_{\textrm{sw}}$ and  $\kappa_{\textrm{turb}}$ are the efficiency factors that indicate the amount of the released vacuum energy converted into the energy of the bubble wall, sound waves and turbulence, respectively, defined in Ref.\cite{Ellis:2019oqb}.
Finally, the peak frequencies read
\bea
f^{\rm{peak}}_{\textrm{coll}}&=&1.65\times 10^{-5}\,\textrm{Hz}\left(\frac{v_w}{\beta/H_\ast}\right)^{-1}\left(\frac{100}{g_\ast}\right)^{-1/6}\left(\frac{T_\ast}{100\,\gev}\right)\left(\frac{0.62v_w}{1.81-0.1v_w+v^2_w}\right)\nonumber\\
f^{\rm{peak}}_{\textrm{sw}}&=&1.90\times 10^{-5}\,\textrm{Hz}\left(\frac{v_w}{\beta/H_\ast}\right)^{-1}\left(\frac{100}{g_\ast}\right)^{-1/6}\left(\frac{T_\ast}{100\,\gev}\right)\nonumber\\
f^{\rm{peak}}_{\textrm{turb}}&=&2.70\times 10^{-5}\,\textrm{Hz}\left(\frac{v_w}{\beta/H_\ast}\right)^{-1}\left(\frac{100}{g_\ast}\right)^{-1/6}\left(\frac{T_\ast}{100\,\gev}\right)\,.
\eea

%%%%%%%%%%%%%%%%%%%%%%%%%%%%%%%%%%%%%%%%%%%%%%%%%%%%%%%%%%%%%%%%%%%%
\bibliographystyle{JHEP}
\bibliography{mybib}

\providecommand{\href}[2]{#2}\begingroup\raggedright\begin{thebibliography}{100}

\bibitem{Minkowski:1977sc}
P.~Minkowski, {\it {$\mu \to e\gamma$ at a Rate of One Out of $10^{9}$ Muon
  Decays?}},  {\em Phys. Lett. B} {\bf 67} (1977) 421--428.

\bibitem{Gell-Mann:1979vob}
M.~Gell-Mann, P.~Ramond, and R.~Slansky, {\it {Complex Spinors and Unified
  Theories}},  {\em Conf. Proc. C} {\bf 790927} (1979) 315--321,
  [\href{http://arxiv.org/abs/1306.4669}{{\tt arXiv:1306.4669}}].

\bibitem{Yanagida:1979as}
T.~Yanagida, {\it {Horizontal gauge symmetry and masses of neutrinos}},  {\em
  Conf. Proc. C} {\bf 7902131} (1979) 95--99.

\bibitem{Glashow:1979nm}
S.~L. Glashow, {\it {The Future of Elementary Particle Physics}},  {\em NATO
  Sci. Ser. B} {\bf 61} (1980) 687.

\bibitem{Mohapatra:1980yp}
R.~N. Mohapatra and G.~Senjanovic, {\it {Neutrino Masses and Mixings in Gauge
  Models with Spontaneous Parity Violation}},  {\em Phys. Rev. D} {\bf 23}
  (1981) 165.

\bibitem{Schechter:1981cv}
J.~Schechter and J.~W.~F. Valle, {\it {Neutrino Decay and Spontaneous Violation
  of Lepton Number}},  {\em Phys. Rev. D} {\bf 25} (1982) 774.

\bibitem{Schechter:1980gr}
J.~Schechter and J.~W.~F. Valle, {\it {Neutrino Masses in SU(2) x U(1)
  Theories}},  {\em Phys. Rev. D} {\bf 22} (1980) 2227.

\bibitem{Cai:2017jrq}
Y.~Cai, J.~Herrero-Garc\'\i{}a, M.~A. Schmidt, A.~Vicente, and R.~R. Volkas,
  {\it {From the trees to the forest: a review of radiative neutrino mass
  models}},  {\em Front. in Phys.} {\bf 5} (2017) 63,
  [\href{http://arxiv.org/abs/1706.08524}{{\tt arXiv:1706.08524}}].

\bibitem{Klein:2019iws}
C.~Klein, M.~Lindner, and S.~Ohmer, {\it {Minimal Radiative Neutrino Masses}},
  {\em JHEP} {\bf 03} (2019) 018, [\href{http://arxiv.org/abs/1901.03225}{{\tt
  arXiv:1901.03225}}].

\bibitem{Kowalska:2022ypk}
K.~Kowalska, S.~Pramanick, and E.~M. Sessolo, {\it {Naturally small Yukawa
  couplings from trans-Planckian asymptotic safety}},  {\em JHEP} {\bf 08}
  (2022) 262, [\href{http://arxiv.org/abs/2204.00866}{{\tt arXiv:2204.00866}}].

\bibitem{Eichhorn:2022vgp}
A.~Eichhorn and A.~Held, {\it {Dynamically vanishing Dirac neutrino mass from
  quantum scale symmetry}},  {\em Phys. Lett. B} {\bf 846} (2023) 138196,
  [\href{http://arxiv.org/abs/2204.09008}{{\tt arXiv:2204.09008}}].

\bibitem{Deppisch:2004fa}
F.~Deppisch and J.~W.~F. Valle, {\it {Enhanced lepton flavor violation in the
  supersymmetric inverse seesaw model}},  {\em Phys. Rev. D} {\bf 72} (2005)
  036001, [\href{http://arxiv.org/abs/hep-ph/0406040}{{\tt hep-ph/0406040}}].

\bibitem{Abada:2014vea}
A.~Abada and M.~Lucente, {\it {Looking for the minimal inverse seesaw
  realisation}},  {\em Nucl. Phys. B} {\bf 885} (2014) 651--678,
  [\href{http://arxiv.org/abs/1401.1507}{{\tt arXiv:1401.1507}}].

\bibitem{Lindner:2014oea}
M.~Lindner, S.~Schmidt, and J.~Smirnov, {\it {Neutrino Masses and Conformal
  Electro-Weak Symmetry Breaking}},  {\em JHEP} {\bf 10} (2014) 177,
  [\href{http://arxiv.org/abs/1405.6204}{{\tt arXiv:1405.6204}}].

\bibitem{inbookWS}
S.~Weinberg, {\em General Relativity}, pp.~790--831.
\newblock S.W.Hawking, W.Israel (Eds.), Cambridge Univ. Press, 1980.

\bibitem{WETTERICH199390}
C.~Wetterich, {\it Exact evolution equation for the effective potential},  {\em
  Physics Letters B} {\bf 301} (1993), no.~1 90 -- 94.

\bibitem{Morris:1993qb}
T.~R. Morris, {\it {The Exact renormalization group and approximate
  solutions}},  {\em Int. J. Mod. Phys. A} {\bf 9} (1994) 2411--2450,
  [\href{http://arxiv.org/abs/hep-ph/9308265}{{\tt hep-ph/9308265}}].

\bibitem{Reuter:1996cp}
M.~Reuter, {\it {Nonperturbative evolution equation for quantum gravity}},
  {\em Phys. Rev. D} {\bf 57} (1998) 971--985,
  [\href{http://arxiv.org/abs/hep-th/9605030}{{\tt hep-th/9605030}}].

\bibitem{Lauscher:2001ya}
O.~Lauscher and M.~Reuter, {\it {Ultraviolet fixed point and generalized flow
  equation of quantum gravity}},  {\em Phys. Rev. D} {\bf 65} (2002) 025013,
  [\href{http://arxiv.org/abs/hep-th/0108040}{{\tt hep-th/0108040}}].

\bibitem{Reuter:2001ag}
M.~Reuter and F.~Saueressig, {\it {Renormalization group flow of quantum
  gravity in the Einstein-Hilbert truncation}},  {\em Phys. Rev. D} {\bf 65}
  (2002) 065016, [\href{http://arxiv.org/abs/hep-th/0110054}{{\tt
  hep-th/0110054}}].

\bibitem{Lauscher:2002sq}
O.~Lauscher and M.~Reuter, {\it {Flow equation of quantum Einstein gravity in a
  higher derivative truncation}},  {\em Phys. Rev. D} {\bf 66} (2002) 025026,
  [\href{http://arxiv.org/abs/hep-th/0205062}{{\tt hep-th/0205062}}].

\bibitem{Litim:2003vp}
D.~F. Litim, {\it {Fixed points of quantum gravity}},  {\em Phys. Rev. Lett.}
  {\bf 92} (2004) 201301, [\href{http://arxiv.org/abs/hep-th/0312114}{{\tt
  hep-th/0312114}}].

\bibitem{Codello:2006in}
A.~Codello and R.~Percacci, {\it {Fixed points of higher derivative gravity}},
  {\em Phys. Rev. Lett.} {\bf 97} (2006) 221301,
  [\href{http://arxiv.org/abs/hep-th/0607128}{{\tt hep-th/0607128}}].

\bibitem{Machado:2007ea}
P.~F. Machado and F.~Saueressig, {\it {On the renormalization group flow of
  f(R)-gravity}},  {\em Phys. Rev. D} {\bf 77} (2008) 124045,
  [\href{http://arxiv.org/abs/0712.0445}{{\tt arXiv:0712.0445}}].

\bibitem{Codello:2008vh}
A.~Codello, R.~Percacci, and C.~Rahmede, {\it {Investigating the Ultraviolet
  Properties of Gravity with a Wilsonian Renormalization Group Equation}},
  {\em Annals Phys.} {\bf 324} (2009) 414--469,
  [\href{http://arxiv.org/abs/0805.2909}{{\tt arXiv:0805.2909}}].

\bibitem{Benedetti:2009rx}
D.~Benedetti, P.~F. Machado, and F.~Saueressig, {\it {Asymptotic safety in
  higher-derivative gravity}},  {\em Mod. Phys. Lett. A} {\bf 24} (2009)
  2233--2241, [\href{http://arxiv.org/abs/0901.2984}{{\tt arXiv:0901.2984}}].

\bibitem{Dietz:2012ic}
J.~A. Dietz and T.~R. Morris, {\it {Asymptotic safety in the f(R)
  approximation}},  {\em JHEP} {\bf 01} (2013) 108,
  [\href{http://arxiv.org/abs/1211.0955}{{\tt arXiv:1211.0955}}].

\bibitem{Falls:2013bv}
K.~Falls, D.~Litim, K.~Nikolakopoulos, and C.~Rahmede, {\it {A bootstrap
  towards asymptotic safety}},  \href{http://arxiv.org/abs/1301.4191}{{\tt
  arXiv:1301.4191}}.

\bibitem{Falls:2014tra}
K.~Falls, D.~F. Litim, K.~Nikolakopoulos, and C.~Rahmede, {\it {Further
  evidence for asymptotic safety of quantum gravity}},  {\em Phys. Rev. D} {\bf
  93} (2016), no.~10 104022, [\href{http://arxiv.org/abs/1410.4815}{{\tt
  arXiv:1410.4815}}].

\bibitem{Oda:2015sma}
K.-y. Oda and M.~Yamada, {\it {Non-minimal coupling in Higgs--Yukawa model with
  asymptotically safe gravity}},  {\em Class. Quant. Grav.} {\bf 33} (2016),
  no.~12 125011, [\href{http://arxiv.org/abs/1510.03734}{{\tt
  arXiv:1510.03734}}].

\bibitem{Hamada:2017rvn}
Y.~Hamada and M.~Yamada, {\it {Asymptotic safety of higher derivative quantum
  gravity non-minimally coupled with a matter system}},  {\em JHEP} {\bf 08}
  (2017) 070, [\href{http://arxiv.org/abs/1703.09033}{{\tt arXiv:1703.09033}}].

\bibitem{Christiansen:2017cxa}
N.~Christiansen, D.~F. Litim, J.~M. Pawlowski, and M.~Reichert, {\it
  {Asymptotic safety of gravity with matter}},  {\em Phys. Rev. D} {\bf 97}
  (2018), no.~10 106012, [\href{http://arxiv.org/abs/1710.04669}{{\tt
  arXiv:1710.04669}}].

\bibitem{Robinson:2005fj}
S.~P. Robinson and F.~Wilczek, {\it {Gravitational correction to running of
  gauge couplings}},  {\em Phys. Rev. Lett.} {\bf 96} (2006) 231601,
  [\href{http://arxiv.org/abs/hep-th/0509050}{{\tt hep-th/0509050}}].

\bibitem{Pietrykowski:2006xy}
A.~R. Pietrykowski, {\it {Gauge dependence of gravitational correction to
  running of gauge couplings}},  {\em Phys. Rev. Lett.} {\bf 98} (2007) 061801,
  [\href{http://arxiv.org/abs/hep-th/0606208}{{\tt hep-th/0606208}}].

\bibitem{Toms:2007sk}
D.~J. Toms, {\it {Quantum gravity and charge renormalization}},  {\em Phys.
  Rev. D} {\bf 76} (2007) 045015, [\href{http://arxiv.org/abs/0708.2990}{{\tt
  arXiv:0708.2990}}].

\bibitem{Tang:2008ah}
Y.~Tang and Y.-L. Wu, {\it {Gravitational Contributions to the Running of Gauge
  Couplings}},  {\em Commun. Theor. Phys.} {\bf 54} (2010) 1040--1044,
  [\href{http://arxiv.org/abs/0807.0331}{{\tt arXiv:0807.0331}}].

\bibitem{Toms:2008dq}
D.~J. Toms, {\it {Cosmological constant and quantum gravitational corrections
  to the running fine structure constant}},  {\em Phys. Rev. Lett.} {\bf 101}
  (2008) 131301, [\href{http://arxiv.org/abs/0809.3897}{{\tt
  arXiv:0809.3897}}].

\bibitem{Rodigast:2009zj}
A.~Rodigast and T.~Schuster, {\it {Gravitational Corrections to Yukawa and
  phi**4 Interactions}},  {\em Phys. Rev. Lett.} {\bf 104} (2010) 081301,
  [\href{http://arxiv.org/abs/0908.2422}{{\tt arXiv:0908.2422}}].

\bibitem{Zanusso:2009bs}
O.~Zanusso, L.~Zambelli, G.~Vacca, and R.~Percacci, {\it {Gravitational
  corrections to Yukawa systems}},  {\em Phys. Lett. B} {\bf 689} (2010)
  90--94, [\href{http://arxiv.org/abs/0904.0938}{{\tt arXiv:0904.0938}}].

\bibitem{Daum:2009dn}
J.-E. Daum, U.~Harst, and M.~Reuter, {\it {Running Gauge Coupling in
  Asymptotically Safe Quantum Gravity}},  {\em JHEP} {\bf 01} (2010) 084,
  [\href{http://arxiv.org/abs/0910.4938}{{\tt arXiv:0910.4938}}].

\bibitem{Daum:2010bc}
J.-E. Daum, U.~Harst, and M.~Reuter, {\it {Non-perturbative QEG Corrections to
  the Yang-Mills Beta Function}},  {\em Gen. Rel. Grav.} {\bf 43} (2011) 2393,
  [\href{http://arxiv.org/abs/1005.1488}{{\tt arXiv:1005.1488}}].

\bibitem{Folkerts:2011jz}
S.~Folkerts, D.~F. Litim, and J.~M. Pawlowski, {\it {Asymptotic freedom of
  Yang-Mills theory with gravity}},  {\em Phys. Lett. B} {\bf 709} (2012)
  234--241, [\href{http://arxiv.org/abs/1101.5552}{{\tt arXiv:1101.5552}}].

\bibitem{Eichhorn:2016esv}
A.~Eichhorn, A.~Held, and J.~M. Pawlowski, {\it {Quantum-gravity effects on a
  Higgs-Yukawa model}},  {\em Phys. Rev. D} {\bf 94} (2016), no.~10 104027,
  [\href{http://arxiv.org/abs/1604.02041}{{\tt arXiv:1604.02041}}].

\bibitem{Eichhorn:2017eht}
A.~Eichhorn and A.~Held, {\it {Viability of quantum-gravity induced ultraviolet
  completions for matter}},  {\em Phys. Rev. D} {\bf 96} (2017), no.~8 086025,
  [\href{http://arxiv.org/abs/1705.02342}{{\tt arXiv:1705.02342}}].

\bibitem{Harst:2011zx}
U.~Harst and M.~Reuter, {\it {QED coupled to QEG}},  {\em JHEP} {\bf 05} (2011)
  119, [\href{http://arxiv.org/abs/1101.6007}{{\tt arXiv:1101.6007}}].

\bibitem{Christiansen:2017gtg}
N.~Christiansen and A.~Eichhorn, {\it {An asymptotically safe solution to the
  U(1) triviality problem}},  {\em Phys. Lett. B} {\bf 770} (2017) 154--160,
  [\href{http://arxiv.org/abs/1702.07724}{{\tt arXiv:1702.07724}}].

\bibitem{Eichhorn:2017lry}
A.~Eichhorn and F.~Versteegen, {\it {Upper bound on the Abelian gauge coupling
  from asymptotic safety}},  {\em JHEP} {\bf 01} (2018) 030,
  [\href{http://arxiv.org/abs/1709.07252}{{\tt arXiv:1709.07252}}].

\bibitem{Shaposhnikov:2009pv}
M.~Shaposhnikov and C.~Wetterich, {\it {Asymptotic safety of gravity and the
  Higgs boson mass}},  {\em Phys. Lett. B} {\bf 683} (2010) 196--200,
  [\href{http://arxiv.org/abs/0912.0208}{{\tt arXiv:0912.0208}}].

\bibitem{Eichhorn:2017als}
A.~Eichhorn, Y.~Hamada, J.~Lumma, and M.~Yamada, {\it {Quantum gravity
  fluctuations flatten the Planck-scale Higgs potential}},  {\em Phys. Rev. D}
  {\bf 97} (2018), no.~8 086004, [\href{http://arxiv.org/abs/1712.00319}{{\tt
  arXiv:1712.00319}}].

\bibitem{Kwapisz:2019wrl}
J.~H. Kwapisz, {\it {Asymptotic safety, the Higgs boson mass, and beyond the
  standard model physics}},  {\em Phys. Rev. D} {\bf 100} (2019), no.~11
  115001, [\href{http://arxiv.org/abs/1907.12521}{{\tt arXiv:1907.12521}}].

\bibitem{Eichhorn:2021tsx}
A.~Eichhorn, M.~Pauly, and S.~Ray, {\it {Towards a Higgs mass determination in
  asymptotically safe gravity with a dark portal}},  {\em JHEP} {\bf 10} (2021)
  100, [\href{http://arxiv.org/abs/2107.07949}{{\tt arXiv:2107.07949}}].

\bibitem{Eichhorn:2017ylw}
A.~Eichhorn and A.~Held, {\it {Top mass from asymptotic safety}},  {\em Phys.
  Lett. B} {\bf 777} (2018) 217--221,
  [\href{http://arxiv.org/abs/1707.01107}{{\tt arXiv:1707.01107}}].

\bibitem{Jenkins:1987ue}
E.~E. Jenkins, {\it {Searching for a ($B^-$l) Gauge Boson in $p \bar{p}$
  Collisions}},  {\em Phys. Lett. B} {\bf 192} (1987) 219--222.

\bibitem{Buchmuller:1991ce}
W.~Buchmuller, C.~Greub, and P.~Minkowski, {\it {Neutrino masses, neutral
  vector bosons and the scale of B-L breaking}},  {\em Phys. Lett. B} {\bf 267}
  (1991) 395--399.

\bibitem{Falls:2010he}
K.~Falls, D.~F. Litim, and A.~Raghuraman, {\it {Black Holes and Asymptotically
  Safe Gravity}},  {\em Int. J. Mod. Phys. A} {\bf 27} (2012) 1250019,
  [\href{http://arxiv.org/abs/1002.0260}{{\tt arXiv:1002.0260}}].

\bibitem{Aharonov:1987tp}
Y.~Aharonov, A.~Casher, and S.~Nussinov, {\it {The Unitarity Puzzle and Planck
  Mass Stable Particles}},  {\em Phys. Lett. B} {\bf 191} (1987) 51.

\bibitem{Banks:1992ba}
T.~Banks, A.~Dabholkar, M.~R. Douglas, and M.~O'Loughlin, {\it {Are horned
  particles the climax of Hawking evaporation?}},  {\em Phys. Rev. D} {\bf 45}
  (1992) 3607--3616, [\href{http://arxiv.org/abs/hep-th/9201061}{{\tt
  hep-th/9201061}}].

\bibitem{PhysRevD.7.1888}
S.~Coleman and E.~Weinberg, {\it Radiative corrections as the origin of
  spontaneous symmetry breaking},  {\em Phys. Rev. D} {\bf 7} (Mar, 1973)
  1888--1910.

\bibitem{PhysRevD.45.4514}
A.~Kosowsky, M.~S. Turner, and R.~Watkins, {\it Gravitational radiation from
  colliding vacuum bubbles},  {\em Phys. Rev. D} {\bf 45} (Jun, 1992)
  4514--4535.

\bibitem{PhysRevLett.69.2026}
A.~Kosowsky, M.~S. Turner, and R.~Watkins, {\it Gravitational waves from
  first-order cosmological phase transitions},  {\em Phys. Rev. Lett.} {\bf 69}
  (Oct, 1992) 2026--2029.

\bibitem{Kosowsky:1992vn}
A.~Kosowsky and M.~S. Turner, {\it {Gravitational radiation from colliding
  vacuum bubbles: envelope approximation to many bubble collisions}},  {\em
  Phys. Rev. D} {\bf 47} (1993) 4372--4391,
  [\href{http://arxiv.org/abs/astro-ph/9211004}{{\tt astro-ph/9211004}}].

\bibitem{Kamionkowski:1993fg}
M.~Kamionkowski, A.~Kosowsky, and M.~S. Turner, {\it {Gravitational radiation
  from first order phase transitions}},  {\em Phys. Rev. D} {\bf 49} (1994)
  2837--2851, [\href{http://arxiv.org/abs/astro-ph/9310044}{{\tt
  astro-ph/9310044}}].

\bibitem{Athron:2023xlk}
P.~Athron, C.~Bal\'azs, A.~Fowlie, L.~Morris, and L.~Wu, {\it {Cosmological
  phase transitions: from perturbative particle physics to gravitational
  waves}},  \href{http://arxiv.org/abs/2305.02357}{{\tt arXiv:2305.02357}}.

\bibitem{Percacci:2002ie}
R.~Percacci and D.~Perini, {\it {Constraints on matter from asymptotic
  safety}},  {\em Phys. Rev. D} {\bf 67} (2003) 081503,
  [\href{http://arxiv.org/abs/hep-th/0207033}{{\tt hep-th/0207033}}].

\bibitem{Percacci:2003jz}
R.~Percacci and D.~Perini, {\it {Asymptotic safety of gravity coupled to
  matter}},  {\em Phys. Rev. D} {\bf 68} (2003) 044018,
  [\href{http://arxiv.org/abs/hep-th/0304222}{{\tt hep-th/0304222}}].

\bibitem{Codello:2007bd}
A.~Codello, R.~Percacci, and C.~Rahmede, {\it {Ultraviolet properties of
  f(R)-gravity}},  {\em Int. J. Mod. Phys. A} {\bf 23} (2008) 143--150,
  [\href{http://arxiv.org/abs/0705.1769}{{\tt arXiv:0705.1769}}].

\bibitem{Narain:2009qa}
G.~Narain and R.~Percacci, {\it {On the scheme dependence of gravitational beta
  functions}},  {\em Acta Phys. Polon. B} {\bf 40} (2009) 3439--3457,
  [\href{http://arxiv.org/abs/0910.5390}{{\tt arXiv:0910.5390}}].

\bibitem{Dona:2013qba}
P.~Donà, A.~Eichhorn, and R.~Percacci, {\it {Matter matters in asymptotically
  safe quantum gravity}},  {\em Phys. Rev. D} {\bf 89} (2014), no.~8 084035,
  [\href{http://arxiv.org/abs/1311.2898}{{\tt arXiv:1311.2898}}].

\bibitem{Falls:2017lst}
K.~Falls, C.~R. King, D.~F. Litim, K.~Nikolakopoulos, and C.~Rahmede, {\it
  {Asymptotic safety of quantum gravity beyond Ricci scalars}},  {\em Phys.
  Rev. D} {\bf 97} (2018), no.~8 086006,
  [\href{http://arxiv.org/abs/1801.00162}{{\tt arXiv:1801.00162}}].

\bibitem{Falls:2018ylp}
K.~G. Falls, D.~F. Litim, and J.~Schröder, {\it {Aspects of asymptotic safety
  for quantum gravity}},  {\em Phys. Rev. D} {\bf 99} (2019), no.~12 126015,
  [\href{http://arxiv.org/abs/1810.08550}{{\tt arXiv:1810.08550}}].

\bibitem{Pastor-Gutierrez:2022nki}
A.~Pastor-Guti\'errez, J.~M. Pawlowski, and M.~Reichert, {\it {The
  Asymptotically Safe Standard Model: From quantum gravity to dynamical chiral
  symmetry breaking}},  {\em SciPost Phys.} {\bf 15} (2023) 105,
  [\href{http://arxiv.org/abs/2207.09817}{{\tt arXiv:2207.09817}}].

\bibitem{Kotlarski:2023mmr}
W.~Kotlarski, K.~Kowalska, D.~Rizzo, and E.~M. Sessolo, {\it {How robust are
  particle physics predictions in asymptotic safety?}},  {\em Eur. Phys. J. C}
  {\bf 83} (2023), no.~7 644, [\href{http://arxiv.org/abs/2304.08959}{{\tt
  arXiv:2304.08959}}].

\bibitem{Eichhorn:2020kca}
A.~Eichhorn and M.~Pauly, {\it {Safety in darkness: Higgs portal to simple
  Yukawa systems}},  {\em Phys. Lett. B} {\bf 819} (2021) 136455,
  [\href{http://arxiv.org/abs/2005.03661}{{\tt arXiv:2005.03661}}].

\bibitem{Eichhorn:2018whv}
A.~Eichhorn and A.~Held, {\it {Mass difference for charged quarks from
  asymptotically safe quantum gravity}},  {\em Phys. Rev. Lett.} {\bf 121}
  (2018), no.~15 151302, [\href{http://arxiv.org/abs/1803.04027}{{\tt
  arXiv:1803.04027}}].

\bibitem{Alkofer:2020vtb}
R.~Alkofer, A.~Eichhorn, A.~Held, C.~M. Nieto, R.~Percacci, and M.~Schr\"ofl,
  {\it {Quark masses and mixings in minimally parameterized UV completions of
  the Standard Model}},  {\em Annals Phys.} {\bf 421} (2020) 168282,
  [\href{http://arxiv.org/abs/2003.08401}{{\tt arXiv:2003.08401}}].

\bibitem{DeBrito:2019rrh}
G.~P. De~Brito, Y.~Hamada, A.~D. Pereira, and M.~Yamada, {\it {On the impact of
  Majorana masses in gravity-matter systems}},  {\em JHEP} {\bf 08} (2019) 142,
  [\href{http://arxiv.org/abs/1905.11114}{{\tt arXiv:1905.11114}}].

\bibitem{Hamada:2020vnf}
Y.~Hamada, K.~Tsumura, and M.~Yamada, {\it {Scalegenesis and fermionic dark
  matters in the flatland scenario}},  {\em Eur. Phys. J. C} {\bf 80} (2020),
  no.~5 368, [\href{http://arxiv.org/abs/2002.03666}{{\tt arXiv:2002.03666}}].

\bibitem{Domenech:2020yjf}
G.~Dom\`enech, M.~Goodsell, and C.~Wetterich, {\it {Neutrino masses, vacuum
  stability and quantum gravity prediction for the mass of the top quark}},
  {\em JHEP} {\bf 01} (2021) 180, [\href{http://arxiv.org/abs/2008.04310}{{\tt
  arXiv:2008.04310}}].

\bibitem{Grabowski:2018fjj}
F.~Grabowski, J.~H. Kwapisz, and K.~A. Meissner, {\it {Asymptotic safety and
  Conformal Standard Model}},  {\em Phys. Rev. D} {\bf 99} (2019), no.~11
  115029, [\href{http://arxiv.org/abs/1810.08461}{{\tt arXiv:1810.08461}}].

\bibitem{Kowalska:2020gie}
K.~Kowalska, E.~M. Sessolo, and Y.~Yamamoto, {\it {Flavor anomalies from
  asymptotically safe gravity}},  {\em Eur. Phys. J. C} {\bf 81} (2021), no.~4
  272, [\href{http://arxiv.org/abs/2007.03567}{{\tt arXiv:2007.03567}}].

\bibitem{Chikkaballi:2022urc}
A.~Chikkaballi, W.~Kotlarski, K.~Kowalska, D.~Rizzo, and E.~M. Sessolo, {\it
  {Constraints on Z' solutions to the flavor anomalies with trans-Planckian
  asymptotic safety}},  {\em JHEP} {\bf 01} (2023) 164,
  [\href{http://arxiv.org/abs/2209.07971}{{\tt arXiv:2209.07971}}].

\bibitem{Kowalska:2020zve}
K.~Kowalska and E.~M. Sessolo, {\it {Minimal models for g-2 and dark matter
  confront asymptotic safety}},  {\em Phys. Rev. D} {\bf 103} (2021), no.~11
  115032, [\href{http://arxiv.org/abs/2012.15200}{{\tt arXiv:2012.15200}}].

\bibitem{Reichert:2019car}
M.~Reichert and J.~Smirnov, {\it {Dark Matter meets Quantum Gravity}},  {\em
  Phys. Rev. D} {\bf 101} (2020), no.~6 063015,
  [\href{http://arxiv.org/abs/1911.00012}{{\tt arXiv:1911.00012}}].

\bibitem{Boos:2022jvc}
J.~Boos, C.~D. Carone, N.~L. Donald, and M.~R. Musser, {\it {Asymptotic safety
  and gauged baryon number}},  {\em Phys. Rev. D} {\bf 106} (2022), no.~3
  035015, [\href{http://arxiv.org/abs/2206.02686}{{\tt arXiv:2206.02686}}].

\bibitem{Boos:2022pyq}
J.~Boos, C.~D. Carone, N.~L. Donald, and M.~R. Musser, {\it {Asymptotically
  safe dark matter with gauged baryon number}},  {\em Phys. Rev. D} {\bf 107}
  (2023), no.~3 035018, [\href{http://arxiv.org/abs/2209.14268}{{\tt
  arXiv:2209.14268}}].

\bibitem{deBrito:2021akp}
G.~P. de~Brito, A.~Eichhorn, and R.~R. Lino~dos Santos, {\it {Are there ALPs in
  the asymptotically safe landscape?}},  {\em JHEP} {\bf 06} (2022) 013,
  [\href{http://arxiv.org/abs/2112.08972}{{\tt arXiv:2112.08972}}].

\bibitem{Eichhorn:2023gat}
A.~Eichhorn, R.~R. Lino~dos Santos, and J.~a.~L. Miqueleto, {\it {From quantum
  gravity to gravitational waves through cosmic strings}},
  \href{http://arxiv.org/abs/2306.17718}{{\tt arXiv:2306.17718}}.

\bibitem{Coriano:2015sea}
C.~Coriano, L.~Delle~Rose, and C.~Marzo, {\it {Constraints on abelian
  extensions of the Standard Model from two-loop vacuum stability and
  $U(1)_{B-L}$}},  {\em JHEP} {\bf 02} (2016) 135,
  [\href{http://arxiv.org/abs/1510.02379}{{\tt arXiv:1510.02379}}].

\bibitem{Lyonnet:2016xiz}
F.~Lyonnet and I.~Schienbein, {\it {PyR@TE 2: A Python tool for computing RGEs
  at two-loop}},  {\em Comput. Phys. Commun.} {\bf 213} (2017) 181--196,
  [\href{http://arxiv.org/abs/1608.07274}{{\tt arXiv:1608.07274}}].

\bibitem{Holdom:1985ag}
B.~Holdom, {\it {Two U(1)'s and Epsilon Charge Shifts}},  {\em Phys. Lett. B}
  {\bf 166} (1986) 196--198.

\bibitem{Babu:1996vt}
K.~S. Babu, C.~F. Kolda, and J.~March-Russell, {\it {Leptophobic U(1) $s$ and
  the R($b$) - R($c$) crisis}},  {\em Phys. Rev. D} {\bf 54} (1996) 4635--4647,
  [\href{http://arxiv.org/abs/hep-ph/9603212}{{\tt hep-ph/9603212}}].

\bibitem{ATLAS:2019erb}
{\bf ATLAS} Collaboration, G.~Aad et~al., {\it {Search for high-mass dilepton
  resonances using 139 fb$^{-1}$ of $pp$ collision data collected at
  $\sqrt{s}=$13 TeV with the ATLAS detector}},  {\em Phys. Lett. B} {\bf 796}
  (2019) 68--87, [\href{http://arxiv.org/abs/1903.06248}{{\tt
  arXiv:1903.06248}}].

\bibitem{CMS:2021ctt}
{\bf CMS} Collaboration, A.~M. Sirunyan et~al., {\it {Search for resonant and
  nonresonant new phenomena in high-mass dilepton final states at $ \sqrt{s} $
  = 13 TeV}},  {\em JHEP} {\bf 07} (2021) 208,
  [\href{http://arxiv.org/abs/2103.02708}{{\tt arXiv:2103.02708}}].

\bibitem{Workman:2022ynf}
{\bf Particle Data Group} Collaboration, R.~L. Workman and Others, {\it {Review
  of Particle Physics}},  {\em PTEP} {\bf 2022} (2022) 083C01.

\bibitem{Hempfling:1996ht}
R.~Hempfling, {\it {The Next-to-minimal Coleman-Weinberg model}},  {\em Phys.
  Lett. B} {\bf 379} (1996) 153--158,
  [\href{http://arxiv.org/abs/hep-ph/9604278}{{\tt hep-ph/9604278}}].

\bibitem{Sher:1996ib}
M.~Sher, {\it {The Coleman-Weinberg phase transition in extended Higgs
  models}},  {\em Phys. Rev. D} {\bf 54} (1996) 7071--7074,
  [\href{http://arxiv.org/abs/hep-ph/9607337}{{\tt hep-ph/9607337}}].

\bibitem{Nishino:2004kb}
H.~Nishino and S.~Rajpoot, {\it {Broken scale invariance in the standard
  model}},  \href{http://arxiv.org/abs/hep-th/0403039}{{\tt hep-th/0403039}}.

\bibitem{Meissner:2006zh}
K.~A. Meissner and H.~Nicolai, {\it {Conformal Symmetry and the Standard
  Model}},  {\em Phys. Lett. B} {\bf 648} (2007) 312--317,
  [\href{http://arxiv.org/abs/hep-th/0612165}{{\tt hep-th/0612165}}].

\bibitem{Iso:2009ss}
S.~Iso, N.~Okada, and Y.~Orikasa, {\it {Classically conformal $B^-$ L extended
  Standard Model}},  {\em Phys. Lett. B} {\bf 676} (2009) 81--87,
  [\href{http://arxiv.org/abs/0902.4050}{{\tt arXiv:0902.4050}}].

\bibitem{Wetterich:2016uxm}
C.~Wetterich and M.~Yamada, {\it {Gauge hierarchy problem in asymptotically
  safe gravity--the resurgence mechanism}},  {\em Phys. Lett. B} {\bf 770}
  (2017) 268--271, [\href{http://arxiv.org/abs/1612.03069}{{\tt
  arXiv:1612.03069}}].

\bibitem{Pawlowski:2018ixd}
J.~M. Pawlowski, M.~Reichert, C.~Wetterich, and M.~Yamada, {\it {Higgs scalar
  potential in asymptotically safe quantum gravity}},  {\em Phys. Rev. D} {\bf
  99} (2019), no.~8 086010, [\href{http://arxiv.org/abs/1811.11706}{{\tt
  arXiv:1811.11706}}].

\bibitem{Wetterich:2019zdo}
C.~Wetterich and M.~Yamada, {\it {Variable Planck mass from the gauge invariant
  flow equation}},  {\em Phys. Rev. D} {\bf 100} (2019), no.~6 066017,
  [\href{http://arxiv.org/abs/1906.01721}{{\tt arXiv:1906.01721}}].

\bibitem{Jinno:2016knw}
R.~Jinno and M.~Takimoto, {\it {Probing a classically conformal B-L model with
  gravitational waves}},  {\em Phys. Rev. D} {\bf 95} (2017), no.~1 015020,
  [\href{http://arxiv.org/abs/1604.05035}{{\tt arXiv:1604.05035}}].

\bibitem{Chao:2017ilw}
W.~Chao, W.-F. Cui, H.-K. Guo, and J.~Shu, {\it {Gravitational wave imprint of
  new symmetry breaking}},  {\em Chin. Phys. C} {\bf 44} (2020), no.~12 123102,
  [\href{http://arxiv.org/abs/1707.09759}{{\tt arXiv:1707.09759}}].

\bibitem{Okada:2018xdh}
N.~Okada and O.~Seto, {\it {Probing the seesaw scale with gravitational
  waves}},  {\em Phys. Rev. D} {\bf 98} (2018), no.~6 063532,
  [\href{http://arxiv.org/abs/1807.00336}{{\tt arXiv:1807.00336}}].

\bibitem{Marzo:2018nov}
C.~Marzo, L.~Marzola, and V.~Vaskonen, {\it {Phase transition and vacuum
  stability in the classically conformal B\textendash{}L model}},  {\em Eur.
  Phys. J. C} {\bf 79} (2019), no.~7 601,
  [\href{http://arxiv.org/abs/1811.11169}{{\tt arXiv:1811.11169}}].

\bibitem{Brdar:2018num}
V.~Brdar, A.~J. Helmboldt, and J.~Kubo, {\it {Gravitational Waves from
  First-Order Phase Transitions: LIGO as a Window to Unexplored Seesaw
  Scales}},  {\em JCAP} {\bf 02} (2019) 021,
  [\href{http://arxiv.org/abs/1810.12306}{{\tt arXiv:1810.12306}}].

\bibitem{Hasegawa:2019amx}
T.~Hasegawa, N.~Okada, and O.~Seto, {\it {Gravitational waves from the minimal
  gauged $U(1)_{B-L}$ model}},  {\em Phys. Rev. D} {\bf 99} (2019), no.~9
  095039, [\href{http://arxiv.org/abs/1904.03020}{{\tt arXiv:1904.03020}}].

\bibitem{PhysRevD.9.3357}
S.~Weinberg, {\it Gauge and global symmetries at high temperature},  {\em Phys.
  Rev. D} {\bf 9} (Jun, 1974) 3357--3378.

\bibitem{PhysRevD.9.3320}
L.~Dolan and R.~Jackiw, {\it Symmetry behavior at finite temperature},  {\em
  Phys. Rev. D} {\bf 9} (Jun, 1974) 3320--3341.

\bibitem{LINDE198137}
A.~Linde, {\it Fate of the false vacuum at finite temperature: Theory and
  applications},  {\em Physics Letters B} {\bf 100} (1981), no.~1 37--40.

\bibitem{LINDE1983421}
A.~Linde, {\it Decay of the false vacuum at finite temperature},  {\em Nuclear
  Physics B} {\bf 216} (1983), no.~2 421--445.

\bibitem{Hambye:2013dgv}
T.~Hambye and A.~Strumia, {\it {Dynamical generation of the weak and Dark
  Matter scale}},  {\em Phys. Rev. D} {\bf 88} (2013) 055022,
  [\href{http://arxiv.org/abs/1306.2329}{{\tt arXiv:1306.2329}}].

\bibitem{Hashino:2016rvx}
K.~Hashino, M.~Kakizaki, S.~Kanemura, and T.~Matsui, {\it {Synergy between
  measurements of gravitational waves and the triple-Higgs coupling in probing
  the first-order electroweak phase transition}},  {\em Phys. Rev. D} {\bf 94}
  (2016), no.~1 015005, [\href{http://arxiv.org/abs/1604.02069}{{\tt
  arXiv:1604.02069}}].

\bibitem{Marzola:2017jzl}
L.~Marzola, A.~Racioppi, and V.~Vaskonen, {\it {Phase transition and
  gravitational wave phenomenology of scalar conformal extensions of the
  Standard Model}},  {\em Eur. Phys. J. C} {\bf 77} (2017), no.~7 484,
  [\href{http://arxiv.org/abs/1704.01034}{{\tt arXiv:1704.01034}}].

\bibitem{Kang:2020jeg}
Z.~Kang and J.~Zhu, {\it {Scale-genesis by Dark Matter and Its Gravitational
  Wave Signal}},  {\em Phys. Rev. D} {\bf 102} (2020), no.~5 053011,
  [\href{http://arxiv.org/abs/2003.02465}{{\tt arXiv:2003.02465}}].

\bibitem{Dasgupta:2022isg}
A.~Dasgupta, P.~S.~B. Dev, A.~Ghoshal, and A.~Mazumdar, {\it {Gravitational
  wave pathway to testable leptogenesis}},  {\em Phys. Rev. D} {\bf 106}
  (2022), no.~7 075027, [\href{http://arxiv.org/abs/2206.07032}{{\tt
  arXiv:2206.07032}}].

\bibitem{Crowder:2005nr}
J.~Crowder and N.~J. Cornish, {\it {Beyond LISA: Exploring future gravitational
  wave missions}},  {\em Phys. Rev. D} {\bf 72} (2005) 083005,
  [\href{http://arxiv.org/abs/gr-qc/0506015}{{\tt gr-qc/0506015}}].

\bibitem{Corbin:2005ny}
V.~Corbin and N.~J. Cornish, {\it {Detecting the cosmic gravitational wave
  background with the big bang observer}},  {\em Class. Quant. Grav.} {\bf 23}
  (2006) 2435--2446, [\href{http://arxiv.org/abs/gr-qc/0512039}{{\tt
  gr-qc/0512039}}].

\bibitem{Reitze:2019iox}
D.~Reitze et~al., {\it {Cosmic Explorer: The U.S. Contribution to
  Gravitational-Wave Astronomy beyond LIGO}},  {\em Bull. Am. Astron. Soc.}
  {\bf 51} (2019), no.~7 035, [\href{http://arxiv.org/abs/1907.04833}{{\tt
  arXiv:1907.04833}}].

\bibitem{Seto:2001qf}
N.~Seto, S.~Kawamura, and T.~Nakamura, {\it {Possibility of direct measurement
  of the acceleration of the universe using 0.1-Hz band laser interferometer
  gravitational wave antenna in space}},  {\em Phys. Rev. Lett.} {\bf 87}
  (2001) 221103, [\href{http://arxiv.org/abs/astro-ph/0108011}{{\tt
  astro-ph/0108011}}].

\bibitem{Musha:2017usi}
{\bf DECIGO Working group} Collaboration, M.~Musha, {\it {Space gravitational
  wave detector DECIGO/pre-DECIGO}},  {\em Proc. SPIE Int. Soc. Opt. Eng.} {\bf
  10562} (2017) 105623T.

\bibitem{Punturo:2010zz}
M.~Punturo et~al., {\it {The Einstein Telescope: A third-generation
  gravitational wave observatory}},  {\em Class. Quant. Grav.} {\bf 27} (2010)
  194002.

\bibitem{Sathyaprakash:2012jk}
B.~Sathyaprakash et~al., {\it {Scientific Objectives of Einstein Telescope}},
  {\em Class. Quant. Grav.} {\bf 29} (2012) 124013,
  [\href{http://arxiv.org/abs/1206.0331}{{\tt arXiv:1206.0331}}]. [Erratum:
  Class.Quant.Grav. 30, 079501 (2013)].

\bibitem{Harry:2010zz}
{\bf LIGO Scientific} Collaboration, G.~M. Harry, {\it {Advanced LIGO: The next
  generation of gravitational wave detectors}},  {\em Class. Quant. Grav.} {\bf
  27} (2010) 084006.

\bibitem{LIGOScientific:2014pky}
{\bf LIGO Scientific} Collaboration, J.~Aasi et~al., {\it {Advanced LIGO}},
  {\em Class. Quant. Grav.} {\bf 32} (2015) 074001,
  [\href{http://arxiv.org/abs/1411.4547}{{\tt arXiv:1411.4547}}].

\bibitem{LIGOScientific:2016wof}
{\bf LIGO Scientific} Collaboration, B.~P. Abbott et~al., {\it {Exploring the
  Sensitivity of Next Generation Gravitational Wave Detectors}},  {\em Class.
  Quant. Grav.} {\bf 34} (2017), no.~4 044001,
  [\href{http://arxiv.org/abs/1607.08697}{{\tt arXiv:1607.08697}}].

\bibitem{LISA:2017pwj}
{\bf LISA} Collaboration, P.~Amaro-Seoane et~al., {\it {Laser Interferometer
  Space Antenna}},  \href{http://arxiv.org/abs/1702.00786}{{\tt
  arXiv:1702.00786}}.

\bibitem{Baker:2019nia}
J.~Baker et~al., {\it {The Laser Interferometer Space Antenna: Unveiling the
  Millihertz Gravitational Wave Sky}},
  \href{http://arxiv.org/abs/1907.06482}{{\tt arXiv:1907.06482}}.

\bibitem{Renzini:2019vmt}
A.~Renzini and C.~Contaldi, {\it {Improved limits on a stochastic
  gravitational-wave background and its anisotropies from Advanced LIGO O1 and
  O2 runs}},  {\em Phys. Rev. D} {\bf 100} (2019), no.~6 063527,
  [\href{http://arxiv.org/abs/1907.10329}{{\tt arXiv:1907.10329}}].

\bibitem{KAGRA:2021kbb}
{\bf KAGRA, Virgo, LIGO Scientific} Collaboration, R.~Abbott et~al., {\it
  {Upper limits on the isotropic gravitational-wave background from Advanced
  LIGO and Advanced Virgo\textquoteright{}s third observing run}},  {\em Phys.
  Rev. D} {\bf 104} (2021), no.~2 022004,
  [\href{http://arxiv.org/abs/2101.12130}{{\tt arXiv:2101.12130}}].

\bibitem{King:2023cgv}
S.~F. King, D.~Marfatia, and M.~H. Rahat, {\it {Towards distinguishing Dirac
  from Majorana neutrino mass with gravitational waves}},
  \href{http://arxiv.org/abs/2306.05389}{{\tt arXiv:2306.05389}}.

\bibitem{Poole:2019kcm}
C.~Poole and A.~E. Thomsen, {\it {Constraints on 3- and 4-loop
  $\beta$-functions in a general four-dimensional Quantum Field Theory}},  {\em
  JHEP} {\bf 09} (2019) 055, [\href{http://arxiv.org/abs/1906.04625}{{\tt
  arXiv:1906.04625}}].

\bibitem{Sartore:2020gou}
L.~Sartore and I.~Schienbein, {\it {PyR@TE 3}},  {\em Comput. Phys. Commun.}
  {\bf 261} (2021) 107819, [\href{http://arxiv.org/abs/2007.12700}{{\tt
  arXiv:2007.12700}}].

\bibitem{Quiros:1999jp}
M.~Quiros, {\it {Finite temperature field theory and phase transitions}},  in
  {\em {ICTP Summer School in High-Energy Physics and Cosmology}},
  pp.~187--259, 1, 1999.
\newblock \href{http://arxiv.org/abs/hep-ph/9901312}{{\tt hep-ph/9901312}}.

\bibitem{Quiros:1994dr}
M.~Quiros, {\it {Field theory at finite temperature and phase transitions}},
  {\em Helv. Phys. Acta} {\bf 67} (1994) 451--583.

\bibitem{Arnold:1992rz}
P.~B. Arnold and O.~Espinosa, {\it {The Effective potential and first order
  phase transitions: Beyond leading-order}},  {\em Phys. Rev. D} {\bf 47}
  (1993) 3546, [\href{http://arxiv.org/abs/hep-ph/9212235}{{\tt
  hep-ph/9212235}}]. [Erratum: Phys.Rev.D 50, 6662 (1994)].

\bibitem{Wainwright:2011kj}
C.~L. Wainwright, {\it {CosmoTransitions: Computing Cosmological Phase
  Transition Temperatures and Bubble Profiles with Multiple Fields}},  {\em
  Comput. Phys. Commun.} {\bf 183} (2012) 2006--2013,
  [\href{http://arxiv.org/abs/1109.4189}{{\tt arXiv:1109.4189}}].

\bibitem{PhysRevD.46.2384}
M.~S. Turner, E.~J. Weinberg, and L.~M. Widrow, {\it Bubble nucleation in
  first-order inflation and other cosmological phase transitions},  {\em Phys.
  Rev. D} {\bf 46} (Sep, 1992) 2384--2403.

\bibitem{Kierkla:2022odc}
M.~Kierkla, A.~Karam, and B.~Swiezewska, {\it {Conformal model for
  gravitational waves and dark matter: a status update}},  {\em JHEP} {\bf 03}
  (2023) 007, [\href{http://arxiv.org/abs/2210.07075}{{\tt arXiv:2210.07075}}].

\bibitem{PhysRevD.23.876}
A.~H. Guth and E.~J. Weinberg, {\it Cosmological consequences of a first-order
  phase transition in the s${\mathrm{u}}_{5}$ grand unified model},  {\em Phys.
  Rev. D} {\bf 23} (Feb, 1981) 876--885.

\bibitem{PhysRevLett.44.631}
A.~H. Guth and S.~H.~H. Tye, {\it Phase transitions and magnetic monopole
  production in the very early universe},  {\em Phys. Rev. Lett.} {\bf 44}
  (Mar, 1980) 631--635.

\bibitem{Ellis:2018mja}
J.~Ellis, M.~Lewicki, and J.~M. No, {\it {On the Maximal Strength of a
  First-Order Electroweak Phase Transition and its Gravitational Wave Signal}},
   {\em JCAP} {\bf 04} (2019) 003, [\href{http://arxiv.org/abs/1809.08242}{{\tt
  arXiv:1809.08242}}].

\bibitem{Rintoul_1997}
M.~D. Rintoul and S.~Torquato, {\it Precise determination of the critical
  threshold and exponents in a three-dimensional continuum percolation model},
  {\em Journal of Physics A: Mathematical and General} {\bf 30} (aug, 1997)
  L585.

\bibitem{Giese:2020rtr}
F.~Giese, T.~Konstandin, and J.~van~de Vis, {\it {Model-independent energy
  budget of cosmological first-order phase transitions\textemdash{}A sound
  argument to go beyond the bag model}},  {\em JCAP} {\bf 07} (2020), no.~07
  057, [\href{http://arxiv.org/abs/2004.06995}{{\tt arXiv:2004.06995}}].

\bibitem{Giese:2020znk}
F.~Giese, T.~Konstandin, K.~Schmitz, and J.~van~de Vis, {\it {Model-independent
  energy budget for LISA}},  {\em JCAP} {\bf 01} (2021) 072,
  [\href{http://arxiv.org/abs/2010.09744}{{\tt arXiv:2010.09744}}].

\bibitem{Eichhorn:2020upj}
A.~Eichhorn, J.~Lumma, J.~M. Pawlowski, M.~Reichert, and M.~Yamada, {\it
  {Universal gravitational-wave signatures from heavy new physics in the
  electroweak sector}},  {\em JCAP} {\bf 05} (2021) 006,
  [\href{http://arxiv.org/abs/2010.00017}{{\tt arXiv:2010.00017}}].

\bibitem{Caprini:2007xq}
C.~Caprini, R.~Durrer, and G.~Servant, {\it {Gravitational wave generation from
  bubble collisions in first-order phase transitions: An analytic approach}},
  {\em Phys. Rev. D} {\bf 77} (2008) 124015,
  [\href{http://arxiv.org/abs/0711.2593}{{\tt arXiv:0711.2593}}].

\bibitem{Gogoberidze:2007an}
G.~Gogoberidze, T.~Kahniashvili, and A.~Kosowsky, {\it {The Spectrum of
  Gravitational Radiation from Primordial Turbulence}},  {\em Phys. Rev. D}
  {\bf 76} (2007) 083002, [\href{http://arxiv.org/abs/0705.1733}{{\tt
  arXiv:0705.1733}}].

\bibitem{Huber:2008hg}
S.~J. Huber and T.~Konstandin, {\it {Gravitational Wave Production by
  Collisions: More Bubbles}},  {\em JCAP} {\bf 09} (2008) 022,
  [\href{http://arxiv.org/abs/0806.1828}{{\tt arXiv:0806.1828}}].

\bibitem{Kahniashvili:2008pe}
T.~Kahniashvili, L.~Campanelli, G.~Gogoberidze, Y.~Maravin, and B.~Ratra, {\it
  {Gravitational Radiation from Primordial Helical Inverse Cascade MHD
  Turbulence}},  {\em Phys. Rev. D} {\bf 78} (2008) 123006,
  [\href{http://arxiv.org/abs/0809.1899}{{\tt arXiv:0809.1899}}]. [Erratum:
  Phys.Rev.D 79, 109901 (2009)].

\bibitem{Kahniashvili:2009mf}
T.~Kahniashvili, L.~Kisslinger, and T.~Stevens, {\it {Gravitational Radiation
  Generated by Magnetic Fields in Cosmological Phase Transitions}},  {\em Phys.
  Rev. D} {\bf 81} (2010) 023004, [\href{http://arxiv.org/abs/0905.0643}{{\tt
  arXiv:0905.0643}}].

\bibitem{Caprini:2009fx}
C.~Caprini, R.~Durrer, T.~Konstandin, and G.~Servant, {\it {General Properties
  of the Gravitational Wave Spectrum from Phase Transitions}},  {\em Phys. Rev.
  D} {\bf 79} (2009) 083519, [\href{http://arxiv.org/abs/0901.1661}{{\tt
  arXiv:0901.1661}}].

\bibitem{Caprini:2009yp}
C.~Caprini, R.~Durrer, and G.~Servant, {\it {The stochastic gravitational wave
  background from turbulence and magnetic fields generated by a first-order
  phase transition}},  {\em JCAP} {\bf 12} (2009) 024,
  [\href{http://arxiv.org/abs/0909.0622}{{\tt arXiv:0909.0622}}].

\bibitem{Espinosa:2010hh}
J.~R. Espinosa, T.~Konstandin, J.~M. No, and G.~Servant, {\it {Energy Budget of
  Cosmological First-order Phase Transitions}},  {\em JCAP} {\bf 06} (2010)
  028, [\href{http://arxiv.org/abs/1004.4187}{{\tt arXiv:1004.4187}}].

\bibitem{Hindmarsh:2013xza}
M.~Hindmarsh, S.~J. Huber, K.~Rummukainen, and D.~J. Weir, {\it {Gravitational
  waves from the sound of a first order phase transition}},  {\em Phys. Rev.
  Lett.} {\bf 112} (2014) 041301, [\href{http://arxiv.org/abs/1304.2433}{{\tt
  arXiv:1304.2433}}].

\bibitem{Giblin:2014qia}
J.~T. Giblin and J.~B. Mertens, {\it {Gravitional radiation from first-order
  phase transitions in the presence of a fluid}},  {\em Phys. Rev. D} {\bf 90}
  (2014), no.~2 023532, [\href{http://arxiv.org/abs/1405.4005}{{\tt
  arXiv:1405.4005}}].

\bibitem{Kalaydzhyan:2014wca}
T.~Kalaydzhyan and E.~Shuryak, {\it {Gravity waves generated by sounds from big
  bang phase transitions}},  {\em Phys. Rev. D} {\bf 91} (2015), no.~8 083502,
  [\href{http://arxiv.org/abs/1412.5147}{{\tt arXiv:1412.5147}}].

\bibitem{Hindmarsh:2015qta}
M.~Hindmarsh, S.~J. Huber, K.~Rummukainen, and D.~J. Weir, {\it {Numerical
  simulations of acoustically generated gravitational waves at a first order
  phase transition}},  {\em Phys. Rev. D} {\bf 92} (2015), no.~12 123009,
  [\href{http://arxiv.org/abs/1504.03291}{{\tt arXiv:1504.03291}}].

\bibitem{Caprini:2015zlo}
C.~Caprini et~al., {\it {Science with the space-based interferometer eLISA. II:
  Gravitational waves from cosmological phase transitions}},  {\em JCAP} {\bf
  04} (2016) 001, [\href{http://arxiv.org/abs/1512.06239}{{\tt
  arXiv:1512.06239}}].

\bibitem{Hindmarsh:2016lnk}
M.~Hindmarsh, {\it {Sound shell model for acoustic gravitational wave
  production at a first-order phase transition in the early Universe}},  {\em
  Phys. Rev. Lett.} {\bf 120} (2018), no.~7 071301,
  [\href{http://arxiv.org/abs/1608.04735}{{\tt arXiv:1608.04735}}].

\bibitem{Jaeckel:2016jlh}
J.~Jaeckel, V.~V. Khoze, and M.~Spannowsky, {\it {Hearing the signal of dark
  sectors with gravitational wave detectors}},  {\em Phys. Rev. D} {\bf 94}
  (2016), no.~10 103519, [\href{http://arxiv.org/abs/1602.03901}{{\tt
  arXiv:1602.03901}}].

\bibitem{Hindmarsh:2017gnf}
M.~Hindmarsh, S.~J. Huber, K.~Rummukainen, and D.~J. Weir, {\it {Shape of the
  acoustic gravitational wave power spectrum from a first order phase
  transition}},  {\em Phys. Rev. D} {\bf 96} (2017), no.~10 103520,
  [\href{http://arxiv.org/abs/1704.05871}{{\tt arXiv:1704.05871}}]. [Erratum:
  Phys.Rev.D 101, 089902 (2020)].

\bibitem{Ai:2023see}
W.-Y. Ai, B.~Laurent, and J.~van~de Vis, {\it {Model-independent bubble wall
  velocities in local thermal equilibrium}},  {\em JCAP} {\bf 07} (2023) 002,
  [\href{http://arxiv.org/abs/2303.10171}{{\tt arXiv:2303.10171}}].

\bibitem{Ellis:2019oqb}
J.~Ellis, M.~Lewicki, J.~M. No, and V.~Vaskonen, {\it {Gravitational wave
  energy budget in strongly supercooled phase transitions}},  {\em JCAP} {\bf
  06} (2019) 024, [\href{http://arxiv.org/abs/1903.09642}{{\tt
  arXiv:1903.09642}}].

\end{thebibliography}\endgroup

\end{document}